\documentclass[a4paper,amsmath,amssymb,aps,bm,nofootinbib,prd,preprint,superscriptaddress,tightenlines]{revtex4-2}

\usepackage[mathscr]{euscript}
\usepackage[normalem]{ulem}
\usepackage{graphicx}
\usepackage{color}
\usepackage{xcolor}
\usepackage{tabularx}
\usepackage{slashed}
\usepackage[colorlinks=true,citecolor=blue,linkcolor=purple,urlcolor=purple]{hyperref}

\begin{document}
\baselineskip=17pt \parskip=3pt

\title{\boldmath$\Sigma^+\to p\ell^+\ell^-$ decays within the standard model and beyond}

\author{Arnab Roy}
\email{arnab.roy1@monash.edu}
\affiliation{School of Physics and Astronomy, Monash University, Wellington Road, Clayton, Victoria 3800, Australia}

\author{Jusak Tandean}
\email{jtandean@yahoo.com}
\affiliation{Institute of High Energy Physics, Chinese Academy of Sciences, Beijing 100049, China}

\author{German Valencia}
\email{german.valencia@monash.edu}
\affiliation{School of Physics and Astronomy, Monash University, Wellington Road, Clayton, Victoria 3800, Australia\bigskip}


\begin{abstract}

Motivated by the LHCb measurement of the hyperon decay mode $\Sigma^+\to p\mu^+\mu^-$ and prospects for improvement, we revisit the estimates for the rate and muon forward-backward asymmetry within the standard model and beyond.
The standard model prediction has a fourfold ambiguity, and we suggest ways to resolve it with other measurements, including possible studies of $\Sigma^+\to p e^+e^-$ in the BESIII and LHCb experiments.
We use the recent BESIII measurements of $\Sigma^+\to p \gamma$ and $\Sigma^+\to N\pi$ to reduce the uncertainty in the long-distance contribution to $\Sigma^+\to p\mu^+\mu^-$.
Beyond the standard model, we consider a general effective Hamiltonian at low energy with ten operators whose Wilson coefficients parametrize the new physics.
We derive expressions for the $\Sigma^+\to p\mu^+\mu^-$ rate and the associated muon forward-backward asymmetry in terms of these coefficients.
Finally, we present the constraints on these Wilson coefficients that result from both kaon and hyperon decays and emphasize their complementarity.

\end{abstract}


\maketitle

\newpage

\section{Introduction}

The LHCb Collaboration published a new measurement of the rate of the rare hyperon decay mode \,$\Sigma^+\to p\mu^+\mu^-$\, in 2018 \cite{LHCb:2017rdd} and is expected to present an improved analysis with more data soon. In combination with the renewed interest in future hyperon and kaon experiments \cite{Li:2016tlt,AlvesJunior:2018ldo,LHCb:2018roe,Cerri:2018ypt,Goudzovski:2022vbt,Anzivino:2023bhp}, this motivates us to revisit the theoretical status of this mode both within the standard model (SM) and beyond.

In the SM the decays \,$\Sigma^+\to p\ell^+\ell^-$\, with \,$\ell=e,\mu$\, each receive short-distance (SD) and long-distance (LD) contributions. The former are known to be significantly smaller than the latter~\cite{He:2005yn}.
We take another look at the SD component for completeness, as previous calculations have partly relied on the value of a Wilson coefficient that predates the discovery of the top quark~\cite{Shifman:1976de}.
We update this number with the formalism of Ref.\,\cite{Buchalla:1995vs}, arriving at a larger value than what has been quoted before, but the corresponding total SD amplitude is still substantially below its LD counterpart.

Although the LD contributions dominate the rate, they suffer from multiple uncertainties, which include a fourfold ambiguity in the prediction \cite{He:2005yn}.
Addressing this is one of the goals of this paper. Critical input to this examination is the measurement of the rate of the radiative channel \,$\Sigma^+\to p\gamma$,\, where there is a recent precise result from the BESIII experiment~\cite{BESIII:2023fhs}.
We employ this information, along with a new fit to the nonleptonic decays \,$\Sigma^+\to p\pi^0$\, and \,$\Sigma^+\to n\pi^+$\, prompted by other recent BESIII data~\cite{BESIII:2020fqg,BESIII:2023sgt}, to reduce the uncertainty in the LD form-factors that derives from experimental input.
The fourfold ambiguity in the SM estimation remains, and we propose possible measurements to resolve or alleviate this issue.
The theoretical uncertainty in the LD calculation is partially quantified by considering extraction of the form factors in relativistic baryon chiral perturbation theory and in its heavy-baryon formulation.

After the complete SM treatment of \,$\Sigma^+\to p\ell^+\ell^-$,\, we predict in the \,$\ell=\mu$\, case the rate, the muon forward-backward asymmetry, the dimuon spectrum, and the additional observable $f_L^{}$ in the angular distribution, as well as the \,$\ell=e$\, rate.
The results are presented for the four solutions of the form factors that can be used for comparison with future experimental analyses.

We also consider the constraints that \,$\Sigma^+\to p\mu^+\mu^-$\, can place on new physics (NP) parametrized in terms of ten Wilson coefficients in the low-energy effective Hamiltonian pertaining to the quark transition \,$s\to d\ell^+\ell^-$\, and how they are complementary to those arising from kaon decay modes.
Even though the restrictions from the kaon sector are much stronger when they exist, we identify the directions in parameter space to which the kaon measurements are blind and constrain them with the available information on the \,$\Sigma^+\to p\mu^+\mu^-$\, rate.

In investigating the NP contributions to \,$s\to d\ell^+\ell^-$,\, we adopt a model-independent approach and do not presuppose that they are directly connected to interactions involving SM neutrinos.
Such a connection could be explored in the framework of SM effective field theory (SMEFT), which has been done in, e.g., Ref.\,\cite{Geng:2021fog}.
Since our focus is on low-energy processes, the potential NP effects on \,$s\to d\ell^+\ell^-$\, may be unrelated to those on \,$s\to d\nu\bar\nu$,\, the neutrinos ($\nu\bar\nu$) being emitted invisibly.
In the latter type of interactions, invisible light particles from beyond the SM might also participate, as has been discussed in, e.g., Refs.\,\cite{Kamenik:2011vy,He:2018uey,Tandean:2019tkm,Li:2019cbk,Su:2019tjn,Geng:2020seh,He:2022ljo}.

The structure of the rest of the paper is as follows.
In Secs.\,\,\ref{sd} and \ref{ld} we deal respectively with the SD and LD amplitudes for \,$\Sigma^+\to p\ell^+\ell^-$\, in the SM and in Sec.\,\,\ref{rates} evaluate the corresponding rates and other observables.
In Sec.\,\ref{bsm} we look at the impact on \,$\Sigma^+\to p\ell^+\ell^-$\, of potential NP manifesting itself via ten low-energy effective \,$s\to d\ell^+\ell^-$\, operators, focusing on the \,$\ell=\mu$\, case.
In Sec.\,\ref{kaonobs} we address the effects of the NP on various kaon observables.
In Sec.\,\ref{constraints} we discuss the complementarity of the hyperon and kaon sectors in probing the NP.
We give our conclusions in Sec.\,\ref{concl}.

\section{SM short-distance contributions\label{sd}}

In the SM, the SD transition \,$s\to d\ell^+\ell^-$\, arises from $Z$-penguin, box, and photon-penguin diagrams~\cite{Shifman:1976de,Inami:1980fz,Buchalla:1995vs}, which involve internal up-type quarks and $W$-boson.
At QCD renormalization scales of order 1 GeV, this is described by the effective Hamiltonian~\cite{Buchalla:1995vs}
\begin{align} \label{smH}
{\cal H}_{\rm eff}^{\textsc{sm}} & \,=\, \frac{G_{\rm F}^{}}{\sqrt{2}} V_{ud}^* V_{us}^{}\,
\Big[ \bigl(z_{7V}^{}+\tau y_{7V}^{}\bigr) O_{7V}^{}+ \tau y_{7A}^{} O_{7A}^{} + C_{7\gamma}^{} O_{7\gamma}^{} \Big] \,, &
\end{align}
where $G_{\rm F}$ is the Fermi coupling constant, $V_{kl}$ denotes an element of the Cabibbo-Kobayashi-Maskawa (CKM) matrix,
$z_{7V}^{}$, $y_{7V,7A}^{}$, and $C_{7\gamma}$ are the Wilson coefficients,
$\,\tau =- V_{td}^*V_{ts}^{}/\bigl(V_{ud}^*V_{us}^{}\bigr)$,\,
\begin{align} \label{ops}
O_{7V}^{} & = 2\,\overline d\gamma^\kappa P_L^{} s\, \overline\ell\gamma_\kappa^{}\ell \,, & O_{7A}^{} & = 2\,\overline d\gamma^\kappa  P_L^{} s\, \overline\ell\gamma_\kappa^{}\gamma_5^{}\ell \,, & O_{7\gamma}^{} & = \frac{{\tt e}F_{\kappa\nu}}{8\pi^2}\, \overline d \sigma^{\kappa\nu} \big( m_s^{} P_R^{} + m_d^{}  P_L^{} \big) s \,,
\nonumber \\
\sigma^{\kappa\nu} & = \tfrac{i}{2} [\gamma^\kappa,\gamma^\nu] \,, &  P_{L,R}^{} & = \tfrac{1}{2} \big(1\mp\gamma_5^{}\big) \,,
\end{align}
with {\tt e} being the proton's electric charge, $m_{d(s)}$ the $d(s)$-quark mass, and $F_{\kappa\nu}^{}$ the electromagnetic field-strength tensor.
The total SD contribution to the amplitude for \,$\Sigma^+\to p\ell^+\ell^-$\, is then
\begin{align} \label{Msd}
{\cal M}_{\rm SM}^{\rm SD} & \,=\, \langle p\ell^+\ell^-|{\cal H}_{\rm eff}^{\textsc{sm}}|\Sigma^+\rangle
\nonumber \\ & \,=\, \frac{G_{\rm F}^{}V_{ud}^*V_{us}^{}}{\sqrt2} \!\! \begin{array}[t]{l} \Bigg\{ \langle p|\overline d\gamma^\kappa(1-\gamma_5^{})s|\Sigma^+\rangle \Big[ \big(z_{7V}^{}+\tau y_{7V}^{}\big)\, \bar u_\ell^{}\gamma_\kappa^{}v_{\bar\ell}^{} + \tau y_{7A}^{}\,  \bar u_\ell^{} \gamma_\kappa^{} \gamma_5^{} v_{\bar\ell}^{} \Big]
\\ \displaystyle ~-\; \frac{i \alpha_{\rm e}^{}\, C_{7\gamma}\, m_s^{}}{2\pi\,q^2} \langle p|\overline d \sigma^{\nu\kappa}\big(1+\gamma_5^{}\big)s|\Sigma^+\rangle q_\kappa^{}\, \bar u_\ell^{}\gamma_\nu^{}v_{\bar\ell}^{}\Bigg\} \,, \end{array} &
\end{align}
where $u_\ell^{}$ $(v_{\bar\ell})$ represents the Dirac spinor of the emitted (anti)lepton, \,$\alpha_{\rm e}={\tt e}^2/(4\pi)$,\, terms linear in $m_d$ have been dropped from the last line because \,$m_d\ll m_s$,\, and \,$q=P_\Sigma^{}-P_p$\, is the difference between the $\Sigma^+$ and proton momenta.

To examine the decay rate from ${\cal M}_{\rm SM}^{\rm SD}$ requires knowing the baryonic matrix elements in it.
The vector and axialvector ones are expressible as~\cite{Cabibbo:2003cu}
\begin{align}
\langle p|\overline d\gamma^\kappa s|\Sigma^+\rangle & \,=\, \bar u_p^{} \big( \gamma^\kappa f_1^{} + i\sigma^{\kappa\nu} f_2^{}\, q_\nu^{} + f_3^{}\, q^\kappa \big) u_\Sigma^{} \,, &
\nonumber \\
\langle p|\overline d\gamma^\kappa\gamma_5^{}s|\Sigma^+\rangle & \,=\, \bar u_p^{} \big( \gamma^\kappa g_1^{} + i\sigma^{\kappa\nu} g_2^{}\, q_\nu^{} + g_3^{}\, q^\kappa \big) \gamma_5^{} u_\Sigma^{} \,, \label{FF}
\end{align}
where $u_{p,\Sigma}$ denote the baryons' Dirac spinors and $f_{1,2,3}^{}$ and $g_{1,2,3}^{}$ are functions of $q^2$.
From the leading-order strong Lagrangian in chiral perturbation theory ($\chi$PT), we take~\cite{Tandean:2019tkm}\footnote{Nonvanishing $f_{2,3}^{}$ and $g_2^{}$ arise from higher orders in $\chi$PT, some of which involve unknown parameters. Whereas good data on $f_2^{}\big(q^2=0\big)$ exists~\cite{Workman:2022ynf}, only rough estimates of $g_2^{}(0)$ are available from lattice studies~\cite{Guadagnoli:2006gj,Sasaki:2017jue}, the outcomes being about twice smaller than, and opposite in sign to, $f_2^{}(0)$.
We have learned that, consequently, nonzero $f_2^{}(0)$ and $g_2^{}(0)$ may produce conflicting effects on some quantities in \,$\Sigma^+\to p\ell^+\ell^-$,\, such as the lepton forward-backward asymmetry.
While awaiting more definitive information on $g_2^{}$, we therefore opt for the leading-order choices in Eq.\,(\ref{<v>}). The impact of $f_3^{}$ and $g_3^{}$ is absent (subdued by the lepton mass) after $q^\kappa$ is contracted with the leptonic vector (axialvector) current.\medskip}
\begin{align} \label{<v>}
f_1^{} & \,=\, -1 \,, & g_1^{} & \,=\, D-F \,, & f_{2,3}^{} & \,=\, g_2^{} \,=\, 0 \,, & g_3^{} & \,=\, (D-F) \frac{m_{\Sigma}+m_p}{q^2-m_{K^0}^2} \,, ~~~
\end{align}
where $m_{p,\Sigma}$ stand for the baryons' masses, $D$ and $F$ are parameters which can be well determined from the measurements on hyperon semileptonic decays~\cite{Cabibbo:2003cu}, and $m_{K^0}$ is the neutral-kaon mass.
The \,$\Sigma^+\to p$\, matrix element of \,$\overline d\sigma^{\kappa\nu}s$\, in $O_{7\gamma}$ involves four $q^2$-dependent form-factors~\cite{Aslam:2008hp}, but after contraction with \,$\bar u_\ell^{}\gamma_\nu^{}v_{\bar\ell}^{}$,\, under the condition \,$q^2=0$,\, only one combination of them remains.
Explicitly
\begin{align} \label{<t>}
\langle p|\overline d \sigma^{\kappa\nu} \big(1,\gamma_5\big) s|\Sigma^+\rangle q_\kappa^{}\, \bar u_\ell^{}\gamma_\nu^{}v_{\bar\ell}^{} & \,=\, c_\sigma^{}\, \bar u_p^{} \sigma^{\kappa\nu} \big(1,\gamma_5\big) u_\Sigma^{}\, q_\kappa^{}\, \bar u_\ell^{} \gamma_\nu^{} v_{\bar\ell}^{} \,, &
\end{align}
where $c_\sigma^{}$ is a constant and \,$\langle p|\overline d \sigma^{\kappa\nu} \gamma_5^{} s|\Sigma^+\rangle$\, is linked to \,$\langle p|\overline d \sigma^{\kappa\nu} s|\Sigma^+\rangle$\, via \,$2i \sigma^{\kappa\nu} \gamma_5^{} = \epsilon^{\kappa\nu\tau\omega} \sigma_{\tau\omega}$.

Numerically, we adopt the CKM parameters available from Ref.\,\cite{Workman:2022ynf}, the Wilson coefficients typically given in the literature, namely
\,$z_{7V}^{} = -0.046\alpha_{\rm e}^{}$,  \,$y_{7V}^{} = 0.73\alpha_{\rm e}^{}$, \,$y_{7A}^{} = -0.68\alpha_{\rm e}^{}$\,~\cite{Buchalla:1995vs,Neshatpour:2022fak}, and \,$C_{7\gamma}=0.72$\,~\cite{Nielsen:1995dp,Tandean:1999mg},\footnote{This is nearly real, results mainly from $O_{7\gamma}$ mixing with \,$O_2=\overline d\gamma^\nu P_L^{}u\,\overline u\gamma_\nu^{}P_L^{}s$\, under QCD renormalization~\cite{Buchalla:1995vs}, and is hugely enhanced relative to the value without QCD corrections, \,$C_{7\gamma}=(-7.1-2.5i)\times10^{-4}$,\, which is dominated by the top contribution, the charm and up ones being four times smaller and negligible, respectively.} and \,$m_s^{}=123\rm\,MeV$,\, all evaluated at a renormalization scale of 1\,\,GeV, with \,$\alpha_{\rm e}^{} = \alpha_{\rm e}^{}(m_Z) = 1/127.95$ \cite{Workman:2022ynf}.
For the $\Sigma^+$ lifetime and hadron masses, we employ the ones supplied by Ref.\,\cite{Workman:2022ynf}.
We also have \,$c_\sigma^{}=0.3$\, from a quark-model computation~\cite{Donoghue:2022wrw} and \,$
D = 0.81\pm0.01$\, and \,$F = 0.47\pm0.01$\, from fitting to baryon semileptonic decay data~\cite{Workman:2022ynf}.

From the foregoing, the SD amplitude induced by $O_{7\gamma}$ ($O_{7V}$ and $O_{7A}$) alone yields the branching fractions \,${\cal B}(\Sigma^+\to pe^+e^-)_{\rm SD}^{} = 2.6\;(0.25)\times10^{-10}$\, and \,${\cal B}(\Sigma^+\to p\mu^+\mu^-)_{\rm SD}^{}=1.3\;(1.1)\times10^{-12}$.\,
They change to \,${\cal B}(\Sigma^+\to pe^+e^-)_{\rm SD}^{} = 2.7\times10^{-10}$\, and \,${\cal B}(\Sigma^+\to p\mu^+\mu^-)_{\rm SD}^{} = 6.5\times10^{-13}$\, if $O_{7\gamma}$ and $O_{7V,7A}$ are present simultaneously, indicating partial cancellation among their contributions in the muon case.
These numbers are 4 to 5 orders of magnitude smaller than their experimental counterparts \cite{Workman:2022ynf,Ang:1969hg,HyperCP:2005mvo,LHCb:2017rdd}.

\section{SM long-distance contributions\label{ld}}

The SD contribution being relatively tiny, the dominant component of \,$\Sigma^+\to p\ell^+\ell^-$\, has been estimated to be the long-distance photon-mediated process \,$\Sigma^+\to p\gamma^*\to p\ell^+\ell^-$.\,
If the photon is on-shell, the amplitude is parametrized by two form factors~\cite{Behrends:1958zz}, $a$ and $b$, that can be partly determined via the radiative weak decay \,$\Sigma^+\to p\gamma$.\,
For the latter mode, the effective Lagrangian has the form
\begin{align} \label{radff}
{\cal L}_{\Sigma^+\to p\gamma}^{} & \,=\, \frac{{\tt e}G_{\rm F}}{2}\, \overline p \big( a +\gamma_5^{} b \big) \sigma^{\kappa\nu} \Sigma^+ F_{\kappa\nu} \,, &
\end{align}
implying a partial width $\Gamma_\gamma$ and distribution $d\Gamma_\gamma/d(\cos\vartheta)$ given by
\begin{align} \label{Ga}
\Gamma_\gamma & \,=\, \frac{{\tt e}^2 G_{\rm F}^2}{\pi} \big(|a|^2+|b|^2\big) E_\gamma^3 \,,
\nonumber \\
\frac{d\Gamma_\gamma}{d(\cos\vartheta)} & \,\sim\, 1+\alpha_\gamma\, \cos\vartheta \,, \hspace{3em} \alpha_\gamma \,=\, \frac{2\,{\rm Re} (ab^*)}{|a|^2+|b|^2} \,, &
\end{align}
where $E_\gamma$ denotes the photon's energy in the rest frame of the $\Sigma^+$ and $\vartheta$ is the angle between its polarization and the proton's three-momentum in this frame~\cite{Workman:2022ynf}.

When the photon is virtual, the amplitude for \,$\Sigma^+\to p\gamma^*$\, contains two additional form-factors~\cite{Lyagin:1962,Bergstrom:1987wr}, $c$ and $d$, and is expressible as
\begin{align} \label{M_BBg}
{\cal M}(\Sigma^+\to p\gamma^*) & \,=\,
- {\tt e}G_{\rm F}^{}\, \bar u_p^{} \big[ i\sigma^{\kappa\nu} q_\kappa^{} \big(a+\gamma_5^{}b\big) + \big( \gamma^\nu q^2 - \slashed q\, q^\nu \big) \big(c+\gamma_5^{}d\big) \big] u_\Sigma^{}\, \varepsilon_\nu^* \,, &
\end{align}
where $q$ stands for the photon's four-momentum.
Its contribution to \,$\Sigma^+\to p\ell^+\ell^-$\, is then found to be
\begin{align} \label{ffabcd}
{\cal M}_{\rm SM}^{\rm LD} & \,=\, {\tt e}^2 G_{\rm F}^{} \Bigg[ \frac{-i}{q^2}\, \bar u_p^{} \big(a+\gamma_5^{}b\big) \sigma^{\kappa\nu} q_\kappa^{} u_\Sigma^{} - \bar u_p^{} \gamma^\nu \big(c+\gamma_5^{}d\big) u_\Sigma^{} \Bigg] \bar u_\ell^{} \gamma_\nu^{} v_{\bar\ell}^{} \,. &
\end{align}
In general  $a$, $b$, $c$, and $d$ depend on $q^2$ and can be complex, and the first two are constrained at \,$q^2=0$\, by the data in the real-photon case, to which we turn next.

There is experimental information available on the branching fraction $\mathscr B_\gamma$ of \,$\Sigma^+\to p\gamma$\, and its decay asymmetry parameter $\alpha_\gamma$, mostly gathered decades ago~\cite{Workman:2022ynf}.
Recently BESIII has reported fresh data on these quantities~\cite{BESIII:2023fhs}:
\begin{align} \label{Spgdatanew}
{\mathscr B}_\gamma & \,=\, \big(0.996\pm 0.021_{\rm stat}\pm0.018_{\rm syst}\big)\times10^{-3} \,, \\
 \alpha_
\gamma & \,=\, -0.652 \pm 0.056_{\rm stat}\pm0.020_{\rm syst} \,. & \nonumber
\end{align}
The former is less than the corresponding previous value of \,$(1.23\pm0.05)\times 10^{-3}$\, from the Particle Data Group (PDG) \cite{Workman:2022ynf} by 4.2 standard deviations and also has a better precision~\cite{BESIII:2023fhs}, whereas the $\alpha_\gamma$ finding is consistent with the prior measurements~\cite{Gershwin:1969fpe,Manz:1980td,Kobayashi:1987yv,E761:1992atm}.
Accordingly, for numerical work, we may simply adopt the $\mathscr B_\gamma$ number in Eq.\,\,(\ref{Spgdatanew}), while for $\alpha_\gamma$ we can take the average of the BESIII and earlier results, which is \,$\alpha_
\gamma=-0.694\pm0.047$.\,
Putting together these choices with Eq.\,(\ref{Ga}) and employing the hadron masses and $\Sigma^+$ lifetime from Ref.\,\cite{Workman:2022ynf}, we then arrive at the \,$q^2=0$\, combinations
\begin{align}  \label{Spgcons}
|a(0)|^{2}+|b(0)|^{2} & \,=\, (13.48\pm 0.19)^{2}{\rm ~MeV}^{2} \,, \nonumber \\
{\rm Re}\bigl(a(0)\,b^*(0)\bigr) & \,=\, (-63.1\pm4.6) {\rm ~MeV}^{2} \,, &
\end{align}
having additionally used \,${\tt e}^2 = 4\pi\alpha_{\rm e}^{}(0)$,\, with \,$\alpha_{\rm e}(0)=1/137.036$\, \cite{Workman:2022ynf}, in $\Gamma_\gamma$.
The LHCb experiment may be able to measure $\mathscr B_\gamma$ and $\alpha_\gamma$ as well, as they have measured similar quantities in the $b$-baryon sector~\cite{LHCb:2021byf}.

As elaborated in Ref.\,\cite{He:2005yn}, to work out how $a$, $b$, $c$, and $d$ depend on $q^2$, first their imaginary parts can be computed by means of relativistic baryon $\chi$PT with the aid of unitarity arguments, and among the main inputs are the empirical S-wave and P-wave amplitudes for the nonleptonic channels \,$\Sigma^+\to p\pi^0,n\pi^+$.\,
Subsequently, with the resulting \,Im$(a,b)$\, evaluated at \,$q^2=0$,\, the real parts of $a$ and $b$ are approximated to be constant and fixed from the observed $\mathscr B_\gamma$ and $\alpha_\gamma$ values of \,$\Sigma^+\to p\gamma$,\, up to a fourfold ambiguity.
For the real parts of $c$ and $d$, which are still empirically inaccessible, they are estimated using the vector-meson-dominance approach.
Since there are new relevant data on  $\Sigma^+\to p\pi^0,n\pi^+$\, from BESIII~\cite{BESIII:2020fqg,BESIII:2023sgt}, this procedure needs to be updated.
Thus, writing the amplitude for \,$\Sigma^+\to N\pi$\, as \,${\cal M}_{\Sigma^+\to N\pi} = iG_{\rm F}^{}m_{\pi^+}^2\, \bar u_N (A_{N\pi} - \gamma_5B_{N\pi}) u_\Sigma$\, and neglecting the small phases of $A_{N\pi}$ and $B_{N\pi}$, we arrive at\footnote{We have used the branching fractions \,${\cal B}(\Sigma^+\to n\pi^+)=(48.31\pm0.30)\%$\, and \,${\cal B}(\Sigma^+\to p\pi^0)=(51.57\pm0.30)\%$  from Ref.\,\cite{Workman:2022ynf}, based on measurements done over 40 years ago,
as well as the decay asymmetry parameters  $\alpha_\pi(\Sigma^+\to p\pi^0)=-0.993\pm0.004$\, and \,$\alpha_\pi(\Sigma^+\to n\pi^+)=0.052\pm0.004$\, from averaging the respective recent BESIII findings on these observables~\cite{BESIII:2020fqg,BESIII:2023sgt} and the corresponding earlier measurements~\cite{Bangerter:1969fta,Harris:1970kq,Bellamy:1972fa,Lipman:1973mz}.}
\begin{align}
A_{n\pi^+} & \,=\, 0.047\pm0.004 \,, & B_{n\pi^+} & \,=\, 18.54\pm0.07 \,, &
\nonumber \\
A_{p\pi^0} & \,=\, -1.383\pm0.021 \,, & B_{p\pi^0} & \,=\, 12.32\pm0.24 \,.
\label{eq:abpars}
\end{align}
With these, we obtain \,Im$\,a(0) = (2.857\pm0.046)$\,MeV\, and \,Im$\,b(0) = (-1.742\pm0.040)$\,MeV,\, which with Eq.\,(\ref{Spgcons}) lead to the four solutions collected in the upper half of Table~\ref{tab:rates}.
Employing heavy-baryon $\chi$PT instead yields \,Im$\,a(0) = (6.98\pm0.10)$\,MeV\, and \,Im$\,b(0) = (-0.43\pm0.03)$\,MeV,\, and hence another four solutions placed in the lower half of Table~\ref{tab:rates}.
To estimate the uncertainties in this table, we have simulated a large number of decays with the parameters in Eqs.\,\,(\ref{Spgcons}) and (\ref{eq:abpars}) assuming that they follow Gaussian distributions, and from this we extract the averages and $1\sigma$ ranges for the rates.
In our figures, we will distinguish between the solutions listed in this table with the color coding: blue for solution 1, yellow solution~2, green solution 3, and red solution 4.

\begin{table}[b] \bigskip
\centering \setlength{\tabcolsep}{1ex}
\begin{tabular}{|cc||c|c|c|c|c|}\hline
$~ {\rm Re}\,a$ (MeV) & $~ {\rm Re}\,b$ (MeV) & $10^8\, {\mathscr B}_{\mu\mu}\vphantom{|_|^{\int}}$ &  $10^8\, {\mathscr B}_{\mu\mu}^{\scriptscriptstyle{\rm Re}(c,d)=0}$ & $10^5\, {\tilde A}_{\rm FB}^{}$ & $10^6\,\mathscr B_{ee}^{}$  & $10^7\, {\mathscr B}_{ee}^{\scriptscriptstyle M_{ee}\ge100\rm\,MeV}$ \\ \hline
$-12.15\pm0.24$ & ~~ $4.78\pm 0.42$ & $2.7\pm 0.2$ & $1.8\pm 0.1$ & $-1.59\pm0.04$& $7.6\pm 0.3$ & ~$8.5\pm0.4$
\\
~$-4.78\pm 0.42$ & ~ $12.15\pm0.24$ & $7.8\pm 0.3$ & $5.8\pm 0.2$ & $-0.32\pm0.02$& $8.3\pm0.3$ & $13.5\pm0.4$
\\
~~ $4.78\pm 0.42$  &  $-12.15\pm0.24$ & $4.2\pm 0.2$ & $5.8\pm 0.2$ & ~ $0.91\pm0.04$& $7.8\pm0.3$ & ~$9.9\pm0.4$
\\
~ $12.15\pm0.24$ & ~$-4.78\pm 0.42$& $1.2\pm 0.1$& $1.8\pm 0.1$ & ~ $4.55\pm0.08$& $7.4\pm0.3$ & ~$6.9\pm 0.3$
\\ \hline \hline
$~{\rm Re}\,a$ (MeV) & $~{\rm Re}\,b$ (MeV) & $10^8\, {\mathscr B}_{\mu\mu}\vphantom{|_|^{\int}}$ & $10^8\, {\mathscr B}_{\mu\mu}^{\scriptscriptstyle{\rm Re}(c,d)=0}$ & $10^5\, {\tilde A}_{\rm FB}^{}$ & $10^6\, \mathscr B_{ee}^{}$  & $10^7\, {\mathscr B}_{ee}^{\scriptscriptstyle M_{ee}\ge100\rm\,MeV}$
\\ \hline
$-9.74\pm0.54$  & ~ $6.17\pm 0.74$ & $3.7\pm 0.5$ & $2.7\pm 0.3$ & $-0.73\pm0.06$  & $7.9\pm 0.5$ & $10.2\pm0.8$
\\
$-6.17\pm 0.74$ & ~ $9.74\pm0.54$  & $6.1\pm0.5$ & $4.5\pm 0.4$  & $-0.29\pm0.05$  & $8.3\pm 0.6$ & $12.6\pm0.8$
\\
~ $6.17\pm 0.74$ &  $-9.74\pm0.54$ & $3.2\pm 0.3$& $4.5\pm 0.4$  & ~ $1.49\pm0.09$ & $7.8\pm 0.6$ & ~$9.7\pm0.7$
\\
~ $9.74\pm0.54$ & $-6.17\pm 0.74$  & $1.9\pm 0.2$ & $2.7\pm 0.3$ & ~ $3.2\pm0.1$   & $7.6\pm 0.5$ & ~$8.3\pm0.6$
\\ \hline
    \end{tabular}
\caption{The four solutions for the real parts of the form factors $a$ and $b$ (columns 1 and 2) obtained in relativistic baryon $\chi$PT (first four rows) and in its heavy-baryon formulation (last four rows) and the corresponding predictions in the SM for each case (columns 3-7).
In the third and fourth columns \,$\mathscr B_{\mu\mu}\equiv{\cal B}(\Sigma^+\to p\mu^+\mu^-)$ is from the complete SM amplitude and ${\mathscr B}_{\mu\mu}^{\scriptscriptstyle{\rm Re}(c,d)=0}$ gets no contribution from the real parts of the form factors $c$ and $d$.
In the fifth column $\tilde A_{\rm FB}$ is the integrated muon forward-backward asymmetry.
The last two columns exhibit \,$\mathscr B_{ee} \equiv {\cal B}(\Sigma^+\to p e^+e^-)$\, without and with a cut in the dielectron invariant mass.}  \label{tab:rates}
\end{table}

It is interesting to mention that, as pointed out long ago \cite{Behrends:1958zz}, the parameter
\begin{align}
\gamma_\gamma & \,=\, \frac{|a(0)|^{2}-|b(0)|^{2} }{|a(0)|^{2}+|b(0)|^{2}} &
\end{align}
may be determined in a weak radiative hyperon decay if the linear polarization of the photon and the spins of the initial and final baryons can all be detected.
If so, $\gamma_\gamma$ in \,$\Sigma^+\to p\gamma$\, could reduce the fourfold ambiguity to a twofold ambiguity.
For the situation illustrated by Table~\ref{tab:rates}, from the top (bottom) half we get \,$\gamma_\gamma=0.72\pm0.04$\, for solutions 1 and 4 and \,$\gamma_\gamma=-0.66\pm0.04$  for solutions 2 and 3 $\big(\gamma_\gamma=0.58\pm0.10$\, for solutions 1 and 4 and \,$\gamma_\gamma=-0.05\pm0.11$\, for solutions 2 and 3$\big)$.
We see that just a measurement of the sign of $\gamma_\gamma$ may suffice to decrease the ambiguity to twofold.

It may be that resolving this ambiguity completely, as well as minimizing other uncertainties in the LD contributions, will only be possible if they are addressed on the lattice.
Initial exploratory efforts along this direction have recently been undertaken, as detailed in Refs.\,\,\cite{Erben:2022tdu,Erben:2022igb}.

\section{Decay rates and observables\label{rates}}

The different contributions to \,$\Sigma^+\to p\ell^+\ell^-$\, can be  incorporated into the general expression for the amplitude
\begin{align} \label{MtotS2pll}
{\cal M} \,= &~ \bigl[ i q_\kappa^{}\, \bar u_p^{} \big( \tilde{\textsc a}
+ \gamma_5^{} \tilde{\textsc b} \big) \sigma^{\nu\kappa} u_\Sigma^{}
- \bar u_p^{}\gamma^\nu \big(\tilde{\textsc c}+\gamma_5^{}\tilde{\textsc d}
\big) u_\Sigma^{} \bigr] \bar u_\ell^{}\gamma_\nu^{} v_{\bar\ell}^{}
\,+\, \bar u_p^{}\gamma^\nu \big(\tilde{\textsc e}+\gamma_5^{}\tilde{\textsc f}
\big) u_\Sigma^{}\, \bar u_\ell^{} \gamma_\nu^{} \gamma_5^{} v_{\bar\ell}^{}
\nonumber \\ & +\,
\bar u_p^{} \big(\tilde{\textsc g}+\gamma_5^{}\tilde{\textsc h}\big) u_\Sigma^{}\,
\bar u_\ell^{}v_{\bar\ell}^{}
+ \bar u_p^{} \big(\tilde{\textsc j}+\gamma_5^{}\tilde{\textsc k}\big) u_\Sigma^{}\,
\bar u_\ell^{}\gamma_5^{}v_{\bar\ell}^{} \,, &
\end{align}
where \,$\tilde{\textsc a},\tilde{\textsc b},.$..$,\tilde{\textsc k}$\, are generally complex form-factors that depend on $q^2$ and  include both SM and NP contributions.
The formula for the rate derived from $\cal M$ can be found in Ref.\,\cite{He:2018yzu} and hence will not be reproduced here.

From Secs.\,\,\ref{sd} and \ref{ld}, the SD and LD amplitudes in the SM yield
\begin{align} \label{abcdefk}
\tilde{\textsc a} & \,=\, \frac{G_{\rm F}^{}}{q^2} \Bigg( a\, {\tt e}^2 - \frac{\lambda_u^*}{\sqrt2} \frac{c_\sigma^{} C_{7\gamma}^{} {\tt e}^2 m_s^{}}{8\pi^2} \Bigg) \,, &
\tilde{\textsc b} \,= &~ \frac{G_{\rm F}^{}}{q^2} \Bigg( b\, {\tt e}^2 - \frac{\lambda_u^*}{\sqrt2} \frac{c_\sigma^{} C_{7\gamma}^{} {\tt e}^2 m_s^{}}{8\pi^2} \Bigg) \,,
\nonumber \\
\tilde{\textsc c} & \,=\, G_{\rm F}^{} \Bigg( c\, {\tt e}^2 + \frac{\lambda_u^*z_{7V}^{}-\lambda_t^*y_{7V}^{}}{\sqrt2} \Bigg) \,, &
\tilde{\textsc d} \,= &~ G_{\rm F}^{} \Bigg[ d\, {\tt e}^2 + (D{-}F) \frac{\lambda_u^*z_{7V}^{}-\lambda_t^*y_{7V}^{}}{\sqrt2} \Bigg] \,,
\nonumber \\
\tilde{\textsc e} & \,=\, \frac{G_{\rm F}^{}}{\sqrt2}\, \lambda_t^*y_{7A}^{} \,, &
\tilde{\textsc f} \,= &~ \frac{D-F}{\sqrt2}~ G_{\rm F}^{}\, \lambda_t^* y_{7A}^{} \,, &
\nonumber \\
\tilde{\textsc k} & \,=\, \sqrt2 (D-F) G_{\rm F}^{}\, \lambda_t^* y_{7A}^{}\, m_\ell^{}~ \frac{m_\Sigma^{}+m_p^{}}{q^2-m_{K^0}^2} \,,
\end{align}
where \,$\lambda_{\tt q}^* = V_{{\tt q}d}^* V_{{\tt q}s}^{}$.\,
These lead to the SM predictions for \,$\mathscr B_{\mu\mu}={\cal B}(\Sigma^+\to p\mu^+\mu^-)$\, and other observables collected in the third through fourth columns of Table~\ref{tab:rates}, where the quoted errors have been translated from the ranges of \,Re$(a,b)$\, in columns 1 and 2 and of the corresponding  Im$\big(a(0),b(0)\big)$  computed in the preceding section.
Parametric uncertainty in other quantities entering these predictions is much smaller, and so we ignore it.

\begin{figure}[b] \bigskip
\includegraphics[width=3.2in]{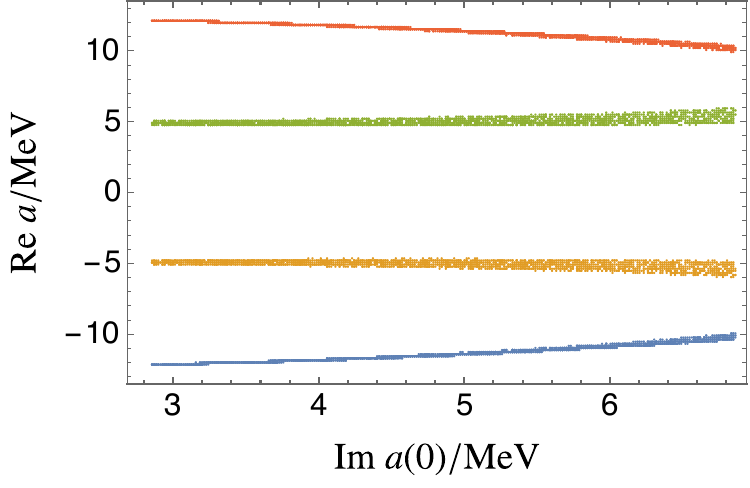} ~ ~ \includegraphics[width=3.2in]{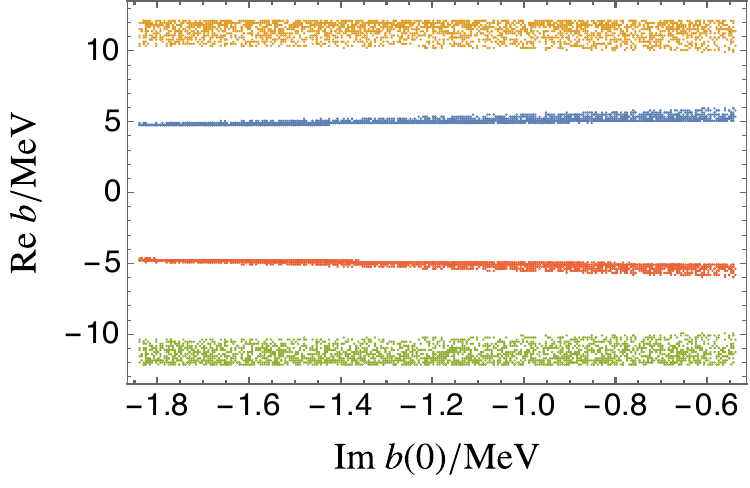}
\bigskip \\
\includegraphics[width=3.2in]{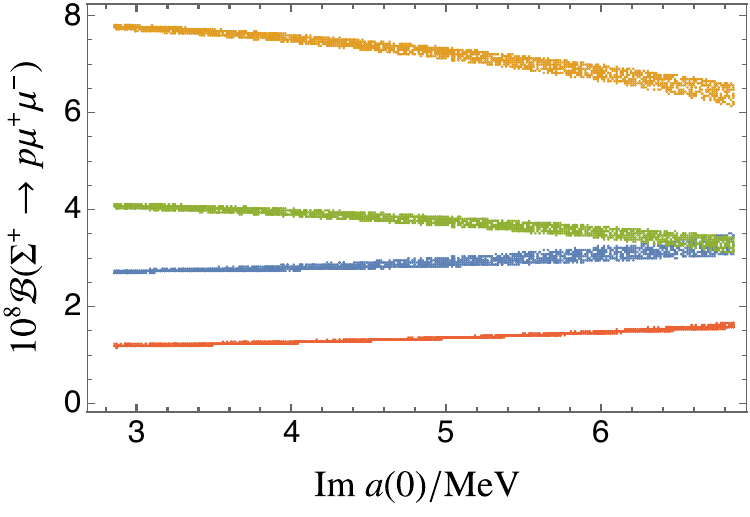} ~ ~  \includegraphics[width=3.2in]{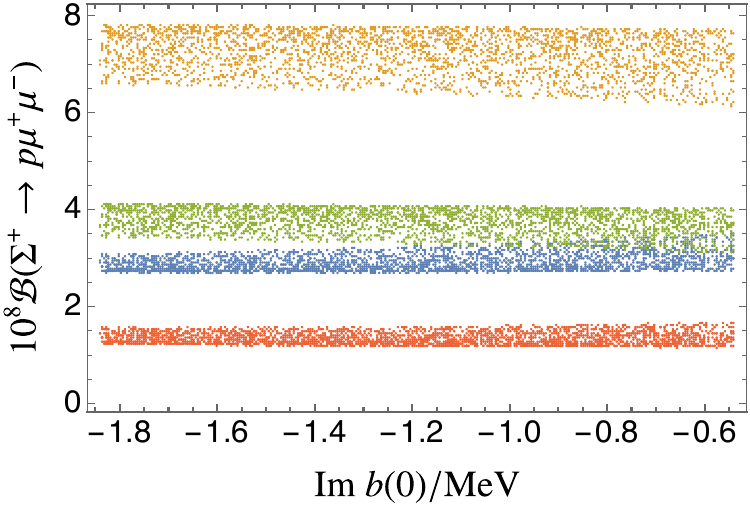}
\bigskip \\
\includegraphics[width=3.2in]{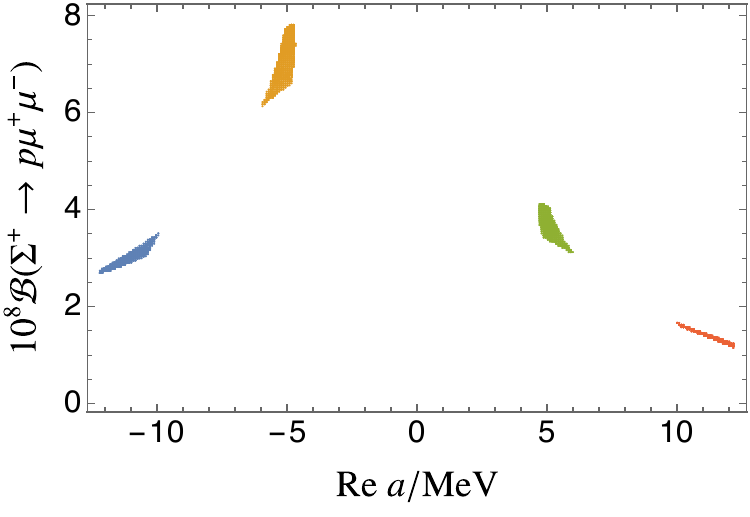} ~ ~ \includegraphics[width=3.2in]{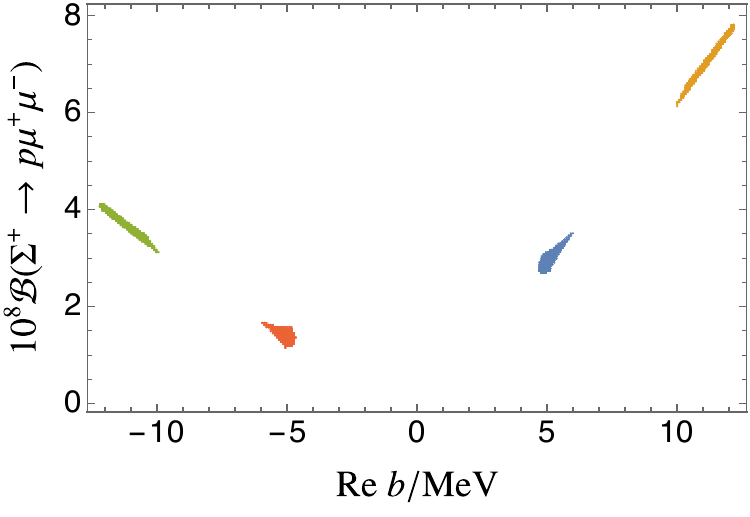} \vspace{-1ex}
\caption{Variation in the real parts of the form factors $a$ and $b$ (top row) and in the resulting branching fraction ${\cal B}(\Sigma^+\to p\mu^+\mu^-)$ (middle row) as \,${\rm Im}\,a(0)$\, and \,${\rm Im}\,b(0)$\, are allowed to vary between the values determined with relativistic baryon $\chi$PT and with its heavy-baryon formulation.
The bottom plots show the corresponding correlations between ${\cal B}(\Sigma^+\to p\mu^+\mu^-)$ and the real parts of $a$ and $b$.} \label{f:ranges}
\end{figure}

Implicit in the results listed in Table~\ref{tab:rates} is the presence of hadronic model dependence which occurs in the treatment of the real parts of $c$ and $d$, discussed in Ref.\,\cite{He:2005yn}.
One can get an indication about the extent of this model dependence in, for instance, the \,$\Sigma^+\to p\mu^+\mu^-$\, branching fraction by comparing the entries in the third and fourth columns of this table.
It is also instructive to estimate the theoretical uncertainty arising from the extraction of the imaginary parts of the form factors at  $q^2=0$\, by comparing the results derived with relativistic baryon $\chi$PT and with its heavy-baryon formulation.
The range of predictions obtained by interpolating between these two  is displayed in Fig.\,\ref{f:ranges}.

The PDG \cite{Workman:2022ynf} currently quotes \,${\cal B}(\Sigma^+\to p\mu^+\mu^-)_{\textsc{pdg}}^{} = \big(2.4^{+1.7}_{-1.3}\big)\times 10^{-8}$,\, based on the findings of the HyperCP~\cite
{HyperCP:2005mvo} and LHCb~\cite
{LHCb:2017rdd} experiments.
It would seem, in view of Table~\ref{tab:rates} and Fig.\,\ref{f:ranges}, to be in conflict with the predictions of solution 2, but one of them, \,${\mathscr B}_{\mu\mu}=(6.1\pm0.5)\times10^{-8}$\, in the third column of the table, is still compatible with the PDG value at the two-sigma level if both of their errors are taken into account.
We expect that the situation will become clearer when an improved result from LHCb becomes available in the near future.

\begin{figure}[b] \bigskip
\includegraphics[width=3.3in]{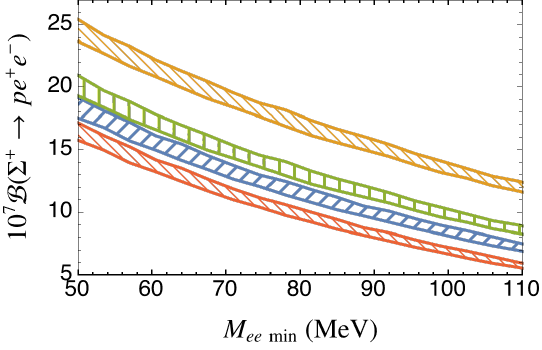} \includegraphics[width=3.3in]{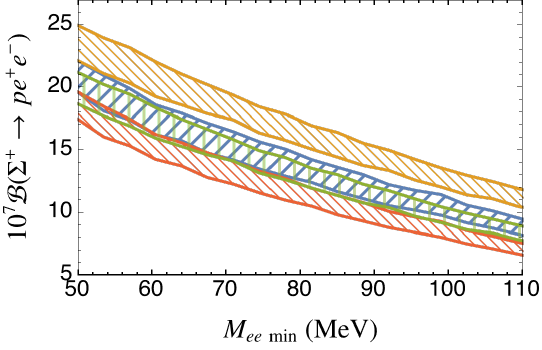}
\caption{Dependence of the SM branching fraction of \,$\Sigma^+\to p e^+e^-$\, on a minimum cut in the dielectron invariant mass for the four solutions obtained with relativistic baryon $\chi$PT (left plot) and with its heavy-baryon formulation (right plot).
In all of the figures in this section, the blue, yellow, green, and red colors correspond to solutions 1, 2, 3, and 4, respectively.
The band of each curve in this and the next three figures reflects the ranges of \,${\rm Re}(a,b)$\, and \,${\rm Im}\big(a(0),b(0)\big)$\, for the corresponding solution.} \label{f:beecut}
\end{figure}

Measurements of the branching fraction of \,$\Sigma^+\to p e^+e^-$\, may offer an additional way to resolve the fourfold ambiguity.
In this mode, the dominance of the LD contributions is more pronounced, owing to the fact that the $a$ and $b$ terms in the amplitude are much enhanced at low values of the dielectron invariant mass, \,$M_{ee}=\sqrt{q^2}\ge2m_e^{}$.\,
As a consequence, the predictions for \,${\mathscr B}_{ee}={\cal B}(\Sigma^+\to p e^+e^-)$\, corresponding to the various solutions are similar to one another, as can be seen in the sixth column of Table~\ref{tab:rates}.
Nevertheless, the difference between solutions can be magnified by raising the minimum of $M_{ee}$.
This might be achieved by applying, for instance, the cut \,$M_{ee}\ge 100$\,MeV,\, leading to the numbers in the last column of Table~\ref{tab:rates}.
Thus, the selection of an appropriate lower bound on $M_{ee}$ could make the predictions of the different solutions more distinguishable.
This point is further illustrated by Fig.~\ref{f:beecut}.
Although the implementation of the $M_{ee\rm\,min}$ cut has a price, which is a decreased rate, the latter may still be within the sensitivity reach of BESIII~\cite{Li:2016tlt} and LHCb~\cite{AlvesJunior:2018ldo,LHCb:2018roe,Cerri:2018ypt}, depending on the choice of $M_{ee\rm\,min}$.

It is worth remarking here that the ${\mathscr B}_{ee}$ range in the sixth column of Table~\ref{tab:rates} is roughly 20\% less than the corresponding predictions made in Refs.\,\cite{He:2005yn,He:2018yzu}.
This is largely because of our adoption of the BESIII finding of ${\mathscr B}_\gamma$ quoted in Eq.\,(\ref{Spgdatanew}), which is also \,{\small$\sim$}\,20\% lower than the previous PDG average~\cite{Workman:2022ynf}.
The empirical information on \,$\Sigma^+\to pe^+e^-$\, is still scarce.
From the only existing data, reported in Ref.\,\cite{Ang:1969hg}, one can infer
\,${\cal B}(\Sigma^+\to pe^+e^-)=(7.7\pm4.6)\times10^{-6}$\,~\cite{He:2005yn},\footnote{The PDG limit \,${\cal B}\big(\Sigma^+\to pe^+e^-\big)<7\times10^{-6}$\, \cite{Workman:2022ynf} is for neutral currents in this mode and not for its full branching fraction \cite{Ang:1969hg}.}
which is far from precise and accommodates well all the ${\mathscr B}_{ee}$ predictions in the sixth column of Table~\ref{tab:rates}.
Thus, it goes without saying that the upcoming measurements of  $\Sigma^+\to p e^+e^-$\, by BESIII and LHCb can be anticipated to provide highly desirable information pertinent to our study.

Returning to the \,$\ell=\mu$\, case, one can write the double differential distribution \cite{LHCb:2015tgy,He:2018yzu}
\begin{align}
    \frac{d^2\Gamma(\Sigma^+{\to} p\mu^+\mu^-)}{dq^2\, d\big(\cos\theta_\mu\big)} & =   \bigg[ \frac{3}{8} \big(1+{\rm cos}^2\theta_\mu\big) \big(1-f_L^{}\big) + {\cal A}_{\rm FB}^{} \cos\theta_\mu + \frac{3}{4} f_L^{}\, {\rm sin}^2\theta_\mu \bigg]
\frac{d\Gamma(\Sigma^+{\to} p\mu^+\mu^-)}{dq^2}
\nonumber \\ & = {\cal F}_0^{} + {\cal F}_1^{}\, \cos\theta_\mu + {\cal F}_2^{}\, \cos^2\theta_\mu \,, & \label{eq:doubledif}
\end{align}
where $\theta_\mu$ stands for the angle between the directions of the $\mu^-$ and $p$ in the rest frame of the $\mu^+\mu^-$ system, $f_L^{}$ is the so-called fraction of longitudinally polarized dimuons~\cite{LHCb:2015tgy}, ${\cal A}_{\rm FB}$ designates the muon forward-backward asymmetry, and $f_L^{}$, ${\cal A}_{\rm FB}$, and ${\cal F}_{0,1,2}$ are all independent of $\theta_\mu$.
The expressions for ${\cal F}_{0,1,2}$ in the general parametrization of Eq.~(\ref{MtotS2pll}) can be found in the appendix of~Ref.\,\cite{He:2018yzu}.

\begin{figure}[t]
\includegraphics[width=3.2in]{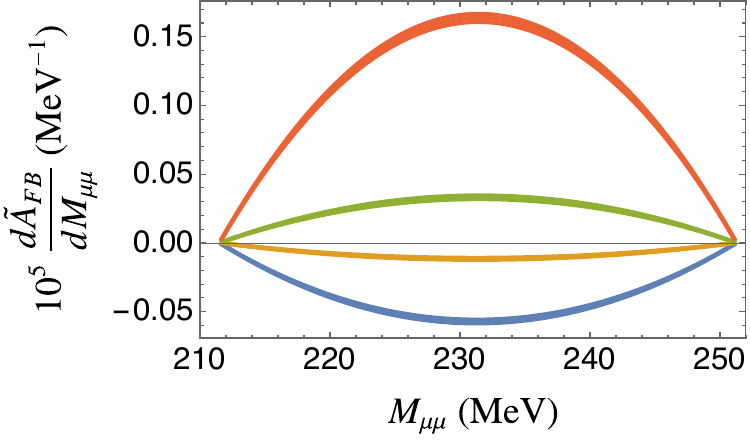} ~ ~ \includegraphics[width=3.2in]{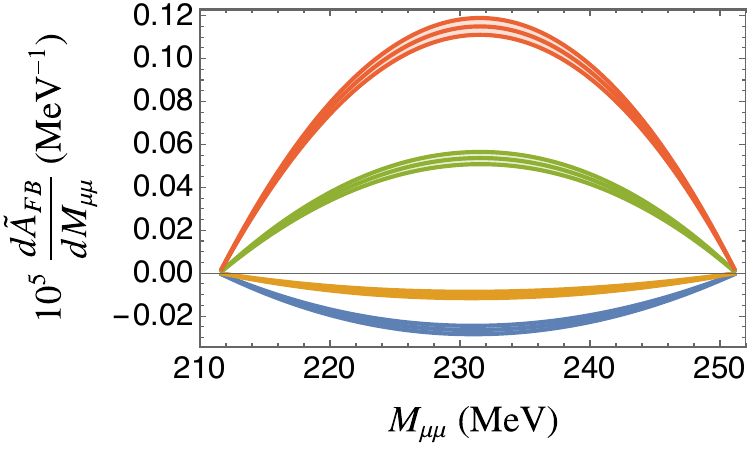} \vspace{-1ex}
\caption{The differential $\tilde A_{\rm FB}$ distribution in \,$\Sigma^+\to p \mu^+\mu^-$\, versus the dimuon invariant mass, $M_{\mu\mu}$, for the four relativistic (left) and heavy-baryon (right) solutions in the SM.} \label{f:afbtil}
\end{figure}

Thus, explicitly we have \cite{He:2018yzu}
\begin{align} \label{AFB}
{\cal A}_{\rm FB}^{} & \,=\, \frac{\beta^2\, \bar\lambda}{64\pi^{3\,}\, \Gamma'\, m_\Sigma^3}\, {\rm Re}\! \begin{array}[t]{l} \Big\{ \big[ {\texttt M}_+^{} \tilde{\textsc a}{}^*\tilde{\textsc f}
- {\texttt M}_-^{} \tilde{\textsc b}{}^* \tilde{\textsc e}
- \big( \tilde{\textsc a}{}^*\tilde{\textsc g}+\tilde{\textsc b}{}^*\tilde{\textsc h}\big) m_\mu
+ \tilde{\textsc c}{}^*\tilde{\textsc f} + {\textsc d}{}^*\tilde{\textsc e} \big] q^2
\vspace{1pt} \\ \;\,
-\; \big( {\texttt M}_+^{} \tilde{\textsc c}{}^* \tilde{\textsc g}
- {\texttt M}_-^{} \tilde{\textsc d}{}^* \tilde{\textsc h} \big) m_\mu \Big\} \,, \end{array} &
\end{align}
where
\begin{align}
\Gamma' & \,\equiv\, \frac{d\Gamma(\Sigma^+\to p \mu^+\mu^-)}{dq^2} \,=\, 2\,{\cal F}_0^{} + \frac{2}{3}\, {\cal F}_2^{} \,, & \beta & \,=\, \sqrt{1-\frac{4m_\mu^2}{q^2}} \,, \nonumber \\
\bar\lambda & \,=\,
\big({\tt M}_+^2-q^2\big) \big({\tt M}_-^2-q^2\big) \,, & {\tt M}_\pm^{} & \,=\, m_\Sigma^{}\pm m_p^{} \,. &
\end{align}
Since $\tilde{\textsc a}$, $\tilde{\textsc b}$, $\tilde{\textsc c}$, and $\tilde{\textsc d}$ are dominated by the LD contributions, while $\tilde{\textsc e}$, $\tilde{\textsc f}$, $\tilde{\textsc g}$, and $\tilde{\textsc h}$ are not, ${\cal A}_{\rm FB}$ is potentially sensitive to NP which can enhance one or more of the latter set of quantities significantly compared to their SM expectations.
Also of interest is the integrated forward-backward asymmetry
\begin{align} \label{eq:afbin}
{\tilde A}_{\rm FB}^{} & \,=\, \frac{1}{\Gamma(\Sigma^+\to p\mu^+\mu^-)} \int \Gamma'\, {\cal A}_{\rm FB}^{}\, dq^2 \,, &
\end{align}
where both the numerator and denominator of ${\tilde A}_{\rm FB}$ are to be evaluated for the same solution.
The SM prediction for ${\tilde A}_{\rm FB}$ can be viewed in the fifth column of Table~\ref{tab:rates}, and its differential distribution,
\begin{align}
\frac{d{\tilde A}_{\rm FB}^{}}{dM_{\mu\mu}} & \,=\, \frac{2\, \Gamma'\, {\cal A}_{\rm FB}^{}\, M_{\mu\mu}}{\Gamma(\Sigma^+\to p\mu^+\mu^-)} \,, &
\end{align}
versus the dimuon invariant mass \,$M_{\mu\mu}=\sqrt{q^2}$\, is depicted in Fig.~\ref{f:afbtil}.
Evidently, there is a sign flip for two of the solutions (distributions 1 and 2 in both the relativistic and heavy-baryon cases become negative).

\begin{figure}[!t] \bigskip
\includegraphics[width=3.2in]{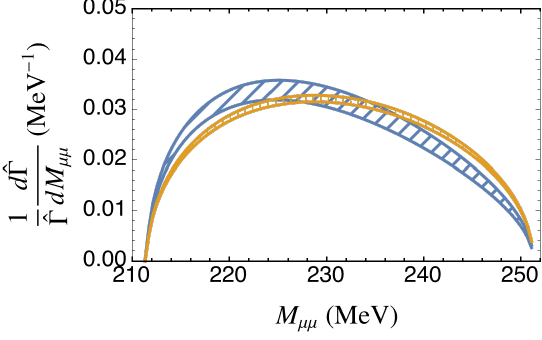} ~ ~ \includegraphics[width=3.2in]{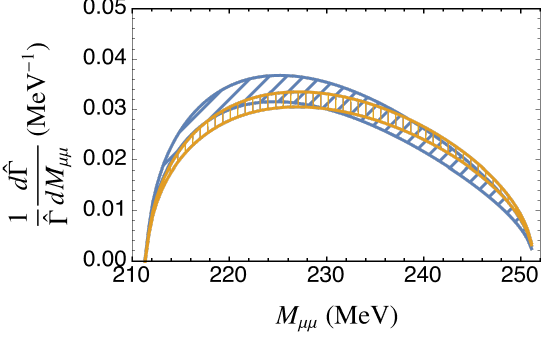} \vspace{1ex} \\ \includegraphics[width=3.2in]{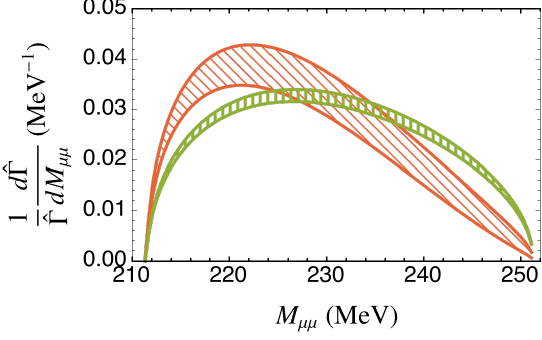} ~ ~ \includegraphics[width=3.1in]{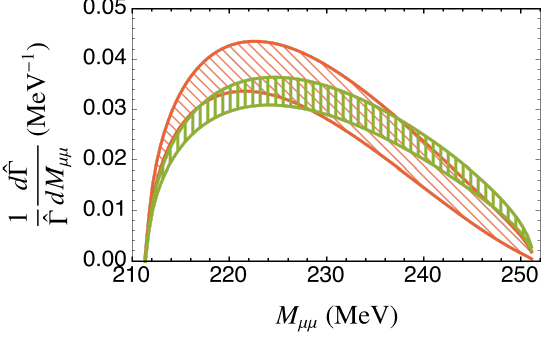} \vspace{-1ex}
\caption{The dimuon spectrum in \,$\Sigma^+\to p \mu^+\mu^-$\, versus the dimuon invariant mass for the four relativistic (left plots) and heavy-baryon (right plots) solutions in the SM.} \label{f:dimuonsp}
\end{figure}

\begin{figure}[t]
\includegraphics[width=4in]{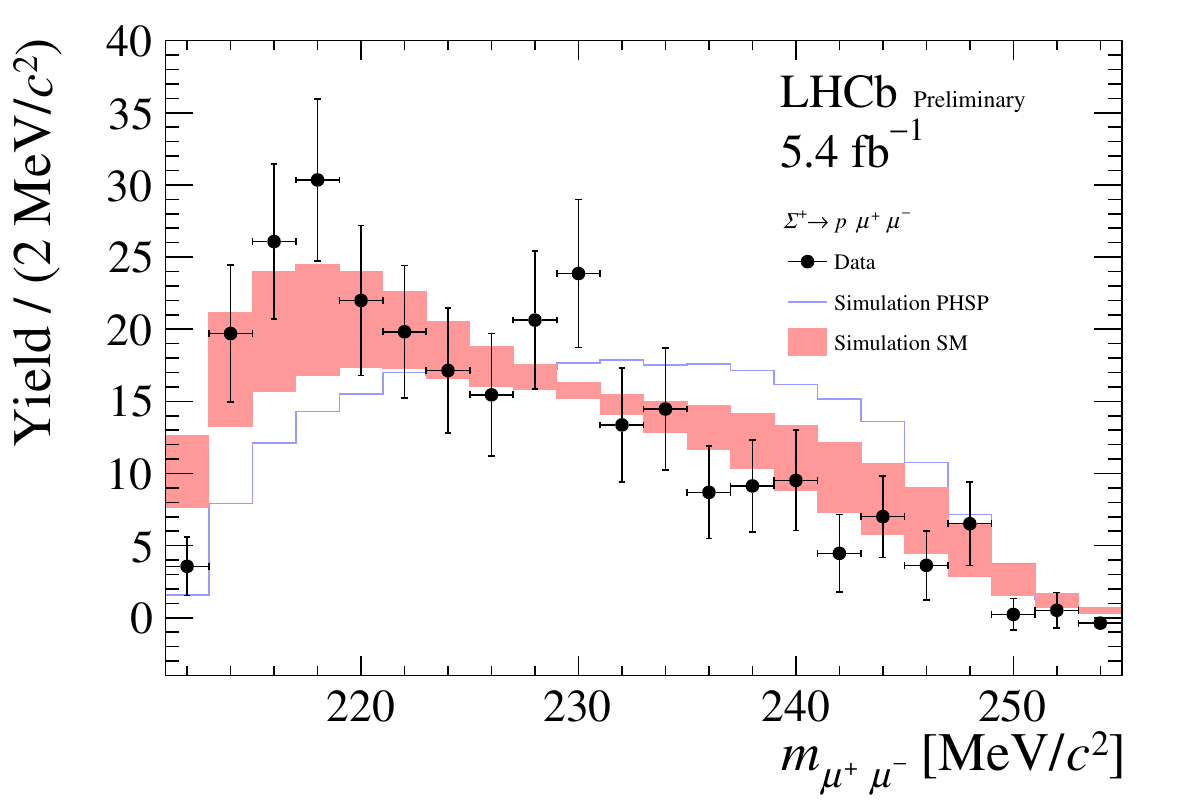}
\caption{The distribution of dimuon invariant-mass for \,$\Sigma^+\to p \mu^+\mu^-$\, candidates in the latest LHCb data compared with simulation, reproduced from Ref.\,\,\cite{LHCb:2024}. The LHCb phase-space is shown as is (blue line) and weighted according to the SM amplitude (red band).} \label{f:lhcbdimuonsp}
\end{figure}

Another important quantity is the dimuon spectrum given by
\begin{align} \label{spectrum}
\frac{1}{\hat\Gamma}\, \frac{d\hat\Gamma}{dM_{\mu\mu}} & \,=\, \frac{2\,\Gamma'\, M_{\mu\mu}}{\hat\Gamma} \,, & \hat\Gamma \,\equiv\, \Gamma(\Sigma^+\to p\mu^+\mu^-) \,, & & &
\end{align}
hence also normalized by the rate.
The dependence of \,$(d\hat\Gamma/dM_{\mu\mu})/\hat\Gamma$ on $M_{\mu\mu}$ for each of the solutions is illustrated in Fig.~\ref{f:dimuonsp}.
The spectrum for solution 4, in both the relativistic and heavy-baryon cases, can be seen to peak at lower $M_{\mu\mu}$ than the other three.
Very recently LHCb~\cite{LHCb:2024} has announced the first observation of this mode, the reported preliminary data on the dimuon invariant-mass distribution being compatible with the SM expectations, as exhibited in Fig.\,\ref{f:lhcbdimuonsp}.
The excess over the pure phase-space spectrum at the lower values of \,$210 \,\mbox{\footnotesize$\lesssim$}\, M_{\mu\mu}/\rm MeV \,\mbox{\footnotesize$\lesssim$}\, 220$, combined with the deficit at the higher values of \,$235 \,\mbox{\footnotesize$\lesssim$}\, M_{\mu\mu}/\rm MeV \,\mbox{\footnotesize$\lesssim$}\, 250$,\, suggests a preference for solution 4.

\begin{figure}[b] \bigskip
\includegraphics[width=3.2in]{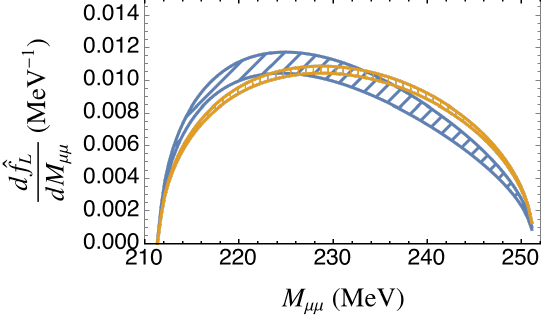} ~ ~ \includegraphics[width=3.2in]{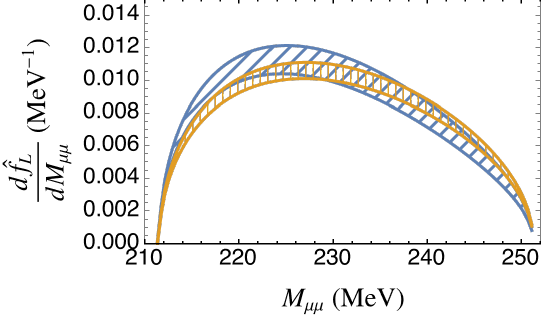}  \vspace{1ex} \\ \includegraphics[width=3.2in]{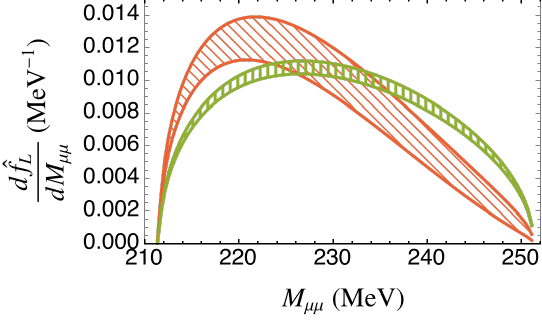} ~ ~ \includegraphics[width=3.2in]{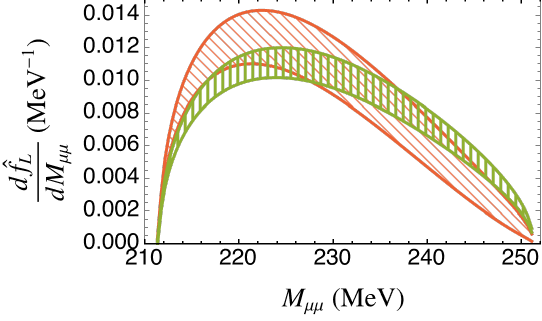}  \vspace{-1ex}
\caption{The $\hat f_L^{}$ distributions in \,$\Sigma^+\to p \mu^+\mu^-$\, in the SM normalized to their corresponding rate for the four relativistic (left plots) and heavy-baryon (right plots) solutions.} \label{f:fldist}
\end{figure}

From the two lines in Eq.\,(\ref{eq:doubledif}), it is straightforward to realize that the fraction $f_L^{}$ of longitudinally polarized dimuons is related to ${\cal F}_{0,2}^{}$ by
\begin{align}
    f_L^{} & \,=\, \frac{2}{3\, \Gamma'} \big({\cal F}_0^{}-{\cal F}_2^{}\big) \,, &
\end{align}
where \,$\Gamma'=2{\cal F}_0^{}+2{\cal F}_2^{}/3$\, with the ${\cal F}_0^{}$ term being numerically dominant, as detailed in Ref.\,\cite{He:2018yzu}.
We can then define the integrated form
\begin{align}
\hat f_L^{} & \,=\, \frac{1}{\hat\Gamma} \int \Gamma'\, f_L^{}\, q^2 \,=\, \frac{4}{3\, \hat\Gamma} \int \big({\cal F}_0^{}-{\cal F}_2^{}\big)\, M_{\mu\mu}^{}\, dM_{\mu\mu}^{} \,. &   \label{fldist}
\end{align}
In Fig.\,\ref{f:fldist} we display $d\hat f_L^{}/dM_{\mu\mu}$ versus $M_{\mu\mu}$.
This distribution mostly mimics the dimuon spectrum in Fig.\,\ref{f:dimuonsp}, which is expected due to the numerical smallness of ${\cal F}_2$ relative to ${\cal F}_0$ within the SM.
It is interesting to mention that experimentally it may be possible to extract $f_L^{}$ from Eq.\,(\ref{eq:doubledif}), as was done for \,$\Lambda_b\to\Lambda\mu^+\mu^-$\, by LHCb~\cite{LHCb:2015tgy}.

\section{New physics contributions\label{bsm}}

To take into account the effects of new physics beyond the SM, we work with the effective Hamiltonian
\begin{align}
{\cal H}_{\rm eff}^{} & \,=\, -\sum_i {\mathscr C}_i{\cal O}_i+ {\rm H.c.} \,, & \label{eq:npeff}
\end{align}
where the operators ${\cal O}_i$ are defined as
\begin{align}
{\cal O}_7^{} & = \frac{\lambda_t^{}G_{\rm F}^{}}{\sqrt2} \frac{{\tt e}\,m_s^{}}{4\pi^2}\, \overline{s_R^{}}\sigma^{\kappa\nu} d_L^{}\, F_{\kappa\nu}^{} \,, &
{\cal O}_{7^\prime}^{} & = \frac{\lambda_t^{}G_{\rm F}^{}}{\sqrt2} \frac{{\tt e}\, m_s^{}}{4\pi^2}\, \overline{s_L^{}}\sigma^{\kappa\nu} d_R^{}\, F_{\kappa\nu}^{} \,,
\nonumber \\
{\cal O}_9^\ell & = \frac{\lambda_t^{}G_{\rm F}}{\sqrt2} \frac{\tt e^2}{4\pi^2}\, \overline{s_L^{}}\gamma^\nu d_L^{}\, \overline\ell\gamma_\nu^{}\ell \,, &
{\cal O}_{9^\prime}^\ell & = \frac{\lambda_t^{}G_{\rm F}^{}}{\sqrt2}  \frac{\tt e^2}{4\pi^2}\, \overline{s_R^{}}\gamma^\nu d_R^{}\, \overline\ell\gamma_\nu^{}\ell \,, \nonumber \\
{\cal O}_{10}^\ell & = \frac{\lambda_t^{}G_{\rm F}^{}}{\sqrt2} \frac{\tt e^2}{4\pi^2}\, \overline{s_L^{}}\gamma^\nu d_L^{}\, \overline\ell \gamma_\nu^{} \gamma_5^{}\ell \,, &
{\cal O}_{10^\prime}^\ell & = \frac{\lambda_t^{}G_{\rm F}^{}}{\sqrt2}  \frac{\tt e^2}{4\pi^2}\, \overline{s_R^{}}\gamma^\nu d_R^{}\, \overline\ell\gamma_\nu^{} \gamma_5^{}\ell \,,
\nonumber \\
{\cal O}_S^{} & = \frac{\lambda_t^{}G_{\rm F}^{}}{\sqrt2} \frac{{\tt e}^2m_s^{}}{4\pi^2}\, \overline{s_L^{}} d_R^{}\, \overline\ell\ell \,, &
{\cal O}_{S^\prime}^{} & = \frac{\lambda_t^{}G_{\rm F}^{}}{\sqrt2} \frac{{\tt e}^2m_s^{}}{4\pi^2}\, \overline{s_R^{}}d_L^{}\, \overline\ell\ell \,,
\nonumber \\
{\cal O}_P^{} & = \frac{\lambda_t^{}G_{\rm F}^{}}{\sqrt2} \frac{{\tt e}^2m_s^{}}{4\pi^2}\, \overline{s_L^{}} d_R^{}\, \overline\ell \gamma_5^{}\ell \,, &
{\cal O}_{P^\prime}^{} & = \frac{\lambda_t^{}G_{\rm F}^{}}{\sqrt2} \frac{{\tt e}^2m_s^{}}{4\pi^2}\, \overline{s_R^{}}d_L^{}\, \overline\ell\gamma_5^{}\ell \,,
 & \label{eq:npdefs}
\end{align}
with \,$\textsl{\tt f}_{L,R}^{}=P_{L,R}^{}\textsl{\tt f}$\, for fermion {\tt f} and \,$\lambda_{\tt q}^{}=V_{{\tt q}d}^{}V_{{\tt q}s}^*$.\,
For the Wilson coefficients ${\mathscr C}_i^{}$ in Eq.\,(\ref{eq:npeff}), we explicitly separate any SM contribution to them by writing \,${\mathscr C}_i^{}=C_i^{\rm SM}+C_i^{}$,\, where ``$C_i$'' designates the part induced by NP.
Since we consider only NP that preserves $CP$ symmetry, we suppose that the products \,$\lambda_t^{}C_i^{}$,\, which occur in ${\cal H}_{\rm eff}$, are real numbers.
As seen below, the $C_i$ enter the kaon observables of interest as either of \,$C_{j\pm}\equiv\big(C_{j}\pm C_{j^\prime}\big)$,\, $j=7,9,10,S,P$,\, and so we will adopt those combinations.

From the H.c. part of ${\cal H}_{\rm eff}$ in Eq.\,(\ref{eq:npeff}), the NP contributions to \,$\Sigma^+\to p\ell^+\ell^-$\, are
\begin{align}
\tilde{\textsc a} & \,=\, \frac{\lambda_t^{}G_{\rm F}^{}}{\sqrt2} \frac{{\tt e}^2m_s^{}}{4\pi^2 q^2}\, c_\sigma^{}\, C_{7+}^{} \,, &
\tilde{\textsc b} & \,=\, \frac{\lambda_t^{}G_{\rm F}^{}}{\sqrt2} \frac{{\tt e}^2m_s^{}}{4\pi^2 q^2}\, c_\sigma^{}\, C_{7-}^{} \,, \nonumber \\
\tilde{\textsc c} & \,=\, \frac{-\lambda_t^{}G_{\rm F}^{}}{\sqrt2} \frac{{\tt e}^2}{8\pi^2}\, C_{9+}^{} \,, &
\tilde{\textsc d} & \,=\, \frac{\lambda_t^{}G_{\rm F}^{}}{\sqrt2} \frac{{\tt e}^2}{8\pi^2} (F-D) C_{9-}^{}  \,, \nonumber \\
\tilde{\textsc e} & \,=\, \frac{\lambda_t^{}G_{\rm F}^{}}{\sqrt2} \frac{{\tt e}^2}{8\pi^2}\, C_{10+}^{} \,, &
\tilde{\textsc f} & \,=\, \frac{\lambda_t^{}G_{\rm F}^{}}{\sqrt2} \frac{{\tt e}^2}{8\pi^2} (D-F) C_{10-}^{}  \,, \nonumber \\
\tilde{\textsc g} & \,=\, \frac{\lambda_t^{}G_{\rm F}^{}}{\sqrt2} \frac{{\tt e}^2m_s^{}}{8\pi^2} \bigg( \frac{m_\Sigma^{}-m_p^{}}{m_s-m_d} \bigg) C_{S+}^{} \,, &
\tilde{\textsc j} & \,=\, \frac{\lambda_t^{}G_{\rm F}^{}}{\sqrt2} \frac{{\tt e}^2m_s^{}}{8\pi^2} \bigg( \frac{m_\Sigma^{}-m_p^{}}{m_d-m_s} \bigg) C_{P+}^{} \,, &
& \label{abcdefgj}
\end{align}

and
\begin{align} \label{hk}
\tilde{\textsc h} & \,=\,  \frac{\lambda_t^{}G_{\rm F}^{}}{\sqrt2} \frac{{\tt e}^2 m_s^{}}{8\pi^2} \Bigg( \frac{m_\Sigma^{}+m_p^{}}{m_{K^0}^2-q^2}\Bigg)\frac{m_{K^0}^2}{m_d^{}+m_s^{}} (D-F) C_{S-}^{} \,, \nonumber \\
\tilde{\textsc k} & \,=\, \frac{\lambda_t^{}G_{\rm F}^{}}{\sqrt2} \frac{{\tt e}^2}{8\pi^2} \Bigg( \frac{m_\Sigma^{}+m_p^{}}{q^2-m_{K^0}^2} \Bigg) (D-F)\Bigg( 2 C_{10-}^{}\, m_\ell^{} + \frac{C_{P-}^{}\, m_{K^0}^2\, m_s^{}}{m_d^{}+m_s^{}} \Bigg) \,, &
\end{align}
where the relation \,$\lambda_t^*C_i^*=\lambda_t^{}C_i^{}$\, has been applied, which is valid for \,${\rm Im}(\lambda_tC_i)=0$.\,
In numerical work we set \,${\tt e}^2 = 4\pi\alpha_{\rm e}^{}(0)$\, for the $a$, $b$, $c$, and $d$ terms in Eq.\,(\ref{abcdefk}), which arise from photon-exchange LD effects, and \,${\tt e}^2 = 4\pi\alpha_{\rm e}^{}(m_Z)$\, in the rest of Eq.\,\,(\ref{abcdefk}) and in Eqs.\,\,(\ref{abcdefgj}) and (\ref{hk}).

\begin{table}[b!] \bigskip
\setlength{\tabcolsep}{1ex}
\begin{tabular}{|c|c|c|c|c|c|}\hline
Solution & SM$_i$ & ~ $C_{7+}$  & ~ $C_{7-}$ & ~ $C_{9+}$ & ~ $C_{9-}$ \\ \hline
Rel 1&2.7& $-$0.070 & ~ 0.34 & ~ 0.046 & ~ 0.37\\
Rel 2&7.8& $-$0.028 & ~ 0.71 & ~ 0.026 & ~ 0.77\\
Rel 3&4.2& ~ 0.025& $-$0.50 & $-0.0008$ & $-$0.55\\
Rel 4&1.2& ~ 0.067& $-$0.14 & $-$0.021 & $-$0.15 \\
 \hline
HB 1&3.6& $-$0.057 & ~ 0.40 & ~ 0.040 & ~ 0.44\\
HB 2&6.2& $-$0.035 & ~ 0.59& ~ 0.029& ~ 0.65\\
HB 3&3.2& ~ 0.032 & $-$0.39& $-$0.0043& $-$0.42\\
HB 4&1.8& ~ 0.054 & $-$0.20& $-$0.015 & $-$0.21\\
\hline
\end{tabular}
\caption{Numerical coefficients SM$_i$ and $n_{ij}^{}$ appearing in Eq.\,(\ref{eq:inter}).
The upper four rows labeled ``Rel'' correspond to the four solutions derived with relativistic baryon $\chi$PT and the bottom four labeled ``HB'' to those derived with the heavy-baryon formulation.}
\label{tab:inter}
\end{table}

\begin{table}[!b] \bigskip
\setlength{\tabcolsep}{1ex}
\begin{tabular}{|c|c|c|c|c|c|c|c|c|c|}\hline
$C_{7+}^2$ & $C_{7-}^2$ & $C_{9+}^2$ & $C_{9-}^2$ & $C_{10+}^2$ & $C_{10-}^2$ & $C_{S+}^2$ & $C_{S-}^2$ & $C_{P+}^2$ & $C_{P-}^2\vphantom{|_|^|}$  \\ \hline
0.194& 1.73& 0.972& 2.06& 5.6&0.41& 0.06 GeV$^2$ & 0.0013 GeV$^2$ & 0.353 GeV$^2$ & 0.01 GeV$^2\vphantom{|^{|^|}}$ \\ \hline
\end{tabular} \smallskip \\
\begin{tabular}{|c|c|c|c|}\hline
 $C_{7+}C_{9+}$  & $C_{7-}C_{9-}$ &$C_{10+}C_{P+}$ & $C_{10-}C_{P-}$ \\ \hline
$-$0.193& 3.75& $-$2.78 GeV & 0.082 GeV \\ \hline
\end{tabular}
\caption{Numerical coefficients $q_{jk}^{}$ appearing in Eq.\,(\ref{eq:inter}).}
\label{tab:sqbl1}
\end{table}

The complete numerical expression for ${\cal B}(\Sigma^+\to p\mu^+\mu^-)$ can be written as
\begin{align}
{\cal B}(\Sigma^+\to p\mu^+\mu^-)_i^{} & \,= \left[ {\rm SM}_i^{} + \sum_j n_{ij}^{}\, \big(\lambda_t^{}C_j\big) + 10^{-2}~\sum_{jk} q_{jk}^{}\, \big(\lambda_t^{} C_j\big) \big(\lambda_t^{} C_k^{}\big) \right] \times 10^{-8} \,, &
\label{eq:inter}
\end{align}
where ${\rm SM}_i$ refers to each of the four solutions derived with relativistic baryon $\chi$PT and with its heavy-baryon formulation.
The terms linear in the Wilson coefficients (WCs) representing NP stem from interference with the SM and are therefore different for each case.
These terms are maximized in size when the WCs interfere with the SM long-distance component.
The part quadratic in the~WCs, $C_j C_k$, originates purely from NP and are thus the same for all the four solutions.
We display our numerical results for $n_{ij}^{}$ and $q_{jk}^{}$ in Tables~\ref{tab:inter} and~\ref{tab:sqbl1},\footnote{Ref.\,\,\cite{Geng:2021fog} gives numerical results for the $C_{9+}^2$ and $C_{9-}^2$ terms.
They differ from our numbers by factors of two, but the authors of Ref.~\cite{Geng:2021fog} have informed us that this is due to a typo that will be fixed in a future version.\medskip} respectively.
To obtain these numbers, we integrate numerically the coefficient of each of the terms in the differential distribution.
These numbers indicate that, with $\lambda_t$ multiplying each WC in Eq.~(\ref{eq:inter}), the dimensionless ones, such as $C_{7-}$ and~$C_{9-}$, must be at least a few times $10^2$ in size to modify the SM rates at the percent level or higher.\footnote{Equivalently, only NP arising from ultraviolet models without a mixing angle suppression similar to $\lambda_t$ can measurably affect this rate.}

We now turn our attention to the integrated forward-backward asymmetry as defined in Eq.\,(\ref{eq:afbin}) in the presence of NP. The decay rate that normalizes this asymmetry also varies when there is NP, as per Eq.~(\ref{eq:inter}), but we present results that include the NP terms only in the numerator of the asymmetry.
The rationale for this choice is that this observable will only be measured after the decay rate is known.
With this in mind, we write for the terms that come from interference with the LD SM contribution as
\begin{align}
{\tilde A}_{\rm FB} & \,=\, \frac{1}{\Gamma(\Sigma^+\to p\mu^+\mu^-)} \int \Gamma'\, {\cal A}_{\rm FB}^{}\, dq^2
\nonumber \\ & \,\supset\, \big[ a_{10+}^{}\, \big(\lambda_t^{}C_{10+}^{}\big) + a_{10-}^{}\, \big(\lambda_t^{}C_{10-}^{}\big) + a_{S+}^{}\, {\rm GeV}\, \big(\lambda_t^{}C_{S+}^{}\big) + a_{S-}^{}\, {\rm GeV}\, \big(\lambda_t^{}C_{S-}^{}\big) \big] \times 10^{-2} \,, \label{eq:afbinnp}
\end{align}
where the coefficients $a_i^{}$ for the four relativistic solutions and their heavy-baryon counterparts are tabulated in Table~\ref{t:afbcoeffs}.\footnote{Numbers corresponding to those in Table~\ref{t:afbcoeffs} are given in Ref.\,\cite{Geng:2021fog} for two of the relativistic solutions, and when converted to our conventions the results disagree with ours. This is partly due to the fact that, unlike our choices in Eq.\,(\ref{<v>}), nonzero $f_1^{}$, $f_2^{}$, and $g_1^{}$ were employed in Ref.\,\cite{Geng:2021fog} along with $g_2^{}=0$.\medskip}
With NP coefficients of order $10^2$, their impact is to cause ${\tilde A}_{\rm FB}$ to exceed its SM expectations by a factor of ten already.

The additional terms in ${\tilde A}_{\rm FB}$, besides those in the second line of Eq.~(\ref{eq:afbinnp}), are expressible as
\begin{align}
  \Gamma(\Sigma^+\to p\mu^+\mu^-)\, {\tilde A}_{\rm FB} & \,\supset\, \sum_{ij}~\tilde{q}_{ij}^{}\, \big(\lambda_t^{}C_i^{}\big) \big(\lambda_t^{}C_j\big)\times 10^{-25} \,, &
  \label{eq:afbinsq}
\end{align}
and therefore the parameters $\tilde{q}_{ij}^{}$ are independent of the SM LD effects.
Our numerical results for $\tilde{q}_{ij}^{}$ can be viewed in Table~\ref{tab:quadafb}.

\begin{table}[h] \bigskip
\setlength{\tabcolsep}{1ex}
\begin{tabular}{|c|c|c|c|c|}  \hline
Solution & $a_{10+}$ & $a_{10-}$ & $a_{S+}$ & $a_{S-}$ \\ \hline
Rel 1 &  $-0.22\pm 0.01$   &  $-1.18\pm0.02$ & ~ $0.082\pm0.001$ & $-0.092\pm0.006$ \\
Rel 2 & $-0.158\pm 0.003$  &  $-0.16\pm0.01$ & ~ $0.020\pm0.001$ & $-0.065\pm0.001$ \\
Rel 3 & ~ $0.214\pm 0.005$ & ~ $0.29\pm0.03$ & ~ $0.015\pm0.001$ & ~ $0.083\pm0.002$ \\
Rel 4 & ~ $0.20\pm 0.03$   & ~ $2.54\pm0.05$ &  $-0.003\pm0.002$ & ~ $0.07\pm0.01$ \\
\hline
HB 1 &  $-0.19\pm 0.02$ &  $-0.70\pm0.04$ & ~$0.054\pm0.001$ & $-0.081\pm0.007$\\
HB 2 &  $-0.16\pm 0.01$ &  $-0.26\pm0.03$ & ~$0.026\pm0.001$ & $-0.067\pm0.003$\\
HB 3 & ~ $0.21\pm 0.01$ & ~ $0.47\pm0.06$ & ~$0.016\pm0.002$ & ~ $0.078\pm0.006$ \\
HB 4 & ~ $0.20\pm 0.03$ & ~ $1.36\pm0.08$ & ~$0.010\pm0.003$ & ~ $0.07\pm0.01$ \\ \hline
\end{tabular}
\caption{Coefficients $a_i^{}$ in Eq.\,(\ref{eq:afbinnp}) for the integrated forward-backward asymmetry.}
\label{t:afbcoeffs}
\end{table}

\begin{table}[h] \bigskip
\setlength{\tabcolsep}{1ex}
\begin{tabular}{|c|c|c|c|c|c|c|c|}\hline
$C_{7+}C_{10-}$ & $C_{7+}C_{S+}$ & $C_{7-}C_{10+}$ & $C_{7-}C_{S-}$ &
$C_{9+}C_{10-}$ & $C_{9+}C_{S+}$ & $C_{9-}C_{10+}$ & $C_{9-}C_{S-}$
\\ \hline
1.5 & $-$0.052 GeV & $-$0.49 & $-$0.20 GeV & $-$0.51 & 1.5 GeV & $-$0.51 & $-$0.24 GeV \\ \hline
\end{tabular}
\caption{Numerical coefficients $\tilde q_{ij}^{}$ appearing in Eq.\,(\ref{eq:afbinsq}).}
\label{tab:quadafb}
\end{table}

\section{Kaon observables\label{kaonobs}}

\subsection{SM Wilson coefficients\label{smwc}}

The kaon modes affected by the quark-level transition \,$s\to d\mu^+\mu^-$\,  are also dominated by long-distance contributions within the SM.
This makes it very difficult to extract precise information on the SM short-distance parameters, or, indeed, new physics. Nevertheless, the corresponding kaon data can serve as useful constraints on NP which is large enough to measurably affect the  $\Sigma^+\to p\mu^+\mu^-$\, rate.
The SM SD contributions enter these modes via the Wilson coefficients, at the scale \,$\mu=1$~GeV,
\begin{align}
    C_7^{\rm SM} & \,=\, \frac{-\lambda_u}{2\lambda_t}\, C_{7\gamma}^{} \,, &
    C_9^{\rm SM} & \,=\,
    \frac{2\pi}{\alpha_{\rm e}} \bigg( y_{7V}^{}-\frac{\lambda_u}{\lambda_t}\, z_{7V}^{} \bigg) \,, &
    C_{10}^{\rm SM} & \,=\,   \frac{2\pi}{\alpha_{\rm e}}\, y_{7A}^{} \,. &
\end{align}
Numerically we adopt \,$z_{7V}^{} = -0.046\alpha_{\rm e}^{}$,  \,$y_{7V}^{} = 0.73\alpha_{\rm e}^{}$,\, and \,$y_{7A}^{} = -0.68\alpha_{\rm e}^{}$,\, as before.\footnote{The error in these coefficients is sometimes quantified as $y_{7V}^{} = (0.73\pm0.04) \alpha_{\rm e}$\, and \,$y_{7A}^{}=(-0.68\pm 0.03) \alpha_{\rm e}^{}$,\, with the uncertainty range arising from taking \,$\bar m_t(m_t)=(167\pm5)$~GeV\, and \,$0.8\leq\mu/{\rm GeV}\leq1.2$\, \cite{Isidori:2004rb}.}

\subsection{Relevant kaon modes}

We begin with the modes \,$K_{L,S}\to \mu^+\mu^-$.\,
For our numerical study, we employ the expressions for all contributions to these two rates that exist in the public {\tt SuperIso} code \cite{Mahmoudi:2009zz,Neshatpour:2021nbn}.
In both cases, the LD contribution is dominated by a two-photon intermediate state.
Its extraction relies on the measured \,$K_{L,S}\to \gamma\gamma$\, rates supplemented with hadronic models.
The LD contributions to \,$K_{L,S}\to \mu^+\mu^-$\, are therefore potentially contaminated by NP entering through $C_{7,7^\prime}$.
This has been considered in the literature, and we will later list the corresponding constraints obtained in~Ref.\,\cite{Mertens:2011ts}, on Table~\ref{t:constraints}.
There are several proposals to improve these measurements \cite{Anzivino:2023bhp}.

\bigskip

1. {\small\boldmath$K_L\to\mu^+\mu^-$} \medskip

This mode receives a SD contribution proportional to \,${\rm Re}~C_{10}^{\rm SM}$,\, but the rate is dominated by the aforesaid LD contribution from a two-photon intermediate state.
The absorptive part of this LD contribution is determined by the measured rate of \,$K_L\to \gamma\gamma$,\, and the difference between this rate and the absorptive contribution can be attributed to a combination of a model-dependent LD dispersive contribution, a SM SD contribution, and NP.
The branching fraction in relation to all of them can be parametrized as  \cite{Chobanova:2017rkj,Neshatpour:2022fak}
\begin{align} \label{eq:brKLmumuComplete}
{\cal B} \big( K^0_{L} \to \mu^+ \mu^- \big) & \,=\, \tau_{K_L}\,  \frac{f_K^2 m_{K^0}^3\, \beta_{\mu,K}^{}}{16\pi} \Big( \big|N_L^{\rm LD}+A_L^{\rm SM}+A_L^{\rm NP} \big|^2 + \beta_{\mu,K}^2\, |B_L|^2 \Big) \,, \nonumber \\
N_L^{\rm LD} & \,=\, \pm   \left[0.54(77) - 3.95 i\right]\times 10^{-11} {\rm~GeV}^{-2} , \nonumber \\
A_L^{\rm SM} & \,=\, \frac{G_{\rm F} \alpha_{\rm e}}{\sqrt{2}\pi}~ \frac{2m_\mu}{m_{K^0}^{}} \Bigg( \frac{ Y_c}{s_W^2}\, {\rm Re}\,\lambda_c+\frac{Y(x_t)}{s_W^2}\, {\rm Re}\,\lambda_t \Bigg) , \nonumber \\
A_L^{\rm NP} & \,=\, \frac{-G_{\rm F} \alpha_{\rm e} }{\sqrt{2}\pi} {\rm~Re} \Bigg[ \frac{2m_\mu}{m_{K^0}^{}} \left(\lambda_tC_{10-}\right) + \frac{m_{K^0}^{}\, m_{s}^{}}{m_d^{}+m_s^{}} \left(\lambda_tC_{P-}\right) \Bigg] \,,
\\ \nonumber
B_L & \,=\, \frac{G_{\rm F}^{} \alpha_{\rm e}^{}}{\sqrt{2}\pi}~ \frac{ m_{K^0}^{}\, m_s^{}}{m_d^{} +m_s^{}}\, {\rm~Im}\left(\lambda_tC_{S-}\right) \,. &
\end{align}
Recall that we are assuming no $CP$ violation in the NP for our numerical study, implying  that we set \,$B_L=0$.\,
We have explicitly separated the SM SD contribution,  $A_L^{\rm SM}$ \cite{Gorbahn:2006bm}, so that $C_{10}$ refers to NP only.

\newpage

2. {\small\boldmath$K_S^{}\to\mu^+\mu^-$} \medskip

This mode also has SD and LD components entering the branching fraction as~\cite{Chobanova:2017rkj,Neshatpour:2022fak}
\begin{align} \label{eq:brKSmumuComplete}
{\cal B} \big( K^0_{S} \to \mu^+ \mu^- \big) & \,=\, \tau_{K_S}\,  \frac{f_K^2 m_{K^0}^3\, \beta_{\mu,K}^{}} {16 \pi} \Big( \beta_{\mu,K}^2\, \big| N_S^{\rm LD}-A_S^{\rm NP} \big|^2 + \big| B_S^{\rm SM}+B_S^{\rm NP} \big|^2 \Big) , \nonumber \\
N_S^{\rm LD} & \,=\, (-2.65 + 1.14 i )\times 10^{-11} {\rm~GeV}^{-2} \,, \nonumber \\
A_S^{\rm NP} & \,=\, \frac{G_{\rm F}^{} \alpha_{\rm e}^{}}{\sqrt2\, \pi}~ \frac{m_{K^0}^{}\, m_s^{}}{m_d^{}+m_s^{}}\, {\rm~Re}\left(\lambda_tC_{S-}\right) , \nonumber \\
B_S^{\rm SM} & \,=\, \frac{G_{\rm F} \alpha_{\rm e}}{\sqrt2\, \pi}~ \frac{2m_\mu}{m_{K^0}^{}}\, {\rm~Im} \Bigg( \frac{-Y_c}{s_W^2}\, \lambda_c-\frac{Y(x_t)}{s_W^2}\, \lambda_t \Bigg) \,, \nonumber \\
B_S^{\rm NP} & \,=\, \frac{G_{\rm F} \alpha_{\rm e}}{\sqrt2\, \pi}\, {\rm~Im} \Bigg[ \frac{2m_\mu}{m_{K^0}^{}} \left(\lambda_t C_{10-}\right) + \frac{m_{K^0}^{}\, m_s^{}}{m_d^{}+m_s^{}} \left(\lambda_t  C_{P-}\right) \Bigg] \,. &
\end{align}
Again, we have  explicitly written the SM SD contribution through $C_{10}$ as $B_S^{\rm SM}$.
In the scenario with no $CP$ violation in the NP, \,$B_S^{\rm NP}=0$.

\bigskip

3. {\small\boldmath$K_{L,S}^{}\to\pi^0\ell^+\ell^-$} \medskip

Within the SM these modes are predominantly $CP$-violating and receive multiple contributions.
There are direct and through-mixing $CP$-violating contributions as well as interference between them,  $C_{\rm dir}^{\ell\ell}$, $C_{\rm mix}^{\ell\ell}$ and $C_{\rm int}^{\ell\ell}$ respectively.
Theoretical considerations suggest that the sign of the interference term is plus \cite{Buchalla:2003sj}.
There is also a LD $CP$-conserving contribution dominated by a two-photon intermediate state,\footnote{For our numerical studies we will use the expressions for these modes as coded for {\tt SuperIso} and provided to us by F. Mahmoudi.\medskip} $C_{\gamma \gamma}^{\ell\ell}$.
Potentially there are NP contributions as well, which we include following Ref.\,\cite{Mescia:2006jd}.
Some enter $C_{\rm dir}^{\ell\ell}$ and others we dub $C_{\rm NP}^{\ell\ell}$.
A practical complication when using these modes to constrain NP is that two contributions are extracted from experiment.
The $CP$-conserving contribution is partially extracted from \,$K_L\to \pi^0\gamma\gamma$\, and as such is potentially contaminated by NP through $C_{7,7^\prime}$.
We partially address this issue by using the constraints on $C_7$ and $C_{7^\prime}$ as obtained from the \,$K\to \pi \pi \gamma$\, and \,$K\to \gamma\gamma$\, modes~\cite{Mertens:2011ts}, respectively.
The $CP$-violating contribution through mixing relies on the measurement of \,$K_S\to \pi^0\ell^+\ell^-$,\, and NP affecting those modes could contaminate the extracted value of $a_S$.
However, an $a_S$ representing LD contributions is the same for both the electron and muon modes, so that the difference between the two extractions can be attributed to NP that breaks lepton flavor universality (LFU).

We follow Refs.\,\cite{Buchalla:2003sj,Isidori:2004rb,Mescia:2006jd,DAmbrosio:2022kvb} to write
\begin{align}
{\cal B}(K_L \to \pi^0 \ell \bar{\ell}) & \,=\, \Big( C_{\rm NP}^{\ell\ell}+C_{\rm dir}^{\ell\ell} \pm C_{\rm int}^{\ell\ell}|a_S| + C_{\rm mix}^{\ell\ell}|a_S|^2 + C_{\gamma \gamma}^{\ell\ell}  \Big) \times 10^{-12} \,. &
\label{eq:klpill}
\end{align}
As the expressions below imply, NP contributions may enter the rate through the term $C^{\ell\ell}_{\rm NP}$ via the coefficients $C_{P+}$ and $C_{S+}$ and also through the term $C^{\ell\ell}_{\rm dir}$ via the coefficients $C_{9+}$ and $C_{10+}$.

\newpage

To exhibit the dependence of the first three terms in Eq.~(\ref{eq:klpill}) on the possible NP, we follow Refs.\,\cite{Mescia:2006jd,DAmbrosio:2022kvb} and write\footnote{It is to be understood that the WCs contributing to the various dielectron and dimuon observables need to have superscripts \,$\ell\ell=ee$\, and \,$\mu\mu$,\, respectively, if LFU is absent and the two types of WCs occur together and are both nonzero, as in a few instances discussed below.
In our numerical analysis in Sec.\,\ref{constraints} we assume that only the \,$\ell=\mu$\, WCs contain NP.}
\begin{align}
     C_{\rm NP}^{\mu\mu} & \,=\, 10^{8}~{\rm Im} (\lambda_t C_{P+\,}{\rm GeV}) \left\{ 0.09~ {\rm Im}\, \lambda_t +\big[0.0039 ~{\rm Im} (\lambda_t C_{P+\,}{\rm GeV}) - 0.02~ {\rm Im}(\lambda_t C_{10+})\big]\right\} , \nonumber \\
     & ~~~ +\, 10^4~ {\rm Re} (\lambda_tC_{S+\,}{\rm GeV}) \big[{-}0.027(7)+0.0019\times 10^4~{\rm Re}(\lambda_t C_{S+\,}{\rm GeV}) \big] , \nonumber \\
    C_{\rm dir}^{\mu\mu} & \,=\, 10^8 \left\{ 0.033 \left[ {\rm Im}(\lambda_t C_{10+})-4.27(19)~ {\rm Im}\,\lambda_t \right]{}^2 + 0.014 \left[{\rm Im}(\lambda_t C_{9+})+4.59(25)~ {\rm Im}\,\lambda_t \right]{}^2 \right\} , \nonumber \\
    C_{\rm int}^{\mu\mu} & \,=\, 10^4 \left[ 1.37(8)~ {\rm Im}\,\lambda_t + 0.3~ {\rm Im}(\lambda_t C_{9+}) \right] , \nonumber \\
    C_{\rm NP}^{ee} & \,=\, 10^{8}~{\rm Im} (\lambda_t C_{P+\,}{\rm GeV}) \left\{ 0.0013~ {\rm Im}\, \lambda_t + \left[ 0.0076 ~{\rm Im} (\lambda_tC_{P+\,}{\rm GeV})-0.0003~ {\rm Im}(\lambda_tC_{10+})\right] \right\} , \nonumber \\
     & ~~~ +\, 10^4~{\rm Re} (\lambda_t C_{S+\,}{\rm GeV})\left[ {-}0.021(4)+0.0077\times 10^4~ {\rm Re}(\lambda_t C_{S+\,}{\rm GeV}) \right] , \nonumber \\
    C_{\rm dir}^{ee} & \,=\, 0.06\times 10^8 \left\{ \left[ {\rm Im}(\lambda_t C_{10+})-4.27(19)~ {\rm Im}\, \lambda_t \right] {}^2 + \left[ {\rm Im}(\lambda_t C_{9+})+4.59(25)~ {\rm Im}\, \lambda_t \right] {}^2 \right\} , \nonumber \\
    C_{\rm int}^{ee} & \,=\, 10^4 \left[ 5.84(35)~ {\rm Im}\, \lambda_t + 1.28~ {\rm Im}(\lambda_tC_{9+}) \right] , & \label{Cllnp}
\end{align}
where the terms linear in \,${\rm Re}~C_{S+}$\, originate from interference with the two-photon contribution.
In Eq.\,(\ref{Cllnp}) the explicit factors of $10^4$ and $10^8$ appear from the practice of writing this result in terms of \,$10^4\,\lambda_t$.\,
In addition, the last two terms in Eq.\,(\ref{eq:klpill}) are given by
\begin{align}
    C_{\rm mix}^{\mu\mu} & = 3.36 \pm 0.20 \,, & C_{\gamma \gamma}^{\mu\mu} & = 5.2 \pm 1.6 \,, &
    C_{\rm mix}^{ee} & = 14.5\pm 0.5 \,, & C_{\gamma \gamma}^{ee} & \approx 0 \,. &
\end{align}
In the scenario we are pursuing here, where the NP is $CP$ conserving, these results indicate that $C_{S+}$ is the only WC that can be constrained by these modes.

However, it is also possible for NP to contaminate the extraction of the parameter $a_S$, and in that way a constrain on $C_{9+}$ arises.
The parameter $a_S$ can be inferred from measurements of $K_S\to \pi^0\ell^+\ell^-$.
Those modes, in turn,  are LD dominated and parametrized in $\chi$PT with the constants $a_S$ and $b_S$ which are then extracted from experiment \cite{Ecker:1987qi,DAmbrosio:1998gur}.
Early extractions assuming LFU suggested the value \,$|a_S| = 1.20 \pm 0.20$\, \cite{Isidori:2004rb}, whereas more recent studies  suggest a larger error, \,$|a_S|=1.3\pm3.2$\, \cite{DAmbrosio:2018ytt}.

A contamination from NP can only be extracted from these if it violates LFU.
The separate measurements of $a_S$ in the muon and electron modes start from the $\chi$PT parametrization of the LD SM branching fractions given by  \cite{Cirigliano:2011ny}
\begin{align}
    {\cal B}(K_S\to\pi^0e^+e^-)=&\left(0.01-0.55a_S-0.17b_S+43.76a_S^2+11.83a_S b_S+1.28b_S^2\right)\times 10^{-10} \,,\nonumber \\
    {\cal B}(K_S\to\pi^0\mu^+\mu^-)=&\left(0.07-3.96a_S-1.34b_S+86.90a_S^2+50.21a_S b_S+7.66b_S^2\right)\times 10^{-11} \,.
\end{align}
Existing measurements do not have enough information to determine both $a_S$ and $b_S$, and so predictions rely on a vector-meson-dominance model leading to \,$b_S/a_S=m_{K}^2/m_\rho^2$\, \cite{DAmbrosio:1998gur}.
Using this form, the experimental collaborations have extracted a value for $a_S$ from each of the two modes~\cite{Batley:2003mu,Batley:2004wg},
\begin{align}
                a_S {\rm~from~} K_S\to\pi^0 e^+e^- ~~~  & \quad1.06^{+0.26}_{-0.21}\pm0.07 \,, \nonumber \\
                a_S {\rm~from~} K_S\to\pi^0\mu^+\mu^- ~~~ & \quad 1.54^{+0.40}_{-0.32}\pm0.06 \,. &
\end{align}
With these two numbers, one can constrain the combination \,$C^{\mu\mu}_{9+}-C^{ee}_{9+}$\, via
\begin{align}\label{eq:K_+FUV}
    a_S^{\mu\mu}-a_S^{ee} & \,=\, -\sqrt{2}\, {\rm Re} \big[ V_{td}^{} V^*_{ts} \big( C^{\mu\mu}_{9+}-C^{ee}_{9+} \big) \big] \,=\, -\sqrt{2}\,|\lambda_t|\big( C^{\mu\mu}_{9+}-C^{ee}_{9+} \big) \,=\, 0.5\pm0.5 \,. &
\end{align}
A forward-backward asymmetry in these modes could also be used to constrain NP \cite{Mescia:2006jd}, but there is no experimental information yet.

\bigskip

4. {\small\boldmath$K^+\to\pi^+\mu^+\mu^-$} \medskip

Within the SM, this decay is also dominated by LD contributions that stem from an intermediate-photon state.
The LD contributions to these modes are parametrized by the low-energy constants $a_+$ and $b_+$ which are themselves extracted from the measurements \cite{Cirigliano:2011ny}:
\begin{align}
    {\cal B}_{\rm LD}^e & \,=\, \big(0.15-3.31a_+-0.90b_++60.51a_+^2+16.36a_+b_++1.77b_+^2\big) \times 10^{-8} \,, \nonumber \\
    {\cal B}_{\rm LD}^\mu & \,=\, \big(1.19-19.97a_+-6.56b_++120.16a_+^2+69.42a_+b_++10.59b_+^2\big) \times 10^{-9} \,. &
\end{align}
The values of $a_+$ and $b_+$  obtained from experiment for the electron mode \cite{NA482:2009pfe} are
\begin{align}
    a_+ & \,=\, -0.578\pm0.016 \,, & b_+ & \,=\, -0.779\pm0.066 &
    \label{abmuon}
\end{align}
and for the muon mode  \cite{NA62:2022qes}
\begin{align}
    a_+ & \,=\, -0.575\pm0.013 \,, & b_+ & \,=\, -0.722\pm0.043 \,. &
    \label{abelectron}
\end{align}
We add possible contributions from NP to the branching ratio resulting, for the muon case, in
\begin{align}
10^8\, {\cal B}(K^+\to\pi^+\mu^+\mu^-)
 & \,=\, {\cal B}_{\rm LD}^\mu + 9.66\, (\lambda_t C_{7+}) + 34\, (\lambda_t C_{9+})+ 17.6\, (\lambda_t C_{7+}) (\lambda_t C_{9+}) \nonumber \\ & ~~~ -\, 48.4\, (\lambda_t C_{10+}) (\lambda_t C_{P+\,}{\rm GeV})
+ 2.5\, (\lambda_t C_{7+})^2 + 31\, (\lambda_t C_{9+})^2 \nonumber \\ & ~~~ +\, 74.3\, (\lambda_t C_{10+})^2 + 9.46\, (\lambda_t C_{P+\,}{\rm GeV})^2 + 4.34\, (\lambda_t C_{S+\,}{\rm GeV})^2 \,.
\label{ktopimm}
\end{align}
This has been partially considered in Ref.\,\cite{Geng:2021fog}, which provides an expression that includes only the term $C_{9+}^2$ with a numerical coefficient approximately twice as large as ours.\footnote{This can be traced to an incorrect factor of $\sqrt{2}$ in the form factor in their Eq.\,(4.16).
The authors of Ref.\,\cite{Geng:2021fog} have confirmed that this is a mistake in their paper.\medskip}

A forward-backward asymmetry  can be defined for this mode in terms of the angle $\theta_{K\mu}$ between the charged-kaon and muon three-momenta in the dilepton center-of-mass frame \cite{Chen:2003nz},
\begin{align}
    A_{{\rm FB},K^+} & \,=\, \frac{{\cal N}(\cos\theta_{K\mu}>0)-{\cal N}(\cos\theta_{K\mu}<0)}{{\cal N}(\cos\theta_{K\mu}>0)+{\cal N}(\cos\theta_{K\mu}<0)} \,. &
\end{align}
In terms of the NP coefficients, the asymmetry is  given by\footnote{Here we differ from Ref.\,\cite{Geng:2021fog} in that we normalize the forward-backward asymmetry to a fixed rate, the current experimental value \,${\cal B}(K^+\to\pi^+\mu^+\mu^-)=(9.17\pm 0.14)\times 10^{-8}$ \cite{Workman:2022ynf}.}
\begin{align}
A_{{\rm FB},K^+} & \,=\, \big[ 0.51 + 0.844~(\lambda_tC_{9+}) + 0.24~ (\lambda_tC_{7+}) \big](\lambda_tC_{S+\,}{\rm GeV}) \,. &
\end{align}
The latest measurements of these observables produced \,${\cal B}(K^+\to\pi^+\mu^+\mu^-)=(9.17\pm 0.08) \times10^{-8}$ and \,$A_{{\rm FB},K^+}=(0.0\pm 0.7)\times 10^{-2}$\, \cite{NA62:2022qes}.

Constraints on NP must rely on comparisons between the muon and electron modes and arise from LFU violation (LFUV).
One possibility  is to take the difference between the branching fraction ${\cal B}(K^+\to\pi^+\mu^+\mu^-)$ evaluated with  $a_+$ and $b_+$ extracted from the muon mode, Eq.~(\ref{abmuon}), and with  $a_+$ and $b_+$ from the electron mode, Eq.~(\ref{abelectron}),
\begin{align}
\Delta{\cal B}(K^+\to\pi^+\mu^+\mu^-)=(-4\pm5)\times 10^{-9},
\end{align}
and to attribute this difference to the NP terms in Eq.~(\ref{ktopimm}).

A second possibility is to compare the parameter $a_+$ as extracted from the muon and electron modes \cite{Crivellin:2016vjc,DAmbrosio:2022kvb},
\begin{align}\label{eq:K_LFUV}
a_+^{\mu\mu}-a_+^{ee} & \,=\, -\sqrt{2}\, \big[ V_{td}^{} V^*_{ts} \big(C^{\mu\mu}_{9+}-C^{ee}_{9+}\big) \big] \,. &
\end{align}
With the current values of $a_+^{\mu\mu,ee}$ \cite{NA482:2009pfe,NA62:2022qes},  this translates into
\begin{align}
a_+^{\mu\mu}-a_+^{ee} & \,=\, 0.003\pm 0.021 \,. &
\end{align}

\section{Organizing the constraints\label{constraints}}

There are ten distinct WCs to describe NP in modes with muons and ten more in modes with electrons if we allow LFUV.
Fortunately, as Table~\ref{t:modescon} indicates, many of the modes depend only on one or two WCs, simplifying the numerical study.

\begin{table}[b] \bigskip \setlength{\tabcolsep}{1ex}
\begin{tabular}{|c|c|c|} \hline
Observable & ~WC---linear~ & WC---quadratic \\ \hline
${\cal B}(K_L\to \mu^+\mu^-)$ & $C_{10-},\,C_{P-}$ & $C_{10-},\,C_{P-}\vphantom{|_|^|}$ \\
${\cal B}(K_S\to \mu^+\mu^-)$ & $C_{S-}$ & $C_{S-}\vphantom{|_|^|}$ \\
${\cal B}(K_L\to \pi^0\mu^+\mu^-)$  & $C_{S+}$ & $C_{S+}\vphantom{|_|^|}$ \\
${\cal B}(K_L\to \pi^0 e^+e^-)$  & $C_{S+}$ & $C_{S+}\vphantom{|_|^|}$ \\
$a_S^{\mu\mu}-a_S^{ee}$ & $C_{9+}\vphantom{|_|^|}$ & $\cdots$ \\
$a_+^{\mu\mu}-a_+^{ee}$ & $C_{9+}\vphantom{|_|^|}$ & $\cdots$ \\
$A_{{\rm FB},K^+}$ & $C_{S+}$ & $C_{7+},\, C_{9+},\, C_{S+}\vphantom{|_|^|}$ \\
$\Delta{\cal B}(K^+\to\pi^+\mu^+\mu^-)$ & $C_{7+},\, C_{9+}$ & $C_{7+},\, C_{9+},\, C_{10+},\, C_{S+},\, C_{P+}\vphantom{|_|^|}$ \\
${\cal B}(\Sigma^+ \to p \mu^+\mu^-) $ & $C_{7\pm},\, C_{9\pm}$ & $C_{9\pm},\, C_{10\pm},\, C_{S\pm},\, C_{P\pm}\vphantom{|_|^|}$
\\ \hline
\end{tabular}
\caption{Wilson coefficients that affect the different kaon and hyperon observables under our scenario of $CP$-conserving NP. The ellipses indicate the absence of quadratic contributions.}
\label{t:modescon}
\end{table}

\begin{table}[t] \setlength{\tabcolsep}{2ex}
\begin{tabular}{|c|c|c|} \hline
Observable & 90\% CL range & Constraint on WC \\ \hline
${\cal B}(K_L\to \mu^+\mu^-)$ \cite{Workman:2022ynf}& $(6.66,7.02)\times 10^{-9}$ & $-6.4\leq C_{KL}\leq-4.7 {\rm ~~or~~} {-}1\leq C_{KL}\leq0.73\vphantom{|_|^{|^|}}$ \\
${\cal B}(K_S\to \mu^+\mu^-)$\cite{LHCb:2020ycd}& $<2.1\times 10^{-10}$ & $|C_{S-}|<40{\rm~GeV}^{-1}\vphantom{|_|^|}$ \\
${\cal B}(K_L\to \pi^0\mu^+\mu^-)$\cite{AlaviHarati:2000hs}& $< 3.8\times 10^{-10}$ & $|C_{S+}| < 124 {\rm~GeV}^{-1}\vphantom{|_|^|}$ \\
${\cal B}(K_L\to \pi^0e^+e^-)$\cite{AlaviHarati:2003mr}& $< 2.8\times 10^{-10}$ & $|C_{S+}|< 52 {\rm~GeV}^{-1}\vphantom{|_|^|}$ \\
$a_S^{\mu\mu}-a_S^{ee}$ & $(-0.3,1.3)$ &  $-603<C_{9+}<2611\vphantom{|_|^|}$ \\
$ |a_+^{\mu\mu}-a_+^{ee}|$ & $< 0.035$ &  $|C_{9+}| < 70\vphantom{|_|^|}$ \\ \hline
Processes & Result from \cite{Mertens:2011ts} & Constraint on WC\\ \hline
$K\to \pi \pi\gamma$ \cite{Mertens:2011ts} & $ | C^-_\gamma|\lesssim 0.1 G_{\rm F}m_K$ &  $|C_{7-}|<278\vphantom{|_|^{|^|}}$ \\
$K\to \gamma\gamma$ \cite{Mertens:2011ts}& $| C^{\pm}_\gamma|\lesssim 0.3 G_{\rm F}m_K$  &  $|C_{7\pm}|<833\vphantom{|_|^|}$ \\ \hline
\end{tabular}
\caption{The 90\%-CL ranges chosen for the different modes along with the restrictions they impose on WCs in cases in which only one WC enters the observable.
\,$C_{KL}\equiv C_{10-}+1.17\;{\rm GeV}~C_{P-}$\, is the combination constrained by $K_L\to\mu^+\mu^-$.
The bounds in the last two rows are taken from Ref.\,\cite{Mertens:2011ts} and are not 90\% CL.}
\label{t:constraints}
\end{table}

Direct constraints on $C_7$ and $C_{7^\prime}$ can be obtained in principle from radiative hyperon and kaon decays \cite{He:1999ik,Tandean:1999mg,Mertens:2011ts}.
However, our method to estimate the LD contributions to hyperon decay involves using the $a$ and $b$ parameters (defined in Sec.\,\ref{ld}) obtained from experiment, entailing that these could be contaminated by any NP of this form.
In principle, a more detailed study of all radiative hyperon modes could be used to isolate any NP here, but we will not pursue it in this paper.
Instead, we rely on the constraints deduced from radiative kaon modes by Ref.\,\cite{Mertens:2011ts}, as listed in Table~\ref{t:constraints}.

From Eq.\,(\ref{eq:inter}) and Tables~\ref{tab:inter} and \ref{tab:sqbl1}, one can infer that NP scenarios with \,$C_{7,7^\prime,9,9^\prime,10,10^\prime}\sim 10^4$\, and \,$C_{P,P^\prime,S,S^\prime}\sim 10^5{\rm\,GeV}^{-1}$\, would be necessary to affect the \,$\Sigma^+\to p \mu^+\mu^-$\, rate at a level that could be observable over the LD contributions.
Since this is significantly larger than the bounds placed by kaon physics, we will first impose the kaon restrictions and then use \,$\Sigma^+\to p \mu^+\mu^-$\, to constrain the directions to which the kaons are blind.
We will not delve into specific ultraviolet completions in this paper, but simply note that WCs of this order would be consistent with dimensional analysis in models without the CKM suppression $\lambda_t$ that occurs in the SM and that was rather arbitrarily introduced by the notation of Eq.\,\,(\ref{eq:npdefs}).
In Table\,\,\ref{t:modescon} we summarize the WCs that can introduce NP into the different kaon observables discussed thus far.

The observables in Table\,\ref{t:modescon} are restricted numerically by the 90\% confidence-level (CL) bounds and ranges \cite{Workman:2022ynf} shown in Table\,\ref{t:constraints}, which also displays the corresponding limits on WCs for the cases in which only one of them is present.
In Eq.\,(\ref{eq:brKLmumuComplete}) we notice that \,$K_L\to\mu^+\mu^-$\, restrains a combination of $C_{10-}$ and $C_{P-}$, which we dub $C_{KL}$,
\begin{align}
C_{KL}^{} & \,=\, C_{10-}^{} + \frac{ m_{K_0}^2\, m_s^{}}{2m_\mu\, (m_d+m_s)}\, C_{P-}^{} \,\simeq\, C_{10-}^{} + 1.17\;{\rm GeV}~C_{P-}^{} \,. &
\end{align}
Interestingly, this is the same combination that appears in the \,$\Sigma^+\to p \mu^+\mu^-$\, form-factor $\tilde{\textsc k}$ in~Eq.\,(\ref{hk}).
However, the hyperon mode also depends on  $C_{10-}$  through  $\tilde{\textsc f}$ in~Eq.\,(\ref{abcdefgj}).
The last two rows in  Table\,\ref{t:constraints} exhibit bounds taken directly from Ref.\,\cite{Mertens:2011ts}.

After employing all the single parameter bounds from Table~\ref{t:constraints}, we are left with the following constraints:
\begin{align}
-1.2 & \,\leq\, 10^8~\Delta{\cal B}(K^+\to\pi^+\mu^+\mu^-) \,\leq\, 0.4 \,, &  &
 \big|A_{{\rm FB},K^+}(K^+\to\pi^+\mu^+\mu^-)\big| \,\leq\, 0.12 \,, ~~~
\nonumber \\
0.26 & \,\leq\, 10^8~{\cal B}(\Sigma^+ \to p \mu^+\mu^-) \,\leq\, 5.2 \,.
    \label{eq:constraints}
\end{align}
Because the restriction from \,$K_L\to \mu^+\mu^-$\, is found to be much stronger than that from the $\Sigma^+$ decay, we present the results for  $C_{10-}$ and $C_{P-}$ in terms of $C_{KL}$ and its orthogonal counterpart on the $C_{10-}$-$C_{P-}$ plane, namely \,$C_{\Sigma}\equiv C_{10-}-0.85 {\rm ~GeV}~C_{P-}$.
The allowed $C_{KL}$ range is quoted in Table~\ref{t:constraints}.

\begin{figure}[b] \bigskip
\includegraphics[width=3.1in]{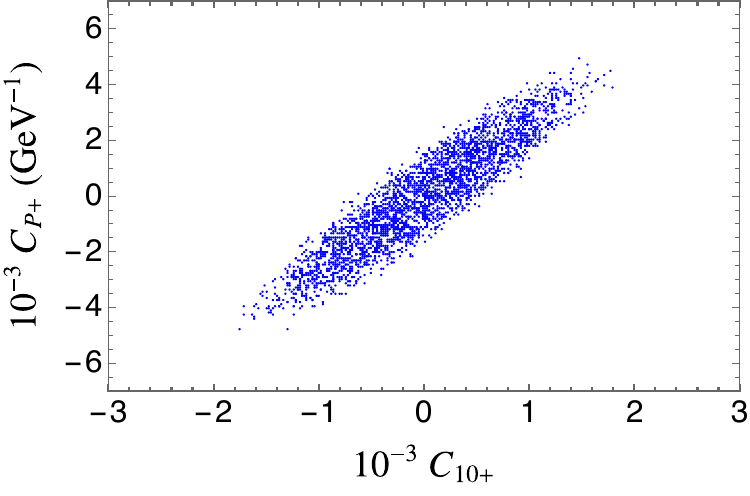} ~ ~ \includegraphics[width=3.2in]{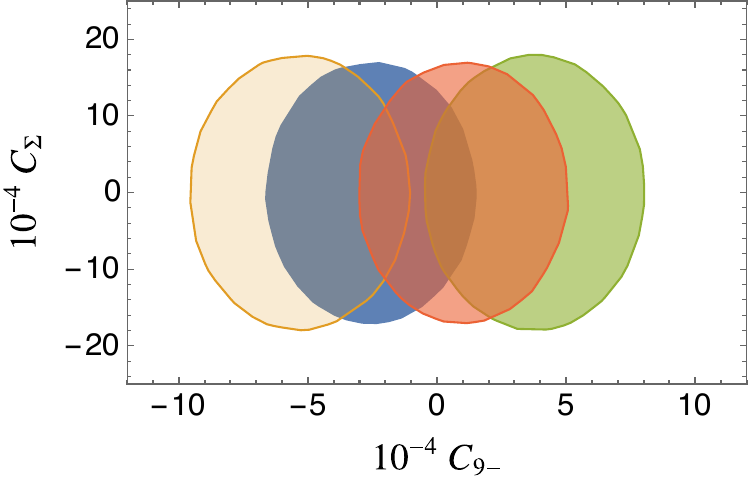}
\caption{Left: allowed $C_{10+}$-$C_{P+}$ parameter space, mostly restricted by $\Delta{\cal B}(K^+\to\pi^+\mu^+\mu^-)$. Right: allowed $C_\Sigma$-$C_{9-}$ parameter space,  restricted by the LHCb measurement of ${\cal B}(\Sigma^+ \to p \mu^+\mu^-)$.  The four relativistic SM LD solutions lead to different allowed regions distinguished by the different colors (blue, yellow, green, red).} \label{f:deltabcon}
\end{figure}

Imposing all the constraints simultaneously leads to allowed regions in $C_{10+}$-$C_{P+}$ and $C_\Sigma$-$C_{9-}$ planes that differ by an order of magnitude.
We therefore display them separately in Fig.\,\ref{f:deltabcon}.
The $C_{10+}$-$C_{P+}$ region is dominantly constrained by $\Delta{\cal B}(K^+\to\pi^+\mu^+\mu^-)$, whereas ${\cal B}(\Sigma^+ \to p \mu^+\mu^-) $ restricts the parameters $C_\Sigma$ and $C_{9-}$.
For the latter, each of the four SM LD solutions leads to slightly different constraints, as illustrated on the right panel of Fig.\,\ref{f:deltabcon} for the relativistic solutions.
The figure shows two-dimensional  projections of the full parameter space, but the numerical scan requires all parameters to satisfy all the constraints in Table\,\ref{t:constraints} and Eq.\,(\ref{eq:constraints}).

\begin{figure}[b!] \bigskip
\includegraphics[width=3.1in]{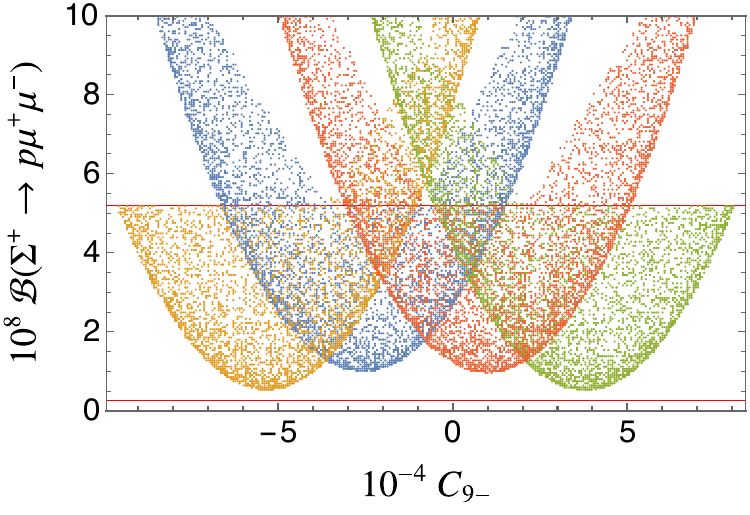} ~ ~ \includegraphics[width=3.2in]{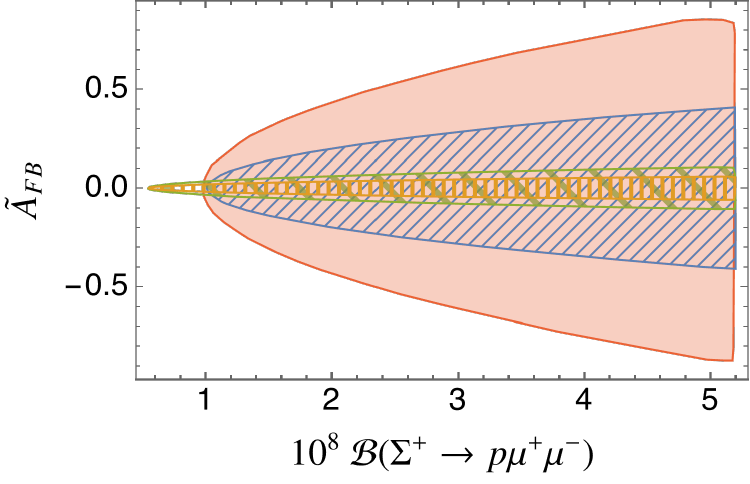}
\caption{Left: the \,$\Sigma^+ \to p \mu^+\mu^-$\, branching fraction as a function of $C_{9-}$ for each of the four relativistic SM LD solutions (blue, yellow, green, red).
The two horizontal red lines span the 90\%-CL empirical range quoted at the bottom of Eq.\,(\ref{eq:constraints}).
Right: the corresponding correlation between the branching fraction and the integrated muon forward-backward asymmetry, $\tilde{A}_{\rm FB}$, in this decay mode.} \label{f:sigmacon}
\end{figure}
\begin{figure}[!b] \bigskip
\includegraphics[width=3.2in]{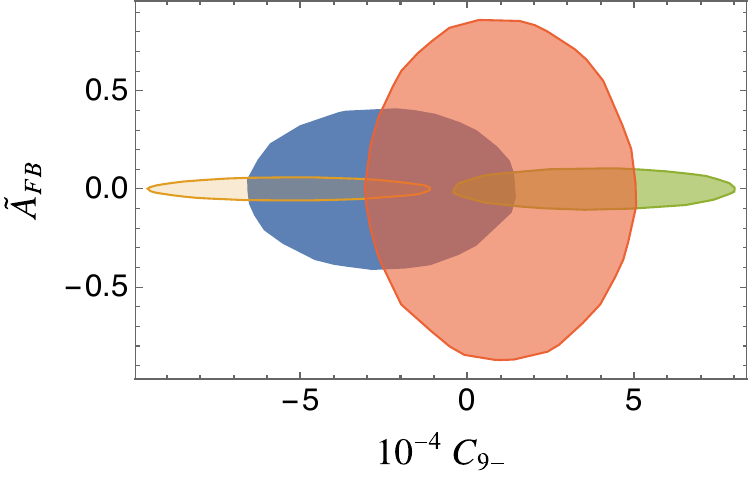} ~ ~ \includegraphics[width=3.2in]{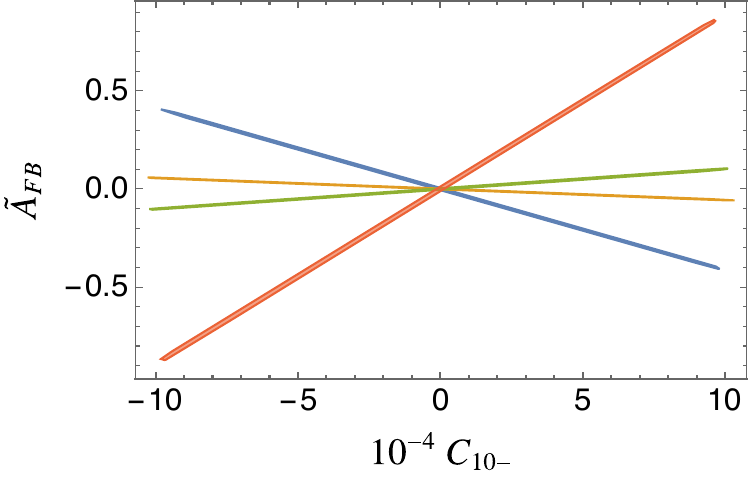}
\caption{The region representing the values of $\tilde{A}_{\rm FB}$ over the allowed range of $C_{9-}$ (left) and $C_{10-}$ (right) for each of the four relativistic SM LD solutions (blue, yellow, green, red).} \label{fig:asym_NP}
\end{figure}

The most important NP contribution to ${\cal B}(\Sigma^+ \to p \mu^+\mu^-)$ enters via $C_{9-}$, as suggested by the numerical entries in Table~\ref{tab:inter}.
The allowed range of $C_{9-}$ corresponding to each of the four relativistic solutions is exhibited on the left panel of Fig.\,\ref{f:sigmacon}.
Its right panel displays the correlation between the  integrated forward-backward asymmetry and the branching fraction for each of the four solutions.

The resulting integrated muon forward-backward asymmetry $\tilde A_{\rm FB}$ for $\Sigma^+\to p \mu^+\mu^-$ across the parameter space that survives all the constraints is also most sensitive to $C_{10-}$, as can be inferred from~Table~\ref{t:afbcoeffs}.
We show this in the right panel of Fig.~\ref{fig:asym_NP} for the four relativistic SM LD solutions.
The left panel exhibits the correlation between the asymmetry and $C_{9-}$, where the variation with $C_{9-}$ is due to the changing rate.
Although Table~\ref{t:afbcoeffs} suggests that the asymmetry would be equally sensitive to $C_{10+}$, this is not observed in the plots because the allowed range of this parameter is considerably smaller.

\begin{figure}[t]
\includegraphics[width=3.2in]{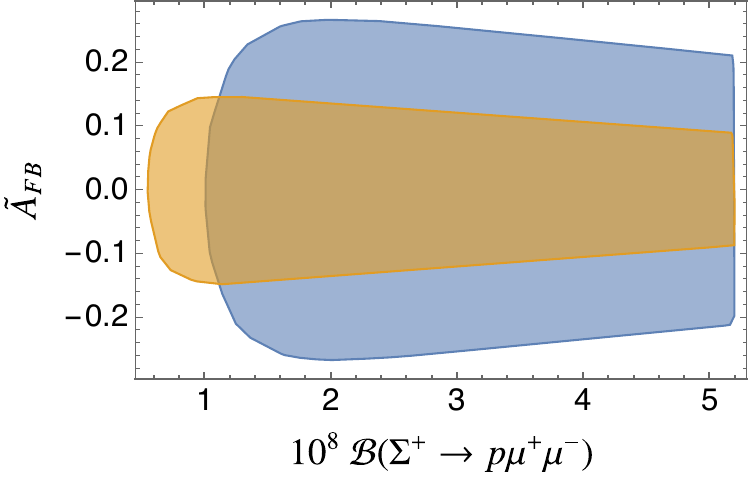} ~ ~ \includegraphics[width=3.2in]{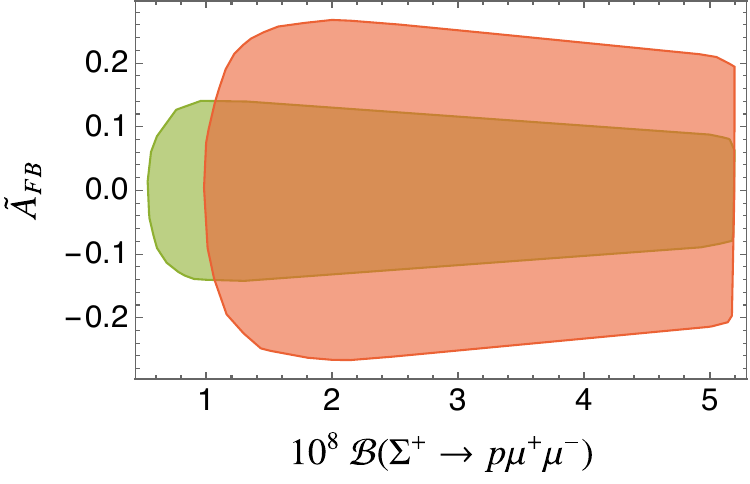} \vspace{3ex} \\
\includegraphics[width=3.2in]{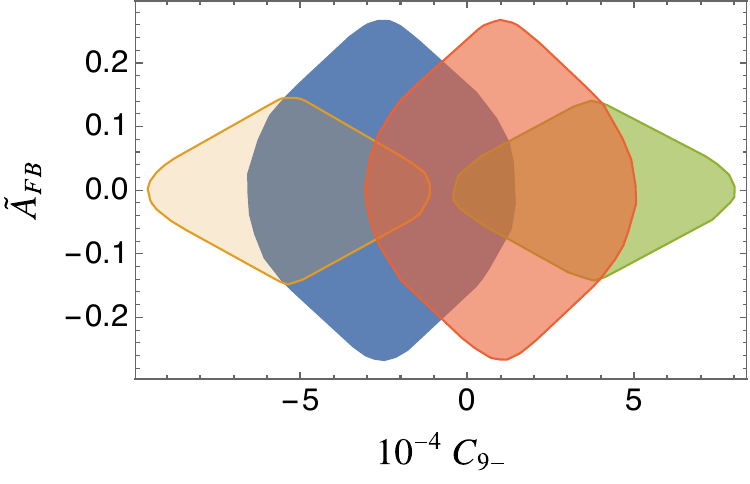} ~ ~ \includegraphics[width=3.2in]{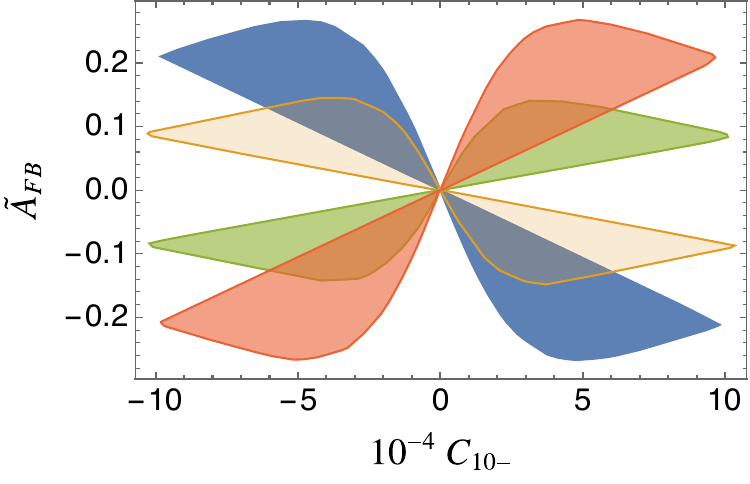}
\caption{Top: correlation between the \,$\Sigma^+\to p \mu^+\mu^-$\, branching fraction and integrated muon forward-backward asymmetry, $\tilde{A}_{\rm FB}$, when new physics is included in the rate normalizing the latter.
Two solutions are plotted on the left panel and two on the right panel, as they overlap.
Bottom: the region representing the values of $\tilde{A}_{\rm FB}$ over the allowed range of $C_{9-}$ (left) and $C_{10-}$ (right) for each of the four relativistic SM LD solutions (blue, yellow, green, red) when new physics contributions are included in the rate normalizing the forward-backward asymmetry.}
    \label{fig:asym_NP2} \bigskip
\end{figure}

It is important to note that our definition for $\tilde A_{\rm FB}$ allows for NP in the numerator, but is normalized to the SM rate for each of the four solutions.
When the contributions of NP are large this can lead to some misleadingly large anomalies seen in Fig.\,\ref{fig:asym_NP}.
If we replace the denominator of the forward-backward asymmetry with the rate including NP contributions, the anomaly changes to what is exhibited in Fig.\,\ref{fig:asym_NP2}.

\section{Summary and conclusions\label{concl}}

We have revisited all aspects of the calculation of the rate and muon forward-backward asymmetry for the mode \,$\Sigma^+\to p \mu^+\mu^-$.\, This was motivated by the expected more precise measurement by LHCb and recent progress in related modes by BESIII.
We now summarize our results and compare them with the existing literature.

Within the SM, we have corrected a long-standing issue with the short-distance contribution.
We updated the evaluation of the Wilson coefficient $C_{7\gamma}$ from Ref.\,\cite{Shifman:1976de}, arriving at a much larger value than previously used.
Nevertheless, its impact on the \,$\Sigma^+\to p\mu^+\mu^-$\, rate is still negligible relative to the long-distance contribution.

We have also updated the experimental input to the determination of the dominant long-distance contribution to \,$\Sigma^+\to p\mu^+\mu^-$.\,
This reduced the uncertainty in the predictions that is attributable to experimental error.
With the new numbers, the estimate for hadronic error obtained by interpolating the results from leading-order relativistic baryon $\chi$PT and from its heavy-baryon formulation is also diminished, to the extent that the four separate solutions remain separate throughout the interpolated range.

In spite of these improvements, the fourfold ambiguity in the prediction for the rate remains.
We discussed several ways which could address this issue from the experimental side.
The preliminary data on the dimuon invariant-mass spectrum reported very recently by LHCb~\cite{LHCb:2024} suggests a preference for solution\,\,4, which was already favored, along with solutions 1 and 3, by the rate measured in 2018~\cite{LHCb:2017rdd}.
The forthcoming new result of LHCb on the rate will hopefully clarify the picture further.
Measurements of the \,$\Sigma^+\to pe^+e^-$\, branching fraction, with a judiciously chosen minimum cut on the dielectron invariant mass, potentially feasible by BESIII and LHCb, could offer another avenue to resolve the fourfold ambiguity.
It might even be possible to reduce the ambiguity to a twofold one if the decay parameter $\gamma_\gamma$ in the radiative mode \,$\Sigma^+\to p\gamma$\, could be measured.
On the theory side, the desired  resolution may be fully attainable in the future via computations on the lattice.

Subsequently, we dealt with possible constraints on new physics in a model-independent manner through a general low-energy effective Hamiltonian for the \,$s\to d \ell^+\ell^-$\, transition.
In this way we examined the restrictions that can be placed on the various Wilson coefficients in light of the latest data on rare kaon decays and \,$\Sigma^+\to p\mu^+\mu^-$,\, emphasizing the usefulness of the hyperon mode in probing the directions in parameter space to which the kaons are blind.

Limits on subsets of these Wilson coefficients from kaon measurements have been derived in the literature, such as Refs.\,\cite{He:1999ik,Tandean:1999mg,Isidori:2004rb,Mescia:2006jd,Mertens:2011ts,Mahmoudi:2009zz,Crivellin:2016vjc,Neshatpour:2021nbn,Chobanova:2017rkj,Geng:2021fog,DAmbrosio:2022kvb,Neshatpour:2022fak}.
We have utilized the results from these papers to update our input parameters of interest.
Moreover, we have identified two more observables in kaon modes that are sensitive to the violation of lepton-flavor universality, namely $\Delta{\cal B}(K^+\to\pi^+\mu^+\mu^-)$ and \,$a_S^{\mu\mu}-a_S^{ee}$.

Generic new-physics contributions to \,$\Sigma^+\to p\mu^+\mu^-$\, were already considered in Ref.\,\cite{He:2018yzu}, albeit not in terms of the Wilson coefficients of the low-energy effective field theory (LEFT).
In Ref.\,\cite{Geng:2021fog} the impact of a subset of them on this mode was studied, specifically the linear effects of $C_{S\pm}$ and $C_{10\pm}$ on the muon forward-backward asymmetry and $C_{9\pm}^2$ on the rate.
In contrast, we have included the effects of all the Wilson coefficients at both the linear and quadratic levels.
In particular, we found that the linear term in the rate, brought about by the interference between SM and NP amplitudes,  provides the most stringent constraint on $C_{9-}$.
Importantly, we further found that the hyperon rate complements the kaon modes by probing the blind direction~$C_\Sigma$, which includes $C_{P-}$, and that the different solutions for the SM long-distance rate are sensitive to different windows of $C_{9-}$.

In this work we have not looked at strangeness-changing neutral current processes with missing energy, on which there is extensive literature as well.
Studies concerning the constraints on new physics from the golden kaon modes, \,$K\to\pi\nu\bar\nu$,\, include Refs.\,\cite{Buras:2004uu,DAmbrosio:2001kut,Buras:2013ooa,Buras:2015yca,Fajfer:2018bfj,He:2018uey,Aebischer:2020mkv,Geng:2021fog}.
The possibility of employing hyperon decays with missing energy to test for new physics has also been investigated before \cite{Tandean:2019tkm,Li:2019cbk,Su:2019tjn,Geng:2021fog}, but empirical information about them is not yet available to date.

We do not pursue connections between the dineutrino and charged-dilepton modes, which arise in the SMEFT or within specific models, for a few reasons.
First, our present focus is on hyperon decay, where no measurements exist for the missing energy modes.
Second, with general beyond-the-SM physics, the connection between these modes implied by the SMEFT is obviated by the fact that the neutrinos are not observed.
This implies that lepton-flavor violating operators, operators with third-family leptons, and operators with new invisible light particles could contribute to the missing energy modes but not to the decays into charged-lepton pairs.

\bigskip

{\bf\small ACKNOWLEDGMENTS} \medskip
\\
We thank Cong Geng and Hong-Fei Shen for experimental information.
The work of A.R. and G.V. was supported in part by an Australian Research Council Discovery Project.
We are grateful to F. Mahmoudi for her help with {\tt SuperIso}, which included providing us with her private code for some of the kaon modes discussed in this paper.
The work of J.T. was supported in part by the National Key R\&D Program of China under Contract No. 2023YFA1606000 and the National Natural Science Foundation of China (NSFC) under Contract No. 11935018.
He would like to thank Hai-Bo Li and the Institute of High Energy Physics, Chinese Academy of Sciences, for kind hospitality and support during the preparation of this paper.

\bigskip

\bibliography{refs.bib}

\begin{thebibliography}{79}%
\makeatletter
\providecommand \@ifxundefined [1]{%
 \@ifx{#1\undefined}
}%
\providecommand \@ifnum [1]{%
 \ifnum #1\expandafter \@firstoftwo
 \else \expandafter \@secondoftwo
 \fi
}%
\providecommand \@ifx [1]{%
 \ifx #1\expandafter \@firstoftwo
 \else \expandafter \@secondoftwo
 \fi
}%
\providecommand \natexlab [1]{#1}%
\providecommand \enquote  [1]{``#1''}%
\providecommand \bibnamefont  [1]{#1}%
\providecommand \bibfnamefont [1]{#1}%
\providecommand \citenamefont [1]{#1}%
\providecommand \href@noop [0]{\@secondoftwo}%
\providecommand \href [0]{\begingroup \@sanitize@url \@href}%
\providecommand \@href[1]{\@@startlink{#1}\@@href}%
\providecommand \@@href[1]{\endgroup#1\@@endlink}%
\providecommand \@sanitize@url [0]{\catcode `\\12\catcode `\$12\catcode
  `\&12\catcode `\#12\catcode `\^12\catcode `\_12\catcode `\%12\relax}%
\providecommand \@@startlink[1]{}%
\providecommand \@@endlink[0]{}%
\providecommand \url  [0]{\begingroup\@sanitize@url \@url }%
\providecommand \@url [1]{\endgroup\@href {#1}{\urlprefix }}%
\providecommand \urlprefix  [0]{URL }%
\providecommand \Eprint [0]{\href }%
\providecommand \doibase [0]{https://doi.org/}%
\providecommand \selectlanguage [0]{\@gobble}%
\providecommand \bibinfo  [0]{\@secondoftwo}%
\providecommand \bibfield  [0]{\@secondoftwo}%
\providecommand \translation [1]{[#1]}%
\providecommand \BibitemOpen [0]{}%
\providecommand \bibitemStop [0]{}%
\providecommand \bibitemNoStop [0]{.\EOS\space}%
\providecommand \EOS [0]{\spacefactor3000\relax}%
\providecommand \BibitemShut  [1]{\csname bibitem#1\endcsname}%
\let\auto@bib@innerbib\@empty
\bibitem [{\citenamefont {Aaij}\ \emph
  {et~al.}(2018{\natexlab{a}})\citenamefont {Aaij} \emph
  {et~al.}}]{LHCb:2017rdd}%
  \BibitemOpen
  \bibfield  {author} {\bibinfo {author} {\bibfnamefont {R.}~\bibnamefont
  {Aaij}} \emph {et~al.} (\bibinfo {collaboration} {LHCb}),\ }\bibfield
  {title} {\bibinfo {title} {{Evidence for the rare decay $\Sigma^+ \to p \mu^+
  \mu^-$}},\ }\href {https://doi.org/10.1103/PhysRevLett.120.221803} {\bibfield
   {journal} {\bibinfo  {journal} {Phys. Rev. Lett.}\ }\textbf {\bibinfo
  {volume} {120}},\ \bibinfo {pages} {221803} (\bibinfo {year}
  {2018}{\natexlab{a}})},\ \Eprint {https://arxiv.org/abs/1712.08606}
  {arXiv:1712.08606 [hep-ex]} \BibitemShut {NoStop}%
\bibitem [{\citenamefont {Li}(2017)}]{Li:2016tlt}%
  \BibitemOpen
  \bibfield  {author} {\bibinfo {author} {\bibfnamefont {H.-B.}\ \bibnamefont
  {Li}},\ }\bibfield  {title} {\bibinfo {title} {{Prospects for rare and
  forbidden hyperon decays at BESIII}},\ }\href
  {https://doi.org/10.1007/s11467-017-0691-9} {\bibfield  {journal} {\bibinfo
  {journal} {Front. Phys. (Beijing)}\ }\textbf {\bibinfo {volume} {12}},\
  \bibinfo {pages} {121301} (\bibinfo {year} {2017})},\ \bibinfo {note}
  {[Erratum: Front. Phys. (Beijing) 14, 64001 (2019)]},\ \Eprint
  {https://arxiv.org/abs/1612.01775} {arXiv:1612.01775 [hep-ex]} \BibitemShut
  {NoStop}%
\bibitem [{\citenamefont {Alves~Junior}\ \emph {et~al.}(2019)\citenamefont
  {Alves~Junior} \emph {et~al.}}]{AlvesJunior:2018ldo}%
  \BibitemOpen
  \bibfield  {author} {\bibinfo {author} {\bibfnamefont {A.~A.}\ \bibnamefont
  {Alves~Junior}} \emph {et~al.},\ }\bibfield  {title} {\bibinfo {title}
  {{Prospects for measurements with strange hadrons at LHCb}},\ }\href
  {https://doi.org/10.1007/JHEP05(2019)048} {\bibfield  {journal} {\bibinfo
  {journal} {J. High Energy Phys.}\ }\textbf {\bibinfo {volume} {05}},\
  \bibinfo {pages} {048 (2019)}},\ \Eprint {https://arxiv.org/abs/1808.03477}
  {arXiv:1808.03477 [hep-ex]} \BibitemShut {NoStop}%
\bibitem [{\citenamefont {Aaij}\ \emph
  {et~al.}(2018{\natexlab{b}})\citenamefont {Aaij} \emph
  {et~al.}}]{LHCb:2018roe}%
  \BibitemOpen
  \bibfield  {author} {\bibinfo {author} {\bibfnamefont {R.}~\bibnamefont
  {Aaij}} \emph {et~al.} (\bibinfo {collaboration} {LHCb}),\ }\bibfield
  {title} {\bibinfo {title} {{Physics case for an LHCb Upgrade II -
  Opportunities in flavour physics, and beyond, in the HL-LHC era}},\
  }\href@noop {} {\  (\bibinfo {year} {2018}{\natexlab{b}})},\ \Eprint
  {https://arxiv.org/abs/1808.08865} {arXiv:1808.08865 [hep-ex]} \BibitemShut
  {NoStop}%
\bibitem [{\citenamefont {Cerri}\ \emph {et~al.}(2019)\citenamefont {Cerri}
  \emph {et~al.}}]{Cerri:2018ypt}%
  \BibitemOpen
  \bibfield  {author} {\bibinfo {author} {\bibfnamefont {A.}~\bibnamefont
  {Cerri}} \emph {et~al.},\ }\bibfield  {title} {\bibinfo {title} {{Report from
  Working Group 4}: {Opportunities in Flavour Physics at the HL-LHC and
  HE-LHC}},\ }\href {https://doi.org/10.23731/CYRM-2019-007.867} {\bibfield
  {journal} {\bibinfo  {journal} {CERN Yellow Rep. Monogr.}\ }\textbf {\bibinfo
  {volume} {7}},\ \bibinfo {pages} {867} (\bibinfo {year} {2019})},\ \Eprint
  {https://arxiv.org/abs/1812.07638} {arXiv:1812.07638 [hep-ph]} \BibitemShut
  {NoStop}%
\bibitem [{\citenamefont {Goudzovski}\ \emph {et~al.}(2023)\citenamefont
  {Goudzovski} \emph {et~al.}}]{Goudzovski:2022vbt}%
  \BibitemOpen
  \bibfield  {author} {\bibinfo {author} {\bibfnamefont {E.}~\bibnamefont
  {Goudzovski}} \emph {et~al.},\ }\bibfield  {title} {\bibinfo {title} {{New
  physics searches at kaon and hyperon factories}},\ }\href
  {https://doi.org/10.1088/1361-6633/ac9cee} {\bibfield  {journal} {\bibinfo
  {journal} {Rept. Prog. Phys.}\ }\textbf {\bibinfo {volume} {86}},\ \bibinfo
  {pages} {016201} (\bibinfo {year} {2023})},\ \Eprint
  {https://arxiv.org/abs/2201.07805} {arXiv:2201.07805 [hep-ph]} \BibitemShut
  {NoStop}%
\bibitem [{\citenamefont {Anzivino}\ \emph {et~al.}(2024)\citenamefont
  {Anzivino} \emph {et~al.}}]{Anzivino:2023bhp}%
  \BibitemOpen
  \bibfield  {author} {\bibinfo {author} {\bibfnamefont {G.}~\bibnamefont
  {Anzivino}} \emph {et~al.},\ }\bibfield  {title} {\bibinfo {title} {{Workshop
  summary: Kaons@CERN 2023}},\ }\href
  {https://doi.org/10.1140/epjc/s10052-024-12565-4} {\bibfield  {journal}
  {\bibinfo  {journal} {Eur. Phys. J. C}\ }\textbf {\bibinfo {volume} {84}},\
  \bibinfo {pages} {377} (\bibinfo {year} {2024})},\ \Eprint
  {https://arxiv.org/abs/2311.02923} {arXiv:2311.02923 [hep-ph]} \BibitemShut
  {NoStop}%
\bibitem [{\citenamefont {He}\ \emph {et~al.}(2005)\citenamefont {He},
  \citenamefont {Tandean},\ and\ \citenamefont {Valencia}}]{He:2005yn}%
  \BibitemOpen
  \bibfield  {author} {\bibinfo {author} {\bibfnamefont {X.-G.}\ \bibnamefont
  {He}}, \bibinfo {author} {\bibfnamefont {J.}~\bibnamefont {Tandean}},\ and\
  \bibinfo {author} {\bibfnamefont {G.}~\bibnamefont {Valencia}},\ }\bibfield
  {title} {\bibinfo {title} {{The decay $\Sigma^+\to p\ell^+\ell^-$ within the
  standard model}},\ }\href {https://doi.org/10.1103/PhysRevD.72.074003}
  {\bibfield  {journal} {\bibinfo  {journal} {Phys. Rev. D}\ }\textbf {\bibinfo
  {volume} {72}},\ \bibinfo {pages} {074003} (\bibinfo {year} {2005})},\
  \Eprint {https://arxiv.org/abs/hep-ph/0506067} {arXiv:hep-ph/0506067}
  \BibitemShut {NoStop}%
\bibitem [{\citenamefont {Shifman}\ \emph {et~al.}(1978)\citenamefont
  {Shifman}, \citenamefont {Vainshtein},\ and\ \citenamefont
  {Zakharov}}]{Shifman:1976de}%
  \BibitemOpen
  \bibfield  {author} {\bibinfo {author} {\bibfnamefont {M.~A.}\ \bibnamefont
  {Shifman}}, \bibinfo {author} {\bibfnamefont {A.~I.}\ \bibnamefont
  {Vainshtein}},\ and\ \bibinfo {author} {\bibfnamefont {V.~I.}\ \bibnamefont
  {Zakharov}},\ }\bibfield  {title} {\bibinfo {title} {{Right-handed currents
  and strong interactions at short distances}},\ }\href
  {https://doi.org/10.1103/PhysRevD.18.2583} {\bibfield  {journal} {\bibinfo
  {journal} {Phys. Rev. D}\ }\textbf {\bibinfo {volume} {18}},\ \bibinfo
  {pages} {2583} (\bibinfo {year} {1978})},\ \bibinfo {note} {[Erratum: Phys.
  Rev. D 19, 2815 (1979)]}\BibitemShut {NoStop}%
\bibitem [{\citenamefont {Buchalla}\ \emph {et~al.}(1996)\citenamefont
  {Buchalla}, \citenamefont {Buras},\ and\ \citenamefont
  {Lautenbacher}}]{Buchalla:1995vs}%
  \BibitemOpen
  \bibfield  {author} {\bibinfo {author} {\bibfnamefont {G.}~\bibnamefont
  {Buchalla}}, \bibinfo {author} {\bibfnamefont {A.~J.}\ \bibnamefont
  {Buras}},\ and\ \bibinfo {author} {\bibfnamefont {M.~E.}\ \bibnamefont
  {Lautenbacher}},\ }\bibfield  {title} {\bibinfo {title} {{Weak decays beyond
  leading logarithms}},\ }\href {https://doi.org/10.1103/RevModPhys.68.1125}
  {\bibfield  {journal} {\bibinfo  {journal} {Rev. Mod. Phys.}\ }\textbf
  {\bibinfo {volume} {68}},\ \bibinfo {pages} {1125} (\bibinfo {year}
  {1996})},\ \Eprint {https://arxiv.org/abs/hep-ph/9512380}
  {arXiv:hep-ph/9512380} \BibitemShut {NoStop}%
\bibitem [{\citenamefont {Ablikim}\ \emph
  {et~al.}(2023{\natexlab{a}})\citenamefont {Ablikim} \emph
  {et~al.}}]{BESIII:2023fhs}%
  \BibitemOpen
  \bibfield  {author} {\bibinfo {author} {\bibfnamefont {M.}~\bibnamefont
  {Ablikim}} \emph {et~al.} (\bibinfo {collaboration} {BESIII}),\ }\bibfield
  {title} {\bibinfo {title} {{Precision Measurement of the Decay $\Sigma^+\to
  p\gamma$ in the Process $J/\psi\to\Sigma^+\Sigma^-$}},\ }\href
  {https://doi.org/10.1103/PhysRevLett.130.211901} {\bibfield  {journal}
  {\bibinfo  {journal} {Phys. Rev. Lett.}\ }\textbf {\bibinfo {volume} {130}},\
  \bibinfo {pages} {211901} (\bibinfo {year} {2023}{\natexlab{a}})},\ \Eprint
  {https://arxiv.org/abs/2302.13568} {arXiv:2302.13568 [hep-ex]} \BibitemShut
  {NoStop}%
\bibitem [{\citenamefont {Ablikim}\ \emph {et~al.}(2020)\citenamefont {Ablikim}
  \emph {et~al.}}]{BESIII:2020fqg}%
  \BibitemOpen
  \bibfield  {author} {\bibinfo {author} {\bibfnamefont {M.}~\bibnamefont
  {Ablikim}} \emph {et~al.} (\bibinfo {collaboration} {BESIII}),\ }\bibfield
  {title} {\bibinfo {title} {{$\Sigma^{+}$ and $\bar{\Sigma}^-$ Polarization in
  the $J/\psi$ and $\psi(3686)$ Decays}},\ }\href
  {https://doi.org/10.1103/PhysRevLett.125.052004} {\bibfield  {journal}
  {\bibinfo  {journal} {Phys. Rev. Lett.}\ }\textbf {\bibinfo {volume} {125}},\
  \bibinfo {pages} {052004} (\bibinfo {year} {2020})},\ \Eprint
  {https://arxiv.org/abs/2004.07701} {arXiv:2004.07701 [hep-ex]} \BibitemShut
  {NoStop}%
\bibitem [{\citenamefont {Ablikim}\ \emph
  {et~al.}(2023{\natexlab{b}})\citenamefont {Ablikim} \emph
  {et~al.}}]{BESIII:2023sgt}%
  \BibitemOpen
  \bibfield  {author} {\bibinfo {author} {\bibfnamefont {M.}~\bibnamefont
  {Ablikim}} \emph {et~al.} (\bibinfo {collaboration} {BESIII}),\ }\bibfield
  {title} {\bibinfo {title} {{Test of $CP$ Symmetry in Hyperon to Neutron
  Decays}},\ }\href {https://doi.org/10.1103/PhysRevLett.131.191802} {\bibfield
   {journal} {\bibinfo  {journal} {Phys. Rev. Lett.}\ }\textbf {\bibinfo
  {volume} {131}},\ \bibinfo {pages} {191802} (\bibinfo {year}
  {2023}{\natexlab{b}})},\ \Eprint {https://arxiv.org/abs/2304.14655}
  {arXiv:2304.14655 [hep-ex]} \BibitemShut {NoStop}%
\bibitem [{\citenamefont {Geng}\ \emph {et~al.}(2022)\citenamefont {Geng},
  \citenamefont {Camalich},\ and\ \citenamefont {Shi}}]{Geng:2021fog}%
  \BibitemOpen
  \bibfield  {author} {\bibinfo {author} {\bibfnamefont {L.-S.}\ \bibnamefont
  {Geng}}, \bibinfo {author} {\bibfnamefont {J.~M.}\ \bibnamefont {Camalich}},\
  and\ \bibinfo {author} {\bibfnamefont {R.-X.}\ \bibnamefont {Shi}},\
  }\bibfield  {title} {\bibinfo {title} {{New physics in $s\to d$ semileptonic
  transitions: rare hyperon vs. kaon decays}},\ }\href
  {https://doi.org/10.1007/JHEP02(2022)178} {\bibfield  {journal} {\bibinfo
  {journal} {J. High Energy Phys.}\ }\textbf {\bibinfo {volume} {02}},\
  \bibinfo {pages} {178 (2022)}},\ \Eprint {https://arxiv.org/abs/2112.11979}
  {arXiv:2112.11979 [hep-ph]} \BibitemShut {NoStop}%
\bibitem [{\citenamefont {Kamenik}\ and\ \citenamefont
  {Smith}(2012)}]{Kamenik:2011vy}%
  \BibitemOpen
  \bibfield  {author} {\bibinfo {author} {\bibfnamefont {J.~F.}\ \bibnamefont
  {Kamenik}}\ and\ \bibinfo {author} {\bibfnamefont {C.}~\bibnamefont
  {Smith}},\ }\bibfield  {title} {\bibinfo {title} {{FCNC portals to the dark
  sector}},\ }\href {https://doi.org/10.1007/JHEP03(2012)090} {\bibfield
  {journal} {\bibinfo  {journal} {J. High Energy Phys.}\ }\textbf {\bibinfo
  {volume} {03}},\ \bibinfo {pages} {090 (2012)}},\ \Eprint
  {https://arxiv.org/abs/1111.6402} {arXiv:1111.6402 [hep-ph]} \BibitemShut
  {NoStop}%
\bibitem [{\citenamefont {He}\ \emph {et~al.}(2018{\natexlab{a}})\citenamefont
  {He}, \citenamefont {Valencia},\ and\ \citenamefont {Wong}}]{He:2018uey}%
  \BibitemOpen
  \bibfield  {author} {\bibinfo {author} {\bibfnamefont {X.-G.}\ \bibnamefont
  {He}}, \bibinfo {author} {\bibfnamefont {G.}~\bibnamefont {Valencia}},\ and\
  \bibinfo {author} {\bibfnamefont {K.}~\bibnamefont {Wong}},\ }\bibfield
  {title} {\bibinfo {title} {{Constraints on new physics from $K \rightarrow
  \pi\nu\bar\nu$}},\ }\href {https://doi.org/10.1140/epjc/s10052-018-5964-0}
  {\bibfield  {journal} {\bibinfo  {journal} {Eur. Phys. J. C}\ }\textbf
  {\bibinfo {volume} {78}},\ \bibinfo {pages} {472} (\bibinfo {year}
  {2018}{\natexlab{a}})},\ \bibinfo {note} {[Erratum: Eur. Phys. J. C 80, 738
  (2020)]},\ \Eprint {https://arxiv.org/abs/1804.07449} {arXiv:1804.07449
  [hep-ph]} \BibitemShut {NoStop}%
\bibitem [{\citenamefont {Tandean}(2019)}]{Tandean:2019tkm}%
  \BibitemOpen
  \bibfield  {author} {\bibinfo {author} {\bibfnamefont {J.}~\bibnamefont
  {Tandean}},\ }\bibfield  {title} {\bibinfo {title} {{Rare hyperon decays with
  missing energy}},\ }\href {https://doi.org/10.1007/JHEP04(2019)104}
  {\bibfield  {journal} {\bibinfo  {journal} {J. High Energy Phys.}\ }\textbf
  {\bibinfo {volume} {04}},\ \bibinfo {pages} {104 (2019)}},\ \Eprint
  {https://arxiv.org/abs/1901.10447} {arXiv:1901.10447 [hep-ph]} \BibitemShut
  {NoStop}%
\bibitem [{\citenamefont {Li}\ \emph {et~al.}(2019)\citenamefont {Li},
  \citenamefont {Su},\ and\ \citenamefont {Tandean}}]{Li:2019cbk}%
  \BibitemOpen
  \bibfield  {author} {\bibinfo {author} {\bibfnamefont {G.}~\bibnamefont
  {Li}}, \bibinfo {author} {\bibfnamefont {J.-Y.}\ \bibnamefont {Su}},\ and\
  \bibinfo {author} {\bibfnamefont {J.}~\bibnamefont {Tandean}},\ }\bibfield
  {title} {\bibinfo {title} {{Flavor-changing hyperon decays with light
  invisible bosons}},\ }\href {https://doi.org/10.1103/PhysRevD.100.075003}
  {\bibfield  {journal} {\bibinfo  {journal} {Phys. Rev. D}\ }\textbf {\bibinfo
  {volume} {100}},\ \bibinfo {pages} {075003} (\bibinfo {year} {2019})},\
  \Eprint {https://arxiv.org/abs/1905.08759} {arXiv:1905.08759 [hep-ph]}
  \BibitemShut {NoStop}%
\bibitem [{\citenamefont {Su}\ and\ \citenamefont
  {Tandean}(2020)}]{Su:2019tjn}%
  \BibitemOpen
  \bibfield  {author} {\bibinfo {author} {\bibfnamefont {J.-Y.}\ \bibnamefont
  {Su}}\ and\ \bibinfo {author} {\bibfnamefont {J.}~\bibnamefont {Tandean}},\
  }\bibfield  {title} {\bibinfo {title} {{Exploring leptoquark effects in
  hyperon and kaon decays with missing energy}},\ }\href
  {https://doi.org/10.1103/PhysRevD.102.075032} {\bibfield  {journal} {\bibinfo
   {journal} {Phys. Rev. D}\ }\textbf {\bibinfo {volume} {102}},\ \bibinfo
  {pages} {075032} (\bibinfo {year} {2020})},\ \Eprint
  {https://arxiv.org/abs/1912.13507} {arXiv:1912.13507 [hep-ph]} \BibitemShut
  {NoStop}%
\bibitem [{\citenamefont {Geng}\ and\ \citenamefont
  {Tandean}(2020)}]{Geng:2020seh}%
  \BibitemOpen
  \bibfield  {author} {\bibinfo {author} {\bibfnamefont {C.-Q.}\ \bibnamefont
  {Geng}}\ and\ \bibinfo {author} {\bibfnamefont {J.}~\bibnamefont {Tandean}},\
  }\bibfield  {title} {\bibinfo {title} {{Probing new physics with the kaon
  decays $K\to\pi\pi\!\not\!\!E$}},\ }\href
  {https://doi.org/10.1103/PhysRevD.102.115021} {\bibfield  {journal} {\bibinfo
   {journal} {Phys. Rev. D}\ }\textbf {\bibinfo {volume} {102}},\ \bibinfo
  {pages} {115021} (\bibinfo {year} {2020})},\ \Eprint
  {https://arxiv.org/abs/2009.00608} {arXiv:2009.00608 [hep-ph]} \BibitemShut
  {NoStop}%
\bibitem [{\citenamefont {He}\ \emph {et~al.}(2023)\citenamefont {He},
  \citenamefont {Ma},\ and\ \citenamefont {Valencia}}]{He:2022ljo}%
  \BibitemOpen
  \bibfield  {author} {\bibinfo {author} {\bibfnamefont {X.-G.}\ \bibnamefont
  {He}}, \bibinfo {author} {\bibfnamefont {X.-D.}\ \bibnamefont {Ma}},\ and\
  \bibinfo {author} {\bibfnamefont {G.}~\bibnamefont {Valencia}},\ }\bibfield
  {title} {\bibinfo {title} {{FCNC $B$ and $K$ meson decays with light bosonic
  Dark Matter}},\ }\href {https://doi.org/10.1007/JHEP03(2023)037} {\bibfield
  {journal} {\bibinfo  {journal} {J. High Energy Phys.}\ }\textbf {\bibinfo
  {volume} {03}},\ \bibinfo {pages} {037 (2023)}},\ \Eprint
  {https://arxiv.org/abs/2209.05223} {arXiv:2209.05223 [hep-ph]} \BibitemShut
  {NoStop}%
\bibitem [{\citenamefont {Inami}\ and\ \citenamefont
  {Lim}(1981)}]{Inami:1980fz}%
  \BibitemOpen
  \bibfield  {author} {\bibinfo {author} {\bibfnamefont {T.}~\bibnamefont
  {Inami}}\ and\ \bibinfo {author} {\bibfnamefont {C.~S.}\ \bibnamefont
  {Lim}},\ }\bibfield  {title} {\bibinfo {title} {{Effects of Superheavy Quarks
  and Leptons in Low-Energy Weak Processes $K_L\to\mu\bar\mu$,
  $K^+\to\pi^+\nu\bar\nu$ and $K^0\leftrightarrow\bar K^0$}},\ }\href
  {https://doi.org/10.1143/PTP.65.297} {\bibfield  {journal} {\bibinfo
  {journal} {Prog. Theor. Phys.}\ }\textbf {\bibinfo {volume} {65}},\ \bibinfo
  {pages} {297} (\bibinfo {year} {1981})},\ \bibinfo {note} {[Erratum: Prog.
  Theor. Phys. 65, 1772 (1981)]}\BibitemShut {NoStop}%
\bibitem [{\citenamefont {Cabibbo}\ \emph {et~al.}(2003)\citenamefont
  {Cabibbo}, \citenamefont {Swallow},\ and\ \citenamefont
  {Winston}}]{Cabibbo:2003cu}%
  \BibitemOpen
  \bibfield  {author} {\bibinfo {author} {\bibfnamefont {N.}~\bibnamefont
  {Cabibbo}}, \bibinfo {author} {\bibfnamefont {E.~C.}\ \bibnamefont
  {Swallow}},\ and\ \bibinfo {author} {\bibfnamefont {R.}~\bibnamefont
  {Winston}},\ }\bibfield  {title} {\bibinfo {title} {{Semileptonic hyperon
  decays}},\ }\href {https://doi.org/10.1146/annurev.nucl.53.013103.155258}
  {\bibfield  {journal} {\bibinfo  {journal} {Ann. Rev. Nucl. Part. Sci.}\
  }\textbf {\bibinfo {volume} {53}},\ \bibinfo {pages} {39} (\bibinfo {year}
  {2003})},\ \Eprint {https://arxiv.org/abs/hep-ph/0307298}
  {arXiv:hep-ph/0307298} \BibitemShut {NoStop}%
\bibitem [{\citenamefont {Workman}\ \emph {et~al.}(2022)\citenamefont {Workman}
  \emph {et~al.}}]{Workman:2022ynf}%
  \BibitemOpen
  \bibfield  {author} {\bibinfo {author} {\bibfnamefont {R.~L.}\ \bibnamefont
  {Workman}} \emph {et~al.} (\bibinfo {collaboration} {Particle Data Group}),\
  }\bibfield  {title} {\bibinfo {title} {{Review of Particle Physics}},\ }\href
  {https://doi.org/10.1093/ptep/ptac097} {\bibfield  {journal} {\bibinfo
  {journal} {PTEP}\ }\textbf {\bibinfo {volume} {2022}},\ \bibinfo {pages}
  {083C01} (\bibinfo {year} {2022})}\BibitemShut {NoStop}%
\bibitem [{\citenamefont {Guadagnoli}\ \emph {et~al.}(2007)\citenamefont
  {Guadagnoli}, \citenamefont {Lubicz}, \citenamefont {Papinutto},\ and\
  \citenamefont {Simula}}]{Guadagnoli:2006gj}%
  \BibitemOpen
  \bibfield  {author} {\bibinfo {author} {\bibfnamefont {D.}~\bibnamefont
  {Guadagnoli}}, \bibinfo {author} {\bibfnamefont {V.}~\bibnamefont {Lubicz}},
  \bibinfo {author} {\bibfnamefont {M.}~\bibnamefont {Papinutto}},\ and\
  \bibinfo {author} {\bibfnamefont {S.}~\bibnamefont {Simula}},\ }\bibfield
  {title} {\bibinfo {title} {{First lattice QCD study of the $\Sigma^-\to n$
  axial and vector form factors with SU(3) breaking corrections}},\ }\href
  {https://doi.org/10.1016/j.nuclphysb.2006.10.022} {\bibfield  {journal}
  {\bibinfo  {journal} {Nucl. Phys. B}\ }\textbf {\bibinfo {volume} {761}},\
  \bibinfo {pages} {63} (\bibinfo {year} {2007})},\ \Eprint
  {https://arxiv.org/abs/hep-ph/0606181} {arXiv:hep-ph/0606181} \BibitemShut
  {NoStop}%
\bibitem [{\citenamefont {Sasaki}(2017)}]{Sasaki:2017jue}%
  \BibitemOpen
  \bibfield  {author} {\bibinfo {author} {\bibfnamefont {S.}~\bibnamefont
  {Sasaki}},\ }\bibfield  {title} {\bibinfo {title} {{Continuum limit of
  hyperon vector coupling $f_1(0)$ from 2+1 flavor domain wall QCD}},\ }\href
  {https://doi.org/10.1103/PhysRevD.96.074509} {\bibfield  {journal} {\bibinfo
  {journal} {Phys. Rev. D}\ }\textbf {\bibinfo {volume} {96}},\ \bibinfo
  {pages} {074509} (\bibinfo {year} {2017})},\ \Eprint
  {https://arxiv.org/abs/1708.04008} {arXiv:1708.04008 [hep-lat]} \BibitemShut
  {NoStop}%
\bibitem [{\citenamefont {Aslam}\ \emph {et~al.}(2008)\citenamefont {Aslam},
  \citenamefont {Wang},\ and\ \citenamefont {Lu}}]{Aslam:2008hp}%
  \BibitemOpen
  \bibfield  {author} {\bibinfo {author} {\bibfnamefont {M.~J.}\ \bibnamefont
  {Aslam}}, \bibinfo {author} {\bibfnamefont {Y.-M.}\ \bibnamefont {Wang}},\
  and\ \bibinfo {author} {\bibfnamefont {C.-D.}\ \bibnamefont {Lu}},\
  }\bibfield  {title} {\bibinfo {title} {{Exclusive semileptonic decays of
  $\Lambda_b\to\Lambda l^+ l^-$ in supersymmetric theories}},\ }\href
  {https://doi.org/10.1103/PhysRevD.78.114032} {\bibfield  {journal} {\bibinfo
  {journal} {Phys. Rev. D}\ }\textbf {\bibinfo {volume} {78}},\ \bibinfo
  {pages} {114032} (\bibinfo {year} {2008})},\ \Eprint
  {https://arxiv.org/abs/0808.2113} {arXiv:0808.2113 [hep-ph]} \BibitemShut
  {NoStop}%
\bibitem [{\citenamefont {Neshatpour}\ and\ \citenamefont
  {Mahmoudi}(2022)}]{Neshatpour:2022fak}%
  \BibitemOpen
  \bibfield  {author} {\bibinfo {author} {\bibfnamefont {S.}~\bibnamefont
  {Neshatpour}}\ and\ \bibinfo {author} {\bibfnamefont {F.}~\bibnamefont
  {Mahmoudi}},\ }\bibfield  {title} {\bibinfo {title} {{Flavour Physics
  Phenomenology with SuperIso}},\ }\href {https://doi.org/10.22323/1.409.0010}
  {\bibfield  {journal} {\bibinfo  {journal} {PoS}\ }\textbf {\bibinfo {volume}
  {CompTools2021}},\ \bibinfo {pages} {010} (\bibinfo {year} {2022})},\ \Eprint
  {https://arxiv.org/abs/2207.04956} {arXiv:2207.04956 [hep-ph]} \BibitemShut
  {NoStop}%
\bibitem [{\citenamefont {Nielsen}\ \emph {et~al.}(1996)\citenamefont
  {Nielsen}, \citenamefont {Barreiro}, \citenamefont {Escobar},\ and\
  \citenamefont {Rosenfeld}}]{Nielsen:1995dp}%
  \BibitemOpen
  \bibfield  {author} {\bibinfo {author} {\bibfnamefont {M.}~\bibnamefont
  {Nielsen}}, \bibinfo {author} {\bibfnamefont {L.~A.}\ \bibnamefont
  {Barreiro}}, \bibinfo {author} {\bibfnamefont {C.~O.}\ \bibnamefont
  {Escobar}},\ and\ \bibinfo {author} {\bibfnamefont {R.}~\bibnamefont
  {Rosenfeld}},\ }\bibfield  {title} {\bibinfo {title} {{A QCD sum rule
  approach to the $s\to d\gamma$ contribution to the $\Omega^-\to\Xi^-\gamma$
  radiative decay}},\ }\href {https://doi.org/10.1103/PhysRevD.53.3620}
  {\bibfield  {journal} {\bibinfo  {journal} {Phys. Rev. D}\ }\textbf {\bibinfo
  {volume} {53}},\ \bibinfo {pages} {3620} (\bibinfo {year} {1996})},\ \Eprint
  {https://arxiv.org/abs/hep-ph/9509297} {arXiv:hep-ph/9509297} \BibitemShut
  {NoStop}%
\bibitem [{\citenamefont {Tandean}(2000)}]{Tandean:1999mg}%
  \BibitemOpen
  \bibfield  {author} {\bibinfo {author} {\bibfnamefont {J.}~\bibnamefont
  {Tandean}},\ }\bibfield  {title} {\bibinfo {title} {{New physics and short
  distance $s\to d\gamma$ transition in $\Omega^-\to\Xi^-\gamma$ decay}},\
  }\href {https://doi.org/10.1103/PhysRevD.61.114022} {\bibfield  {journal}
  {\bibinfo  {journal} {Phys. Rev. D}\ }\textbf {\bibinfo {volume} {61}},\
  \bibinfo {pages} {114022} (\bibinfo {year} {2000})},\ \Eprint
  {https://arxiv.org/abs/hep-ph/9912497} {arXiv:hep-ph/9912497} \BibitemShut
  {NoStop}%
\bibitem [{\citenamefont {Donoghue}\ \emph {et~al.}(2022)\citenamefont
  {Donoghue}, \citenamefont {Golowich},\ and\ \citenamefont
  {Holstein}}]{Donoghue:2022wrw}%
  \BibitemOpen
  \bibfield  {author} {\bibinfo {author} {\bibfnamefont {J.~F.}\ \bibnamefont
  {Donoghue}}, \bibinfo {author} {\bibfnamefont {E.}~\bibnamefont {Golowich}},\
  and\ \bibinfo {author} {\bibfnamefont {B.~R.}\ \bibnamefont {Holstein}},\
  }\href {https://doi.org/10.1017/9781009291033} {\emph {\bibinfo {title}
  {{Dynamics of the Standard Model}: {Second edition}}}}\ (\bibinfo
  {publisher} {Cambridge University Press},\ \bibinfo {year}
  {2022})\BibitemShut {NoStop}%
\bibitem [{\citenamefont {Ang}\ \emph {et~al.}(1969)\citenamefont {Ang} \emph
  {et~al.}}]{Ang:1969hg}%
  \BibitemOpen
  \bibfield  {author} {\bibinfo {author} {\bibfnamefont {G.}~\bibnamefont
  {Ang}} \emph {et~al.},\ }\bibfield  {title} {\bibinfo {title} {{Radiative
  $\Sigma^\pm$ decays and search for neutral currents}},\ }\href
  {https://doi.org/10.1007/BF01397536} {\bibfield  {journal} {\bibinfo
  {journal} {Z. Phys.}\ }\textbf {\bibinfo {volume} {228}},\ \bibinfo {pages}
  {151} (\bibinfo {year} {1969})}\BibitemShut {NoStop}%
\bibitem [{\citenamefont {Park}\ \emph {et~al.}(2005)\citenamefont {Park} \emph
  {et~al.}}]{HyperCP:2005mvo}%
  \BibitemOpen
  \bibfield  {author} {\bibinfo {author} {\bibfnamefont {H.}~\bibnamefont
  {Park}} \emph {et~al.} (\bibinfo {collaboration} {HyperCP}),\ }\bibfield
  {title} {\bibinfo {title} {{Evidence for the Decay $\Sigma^+\to
  p\mu^+\mu^-$}},\ }\href {https://doi.org/10.1103/PhysRevLett.94.021801}
  {\bibfield  {journal} {\bibinfo  {journal} {Phys. Rev. Lett.}\ }\textbf
  {\bibinfo {volume} {94}},\ \bibinfo {pages} {021801} (\bibinfo {year}
  {2005})},\ \Eprint {https://arxiv.org/abs/hep-ex/0501014}
  {arXiv:hep-ex/0501014} \BibitemShut {NoStop}%
\bibitem [{\citenamefont {Behrends}(1958)}]{Behrends:1958zz}%
  \BibitemOpen
  \bibfield  {author} {\bibinfo {author} {\bibfnamefont {R.~E.}\ \bibnamefont
  {Behrends}},\ }\bibfield  {title} {\bibinfo {title} {{Photon Decay of
  Hyperons}},\ }\href {https://doi.org/10.1103/PhysRev.111.1691} {\bibfield
  {journal} {\bibinfo  {journal} {Phys. Rev.}\ }\textbf {\bibinfo {volume}
  {111}},\ \bibinfo {pages} {1691} (\bibinfo {year} {1958})}\BibitemShut
  {NoStop}%
\bibitem [{\citenamefont {Lyagin}\ and\ \citenamefont
  {Ginzburg}(1962)}]{Lyagin:1962}%
  \BibitemOpen
  \bibfield  {author} {\bibinfo {author} {\bibfnamefont {I.~V.}\ \bibnamefont
  {Lyagin}}\ and\ \bibinfo {author} {\bibfnamefont {E.~K.}\ \bibnamefont
  {Ginzburg}},\ }\bibfield  {title} {\bibinfo {title} {{On $\Sigma^+\to
  pe^+e^-$ and $\Sigma^+\to p\mu^+\mu^-$ Decays}},\ }\href
  {{http://jetp.ras.ru/cgi-bin/e/index/e/14/3/p653?a=list}} {\bibfield
  {journal} {\bibinfo  {journal} {Sov. Phys. JETP}\ }\textbf {\bibinfo {volume}
  {14}},\ \bibinfo {pages} {653} (\bibinfo {year} {1962})}\BibitemShut
  {NoStop}%
\bibitem [{\citenamefont {Bergstrom}\ \emph {et~al.}(1988)\citenamefont
  {Bergstrom}, \citenamefont {Safadi},\ and\ \citenamefont
  {Singer}}]{Bergstrom:1987wr}%
  \BibitemOpen
  \bibfield  {author} {\bibinfo {author} {\bibfnamefont {L.}~\bibnamefont
  {Bergstrom}}, \bibinfo {author} {\bibfnamefont {R.}~\bibnamefont {Safadi}},\
  and\ \bibinfo {author} {\bibfnamefont {P.}~\bibnamefont {Singer}},\
  }\bibfield  {title} {\bibinfo {title} {{Phenomenology of $\Sigma^+ \to p
  \ell^+ \ell^-$ and the structure of the weak non-leptonic Hamiltonian}},\
  }\href {https://doi.org/10.1007/BF01579914} {\bibfield  {journal} {\bibinfo
  {journal} {Z. Phys. C}\ }\textbf {\bibinfo {volume} {37}},\ \bibinfo {pages}
  {281} (\bibinfo {year} {1988})}\BibitemShut {NoStop}%
\bibitem [{\citenamefont {Gershwin}\ \emph {et~al.}(1969)\citenamefont
  {Gershwin}, \citenamefont {Alston-Garnjost}, \citenamefont {Bangerter},
  \citenamefont {Barbaro-Galtieri}, \citenamefont {Mast}, \citenamefont
  {Solmitz},\ and\ \citenamefont {Tripp}}]{Gershwin:1969fpe}%
  \BibitemOpen
  \bibfield  {author} {\bibinfo {author} {\bibfnamefont {L.~K.}\ \bibnamefont
  {Gershwin}}, \bibinfo {author} {\bibfnamefont {M.}~\bibnamefont
  {Alston-Garnjost}}, \bibinfo {author} {\bibfnamefont {R.~O.}\ \bibnamefont
  {Bangerter}}, \bibinfo {author} {\bibfnamefont {A.}~\bibnamefont
  {Barbaro-Galtieri}}, \bibinfo {author} {\bibfnamefont {T.~S.}\ \bibnamefont
  {Mast}}, \bibinfo {author} {\bibfnamefont {F.~T.}\ \bibnamefont {Solmitz}},\
  and\ \bibinfo {author} {\bibfnamefont {R.~D.}\ \bibnamefont {Tripp}},\
  }\bibfield  {title} {\bibinfo {title} {{Asymmetry parameter and branching
  ratio of $\Sigma^+\to p\gamma$}},\ }\href
  {https://doi.org/10.1103/PhysRev.188.2077} {\bibfield  {journal} {\bibinfo
  {journal} {Phys. Rev.}\ }\textbf {\bibinfo {volume} {188}},\ \bibinfo {pages}
  {2077} (\bibinfo {year} {1969})}\BibitemShut {NoStop}%
\bibitem [{\citenamefont {Manz}\ \emph {et~al.}(1980)\citenamefont {Manz},
  \citenamefont {Reucroft}, \citenamefont {Settles}, \citenamefont {Wolf},
  \citenamefont {Marraffino}, \citenamefont {Roos}, \citenamefont {Waters},\
  and\ \citenamefont {Webster}}]{Manz:1980td}%
  \BibitemOpen
  \bibfield  {author} {\bibinfo {author} {\bibfnamefont {A.}~\bibnamefont
  {Manz}}, \bibinfo {author} {\bibfnamefont {S.}~\bibnamefont {Reucroft}},
  \bibinfo {author} {\bibfnamefont {R.}~\bibnamefont {Settles}}, \bibinfo
  {author} {\bibfnamefont {G.}~\bibnamefont {Wolf}}, \bibinfo {author}
  {\bibfnamefont {J.}~\bibnamefont {Marraffino}}, \bibinfo {author}
  {\bibfnamefont {C.}~\bibnamefont {Roos}}, \bibinfo {author} {\bibfnamefont
  {J.}~\bibnamefont {Waters}},\ and\ \bibinfo {author} {\bibfnamefont
  {M.}~\bibnamefont {Webster}},\ }\bibfield  {title} {\bibinfo {title} {{A new
  measurement of $\Sigma^+\to p\gamma$ decay properties}},\ }\href
  {https://doi.org/10.1016/0370-2693(80)90248-8} {\bibfield  {journal}
  {\bibinfo  {journal} {Phys. Lett. B}\ }\textbf {\bibinfo {volume} {96}},\
  \bibinfo {pages} {217} (\bibinfo {year} {1980})}\BibitemShut {NoStop}%
\bibitem [{\citenamefont {Kobayashi}\ \emph {et~al.}(1987)\citenamefont
  {Kobayashi}, \citenamefont {Haba}, \citenamefont {Homma}, \citenamefont
  {Kawai}, \citenamefont {Miyake}, \citenamefont {Nakamura}, \citenamefont
  {Sasao},\ and\ \citenamefont {Sugimoto}}]{Kobayashi:1987yv}%
  \BibitemOpen
  \bibfield  {author} {\bibinfo {author} {\bibfnamefont {M.}~\bibnamefont
  {Kobayashi}}, \bibinfo {author} {\bibfnamefont {J.}~\bibnamefont {Haba}},
  \bibinfo {author} {\bibfnamefont {T.}~\bibnamefont {Homma}}, \bibinfo
  {author} {\bibfnamefont {H.}~\bibnamefont {Kawai}}, \bibinfo {author}
  {\bibfnamefont {K.}~\bibnamefont {Miyake}}, \bibinfo {author} {\bibfnamefont
  {T.~S.}\ \bibnamefont {Nakamura}}, \bibinfo {author} {\bibfnamefont
  {N.}~\bibnamefont {Sasao}},\ and\ \bibinfo {author} {\bibfnamefont
  {Y.}~\bibnamefont {Sugimoto}},\ }\bibfield  {title} {\bibinfo {title} {{New
  measurement of the asymmetry parameter for the $\Sigma^+\to p\gamma$
  decay}},\ }\href {https://doi.org/10.1103/PhysRevLett.59.868} {\bibfield
  {journal} {\bibinfo  {journal} {Phys. Rev. Lett.}\ }\textbf {\bibinfo
  {volume} {59}},\ \bibinfo {pages} {868} (\bibinfo {year} {1987})}\BibitemShut
  {NoStop}%
\bibitem [{\citenamefont {Foucher}\ \emph {et~al.}(1992)\citenamefont {Foucher}
  \emph {et~al.}}]{E761:1992atm}%
  \BibitemOpen
  \bibfield  {author} {\bibinfo {author} {\bibfnamefont {M.}~\bibnamefont
  {Foucher}} \emph {et~al.} (\bibinfo {collaboration} {E761}),\ }\bibfield
  {title} {\bibinfo {title} {{Measurement of the asymmetry parameter in the
  hyperon radiative decay $\Sigma^+\to p\gamma$}},\ }\href
  {https://doi.org/10.1103/PhysRevLett.68.3004} {\bibfield  {journal} {\bibinfo
   {journal} {Phys. Rev. Lett.}\ }\textbf {\bibinfo {volume} {68}},\ \bibinfo
  {pages} {3004} (\bibinfo {year} {1992})}\BibitemShut {NoStop}%
\bibitem [{\citenamefont {Aaij}\ \emph {et~al.}(2022)\citenamefont {Aaij} \emph
  {et~al.}}]{LHCb:2021byf}%
  \BibitemOpen
  \bibfield  {author} {\bibinfo {author} {\bibfnamefont {R.}~\bibnamefont
  {Aaij}} \emph {et~al.} (\bibinfo {collaboration} {LHCb}),\ }\bibfield
  {title} {\bibinfo {title} {{Measurement of the photon polarization in
  $\Lambda_{b}^{0}\to\Lambda\gamma$ decays}},\ }\href
  {https://doi.org/10.1103/PhysRevD.105.L051104} {\bibfield  {journal}
  {\bibinfo  {journal} {Phys. Rev. D}\ }\textbf {\bibinfo {volume} {105}},\
  \bibinfo {pages} {L051104} (\bibinfo {year} {2022})},\ \Eprint
  {https://arxiv.org/abs/2111.10194} {arXiv:2111.10194 [hep-ex]} \BibitemShut
  {NoStop}%
\bibitem [{\citenamefont {Bangerter}(1969)}]{Bangerter:1969fta}%
  \BibitemOpen
  \bibfield  {author} {\bibinfo {author} {\bibfnamefont {R.~O.}\ \bibnamefont
  {Bangerter}},\ }\emph {\bibinfo {title} {{Nonleptonic decay of Sigma
  hyperons}}},\ \href@noop {} {Ph.D. thesis},\ \bibinfo  {school} {Calif. U.
  Berkeley} (\bibinfo {year} {1969})\BibitemShut {NoStop}%
\bibitem [{\citenamefont {Harris}\ \emph {et~al.}(1970)\citenamefont {Harris},
  \citenamefont {Overseth}, \citenamefont {Pondrom},\ and\ \citenamefont
  {Dettmann}}]{Harris:1970kq}%
  \BibitemOpen
  \bibfield  {author} {\bibinfo {author} {\bibfnamefont {F.}~\bibnamefont
  {Harris}}, \bibinfo {author} {\bibfnamefont {O.~E.}\ \bibnamefont
  {Overseth}}, \bibinfo {author} {\bibfnamefont {L.}~\bibnamefont {Pondrom}},\
  and\ \bibinfo {author} {\bibfnamefont {E.}~\bibnamefont {Dettmann}},\
  }\bibfield  {title} {\bibinfo {title} {{Proton Polarization in $\Sigma^+\to
  p\pi^0$}},\ }\href {https://doi.org/10.1103/PhysRevLett.24.165} {\bibfield
  {journal} {\bibinfo  {journal} {Phys. Rev. Lett.}\ }\textbf {\bibinfo
  {volume} {24}},\ \bibinfo {pages} {165} (\bibinfo {year} {1970})}\BibitemShut
  {NoStop}%
\bibitem [{\citenamefont {Bellamy}\ \emph {et~al.}(1972)\citenamefont {Bellamy}
  \emph {et~al.}}]{Bellamy:1972fa}%
  \BibitemOpen
  \bibfield  {author} {\bibinfo {author} {\bibfnamefont {E.~H.}\ \bibnamefont
  {Bellamy}} \emph {et~al.},\ }\bibfield  {title} {\bibinfo {title}
  {{Polarisation in the reaction $\pi^+ p\to K^+ \Sigma^+$ at 1.11 GeV/$c$}},\
  }\href {https://doi.org/10.1016/0370-2693(72)90801-5} {\bibfield  {journal}
  {\bibinfo  {journal} {Phys. Lett. B}\ }\textbf {\bibinfo {volume} {39}},\
  \bibinfo {pages} {299} (\bibinfo {year} {1972})}\BibitemShut {NoStop}%
\bibitem [{\citenamefont {Lipman}\ \emph {et~al.}(1973)\citenamefont {Lipman}
  \emph {et~al.}}]{Lipman:1973mz}%
  \BibitemOpen
  \bibfield  {author} {\bibinfo {author} {\bibfnamefont {N.~H.}\ \bibnamefont
  {Lipman}} \emph {et~al.},\ }\bibfield  {title} {\bibinfo {title} {{A test of
  the $\Delta I=1/2$ rule and the Lee-Sugawara relation in the decay
  $\Sigma^+\to p\pi^0$}},\ }\href
  {https://doi.org/10.1016/0370-2693(73)90551-0} {\bibfield  {journal}
  {\bibinfo  {journal} {Phys. Lett. B}\ }\textbf {\bibinfo {volume} {43}},\
  \bibinfo {pages} {89} (\bibinfo {year} {1973})}\BibitemShut {NoStop}%
\bibitem [{\citenamefont {Erben}\ \emph
  {et~al.}(2023{\natexlab{a}})\citenamefont {Erben}, \citenamefont {G\"ulpers},
  \citenamefont {Hansen}, \citenamefont {Hodgson},\ and\ \citenamefont
  {Portelli}}]{Erben:2022tdu}%
  \BibitemOpen
  \bibfield  {author} {\bibinfo {author} {\bibfnamefont {F.}~\bibnamefont
  {Erben}}, \bibinfo {author} {\bibfnamefont {V.}~\bibnamefont {G\"ulpers}},
  \bibinfo {author} {\bibfnamefont {M.~T.}\ \bibnamefont {Hansen}}, \bibinfo
  {author} {\bibfnamefont {R.}~\bibnamefont {Hodgson}},\ and\ \bibinfo {author}
  {\bibfnamefont {A.}~\bibnamefont {Portelli}},\ }\bibfield  {title} {\bibinfo
  {title} {{Prospects for a lattice calculation of the rare decay $\Sigma^+\to
  p\ell^+\ell^-$}},\ }\href {https://doi.org/10.1007/JHEP04(2023)108}
  {\bibfield  {journal} {\bibinfo  {journal} {J. High Energy Phys.}\ }\textbf
  {\bibinfo {volume} {04}},\ \bibinfo {pages} {108 (2023)}},\ \Eprint
  {https://arxiv.org/abs/2209.15460} {arXiv:2209.15460 [hep-lat]} \BibitemShut
  {NoStop}%
\bibitem [{\citenamefont {Erben}\ \emph
  {et~al.}(2023{\natexlab{b}})\citenamefont {Erben}, \citenamefont {G\"ulpers},
  \citenamefont {Hansen}, \citenamefont {Hodgson},\ and\ \citenamefont
  {Portelli}}]{Erben:2022igb}%
  \BibitemOpen
  \bibfield  {author} {\bibinfo {author} {\bibfnamefont {F.}~\bibnamefont
  {Erben}}, \bibinfo {author} {\bibfnamefont {V.}~\bibnamefont {G\"ulpers}},
  \bibinfo {author} {\bibfnamefont {M.~T.}\ \bibnamefont {Hansen}}, \bibinfo
  {author} {\bibfnamefont {R.}~\bibnamefont {Hodgson}},\ and\ \bibinfo {author}
  {\bibfnamefont {A.}~\bibnamefont {Portelli}},\ }\bibfield  {title} {\bibinfo
  {title} {{Progress on the exploratory calculation of the rare Hyperon decay
  $\Sigma^+\to p\ell^+\ell^-$}},\ }\href {https://doi.org/10.22323/1.430.0315}
  {\bibfield  {journal} {\bibinfo  {journal} {PoS}\ }\textbf {\bibinfo {volume}
  {LATTICE2022}},\ \bibinfo {pages} {315} (\bibinfo {year}
  {2023}{\natexlab{b}})},\ \Eprint {https://arxiv.org/abs/2212.09595}
  {arXiv:2212.09595 [hep-lat]} \BibitemShut {NoStop}%
\bibitem [{\citenamefont {He}\ \emph {et~al.}(2018{\natexlab{b}})\citenamefont
  {He}, \citenamefont {Tandean},\ and\ \citenamefont {Valencia}}]{He:2018yzu}%
  \BibitemOpen
  \bibfield  {author} {\bibinfo {author} {\bibfnamefont {X.-G.}\ \bibnamefont
  {He}}, \bibinfo {author} {\bibfnamefont {J.}~\bibnamefont {Tandean}},\ and\
  \bibinfo {author} {\bibfnamefont {G.}~\bibnamefont {Valencia}},\ }\bibfield
  {title} {\bibinfo {title} {{Decay rate and asymmetries of $\Sigma^+\to
  p\mu^+\mu^-$}},\ }\href {https://doi.org/10.1007/JHEP10(2018)040} {\bibfield
  {journal} {\bibinfo  {journal} {J. High Energy Phys.}\ }\textbf {\bibinfo
  {volume} {10}},\ \bibinfo {pages} {040 (2018)}},\ \Eprint
  {https://arxiv.org/abs/1806.08350} {arXiv:1806.08350 [hep-ph]} \BibitemShut
  {NoStop}%
\bibitem [{\citenamefont {Aaij}\ \emph {et~al.}(2015)\citenamefont {Aaij} \emph
  {et~al.}}]{LHCb:2015tgy}%
  \BibitemOpen
  \bibfield  {author} {\bibinfo {author} {\bibfnamefont {R.}~\bibnamefont
  {Aaij}} \emph {et~al.} (\bibinfo {collaboration} {LHCb}),\ }\bibfield
  {title} {\bibinfo {title} {{Differential branching fraction and angular
  analysis of $\Lambda^{0}_{b} \to \Lambda \mu^+\mu^-$ decays}},\ }\href
  {https://doi.org/10.1007/JHEP06(2015)115} {\bibfield  {journal} {\bibinfo
  {journal} {J. High Energy Phys.}\ }\textbf {\bibinfo {volume} {06}},\
  \bibinfo {pages} {115 (2015)}},\ \bibinfo {note} {[Erratum: J. High Energy
  Phys. 09, 145 (2018)]},\ \Eprint {https://arxiv.org/abs/1503.07138}
  {arXiv:1503.07138 [hep-ex]} \BibitemShut {NoStop}%
\bibitem [{\citenamefont {LHCb}(2024)}]{LHCb:2024}%
  \BibitemOpen
  \bibfield  {author} {\bibinfo {author} {\bibnamefont {LHCb}},\ }\bibfield
  {title} {\bibinfo {title} {{Observation of the $\Sigma^+ \to p \mu^+ \mu^-$
  rare decay at LHCb, LHCb-CONF-2024-002}},\ }\href@noop {} {\  (\bibinfo
  {year} {2024})}\BibitemShut {NoStop}%
\bibitem [{\citenamefont {Isidori}\ \emph {et~al.}(2004)\citenamefont
  {Isidori}, \citenamefont {Smith},\ and\ \citenamefont
  {Unterdorfer}}]{Isidori:2004rb}%
  \BibitemOpen
  \bibfield  {author} {\bibinfo {author} {\bibfnamefont {G.}~\bibnamefont
  {Isidori}}, \bibinfo {author} {\bibfnamefont {C.}~\bibnamefont {Smith}},\
  and\ \bibinfo {author} {\bibfnamefont {R.}~\bibnamefont {Unterdorfer}},\
  }\bibfield  {title} {\bibinfo {title} {{The rare decay $K_{\rm L}\to \pi^0
  \mu^+ \mu^-$ within the SM}},\ }\href
  {https://doi.org/10.1140/epjc/s2004-01879-0} {\bibfield  {journal} {\bibinfo
  {journal} {Eur. Phys. J. C}\ }\textbf {\bibinfo {volume} {36}},\ \bibinfo
  {pages} {57} (\bibinfo {year} {2004})},\ \Eprint
  {https://arxiv.org/abs/hep-ph/0404127} {arXiv:hep-ph/0404127} \BibitemShut
  {NoStop}%
\bibitem [{\citenamefont {Mahmoudi}(2009)}]{Mahmoudi:2009zz}%
  \BibitemOpen
  \bibfield  {author} {\bibinfo {author} {\bibfnamefont {F.}~\bibnamefont
  {Mahmoudi}},\ }\bibfield  {title} {\bibinfo {title} {{SuperIso v3.0, flavor
  physics observables calculations: Extension to NMSSM}},\ }\href
  {https://doi.org/10.1016/j.cpc.2009.05.001} {\bibfield  {journal} {\bibinfo
  {journal} {Comput. Phys. Commun.}\ }\textbf {\bibinfo {volume} {180}},\
  \bibinfo {pages} {1718} (\bibinfo {year} {2009})}\BibitemShut {NoStop}%
\bibitem [{\citenamefont {Neshatpour}\ and\ \citenamefont
  {Mahmoudi}(2021)}]{Neshatpour:2021nbn}%
  \BibitemOpen
  \bibfield  {author} {\bibinfo {author} {\bibfnamefont {S.}~\bibnamefont
  {Neshatpour}}\ and\ \bibinfo {author} {\bibfnamefont {F.}~\bibnamefont
  {Mahmoudi}},\ }\bibfield  {title} {\bibinfo {title} {{Flavour Physics with
  SuperIso}},\ }\href {https://doi.org/10.22323/1.392.0036} {\bibfield
  {journal} {\bibinfo  {journal} {PoS}\ }\textbf {\bibinfo {volume}
  {TOOLS2020}},\ \bibinfo {pages} {036} (\bibinfo {year} {2021})},\ \Eprint
  {https://arxiv.org/abs/2105.03428} {arXiv:2105.03428 [hep-ph]} \BibitemShut
  {NoStop}%
\bibitem [{\citenamefont {Mertens}\ and\ \citenamefont
  {Smith}(2011)}]{Mertens:2011ts}%
  \BibitemOpen
  \bibfield  {author} {\bibinfo {author} {\bibfnamefont {P.}~\bibnamefont
  {Mertens}}\ and\ \bibinfo {author} {\bibfnamefont {C.}~\bibnamefont
  {Smith}},\ }\bibfield  {title} {\bibinfo {title} {{The $s\to d\gamma$ decay
  in and beyond the Standard Model}},\ }\href
  {https://doi.org/10.1007/JHEP08(2011)069} {\bibfield  {journal} {\bibinfo
  {journal} {J. High Energy Phys.}\ }\textbf {\bibinfo {volume} {08}},\
  \bibinfo {pages} {069 (2011)}},\ \Eprint {https://arxiv.org/abs/1103.5992}
  {arXiv:1103.5992 [hep-ph]} \BibitemShut {NoStop}%
\bibitem [{\citenamefont {Chobanova}\ \emph {et~al.}(2018)\citenamefont
  {Chobanova}, \citenamefont {D'Ambrosio}, \citenamefont {Kitahara},
  \citenamefont {Lucio~Martinez}, \citenamefont {Martinez~Santos},
  \citenamefont {Fernandez},\ and\ \citenamefont
  {Yamamoto}}]{Chobanova:2017rkj}%
  \BibitemOpen
  \bibfield  {author} {\bibinfo {author} {\bibfnamefont {V.}~\bibnamefont
  {Chobanova}}, \bibinfo {author} {\bibfnamefont {G.}~\bibnamefont
  {D'Ambrosio}}, \bibinfo {author} {\bibfnamefont {T.}~\bibnamefont
  {Kitahara}}, \bibinfo {author} {\bibfnamefont {M.}~\bibnamefont
  {Lucio~Martinez}}, \bibinfo {author} {\bibfnamefont {D.}~\bibnamefont
  {Martinez~Santos}}, \bibinfo {author} {\bibfnamefont {I.~S.}\ \bibnamefont
  {Fernandez}},\ and\ \bibinfo {author} {\bibfnamefont {K.}~\bibnamefont
  {Yamamoto}},\ }\bibfield  {title} {\bibinfo {title} {{Probing SUSY effects in
  $K_S^0\rightarrow\mu^+\mu^-$}},\ }\href
  {https://doi.org/10.1007/JHEP05(2018)024} {\bibfield  {journal} {\bibinfo
  {journal} {J. High Energy Phys.}\ }\textbf {\bibinfo {volume} {05}},\
  \bibinfo {pages} {024 (2018)}},\ \Eprint {https://arxiv.org/abs/1711.11030}
  {arXiv:1711.11030 [hep-ph]} \BibitemShut {NoStop}%
\bibitem [{\citenamefont {Gorbahn}\ and\ \citenamefont
  {Haisch}(2006)}]{Gorbahn:2006bm}%
  \BibitemOpen
  \bibfield  {author} {\bibinfo {author} {\bibfnamefont {M.}~\bibnamefont
  {Gorbahn}}\ and\ \bibinfo {author} {\bibfnamefont {U.}~\bibnamefont
  {Haisch}},\ }\bibfield  {title} {\bibinfo {title} {{Charm Quark Contribution
  to $K_L\to\mu^+\mu^-$ at Next-to-Next-to-Leading Order}},\ }\href
  {https://doi.org/10.1103/PhysRevLett.97.122002} {\bibfield  {journal}
  {\bibinfo  {journal} {Phys. Rev. Lett.}\ }\textbf {\bibinfo {volume} {97}},\
  \bibinfo {pages} {122002} (\bibinfo {year} {2006})},\ \Eprint
  {https://arxiv.org/abs/hep-ph/0605203} {arXiv:hep-ph/0605203} \BibitemShut
  {NoStop}%
\bibitem [{\citenamefont {Buchalla}\ \emph {et~al.}(2003)\citenamefont
  {Buchalla}, \citenamefont {D'Ambrosio},\ and\ \citenamefont
  {Isidori}}]{Buchalla:2003sj}%
  \BibitemOpen
  \bibfield  {author} {\bibinfo {author} {\bibfnamefont {G.}~\bibnamefont
  {Buchalla}}, \bibinfo {author} {\bibfnamefont {G.}~\bibnamefont
  {D'Ambrosio}},\ and\ \bibinfo {author} {\bibfnamefont {G.}~\bibnamefont
  {Isidori}},\ }\bibfield  {title} {\bibinfo {title} {{Extracting short
  distance physics from $K_{L,S}\to\pi^0 e^+ e^-$ decays}},\ }\href
  {https://doi.org/10.1016/j.nuclphysb.2003.09.010} {\bibfield  {journal}
  {\bibinfo  {journal} {Nucl. Phys. B}\ }\textbf {\bibinfo {volume} {672}},\
  \bibinfo {pages} {387} (\bibinfo {year} {2003})},\ \Eprint
  {https://arxiv.org/abs/hep-ph/0308008} {arXiv:hep-ph/0308008} \BibitemShut
  {NoStop}%
\bibitem [{\citenamefont {Mescia}\ \emph {et~al.}(2006)\citenamefont {Mescia},
  \citenamefont {Smith},\ and\ \citenamefont {Trine}}]{Mescia:2006jd}%
  \BibitemOpen
  \bibfield  {author} {\bibinfo {author} {\bibfnamefont {F.}~\bibnamefont
  {Mescia}}, \bibinfo {author} {\bibfnamefont {C.}~\bibnamefont {Smith}},\ and\
  \bibinfo {author} {\bibfnamefont {S.}~\bibnamefont {Trine}},\ }\bibfield
  {title} {\bibinfo {title} {{$K_L\to\pi^0e^+e^-$ and $K_L\to\pi^0\mu^+\mu^-$:
  a binary star on the stage of flavor physics}},\ }\href
  {https://doi.org/10.1088/1126-6708/2006/08/088} {\bibfield  {journal}
  {\bibinfo  {journal} {J. High Energy Phys.}\ }\textbf {\bibinfo {volume}
  {08}},\ \bibinfo {pages} {088 (2006)}},\ \Eprint
  {https://arxiv.org/abs/hep-ph/0606081} {arXiv:hep-ph/0606081} \BibitemShut
  {NoStop}%
\bibitem [{\citenamefont {D'Ambrosio}\ \emph {et~al.}(2022)\citenamefont
  {D'Ambrosio}, \citenamefont {Iyer}, \citenamefont {Mahmoudi},\ and\
  \citenamefont {Neshatpour}}]{DAmbrosio:2022kvb}%
  \BibitemOpen
  \bibfield  {author} {\bibinfo {author} {\bibfnamefont {G.}~\bibnamefont
  {D'Ambrosio}}, \bibinfo {author} {\bibfnamefont {A.~M.}\ \bibnamefont
  {Iyer}}, \bibinfo {author} {\bibfnamefont {F.}~\bibnamefont {Mahmoudi}},\
  and\ \bibinfo {author} {\bibfnamefont {S.}~\bibnamefont {Neshatpour}},\
  }\bibfield  {title} {\bibinfo {title} {{Anatomy of kaon decays and prospects
  for lepton flavour universality violation}},\ }\href
  {https://doi.org/10.1007/JHEP09(2022)148} {\bibfield  {journal} {\bibinfo
  {journal} {J. High Energy Phys.}\ }\textbf {\bibinfo {volume} {09}},\
  \bibinfo {pages} {148 (2022)}},\ \Eprint {https://arxiv.org/abs/2206.14748}
  {arXiv:2206.14748 [hep-ph]} \BibitemShut {NoStop}%
\bibitem [{\citenamefont {Ecker}\ \emph {et~al.}(1987)\citenamefont {Ecker},
  \citenamefont {Pich},\ and\ \citenamefont {de~Rafael}}]{Ecker:1987qi}%
  \BibitemOpen
  \bibfield  {author} {\bibinfo {author} {\bibfnamefont {G.}~\bibnamefont
  {Ecker}}, \bibinfo {author} {\bibfnamefont {A.}~\bibnamefont {Pich}},\ and\
  \bibinfo {author} {\bibfnamefont {E.}~\bibnamefont {de~Rafael}},\ }\bibfield
  {title} {\bibinfo {title} {{$K\to\pi\ell^+\ell^-$ decays in the effective
  chiral lagrangian of the standard model}},\ }\href
  {https://doi.org/10.1016/0550-3213(87)90491-3} {\bibfield  {journal}
  {\bibinfo  {journal} {Nucl. Phys. B}\ }\textbf {\bibinfo {volume} {291}},\
  \bibinfo {pages} {692} (\bibinfo {year} {1987})}\BibitemShut {NoStop}%
\bibitem [{\citenamefont {D'Ambrosio}\ \emph {et~al.}(1998)\citenamefont
  {D'Ambrosio}, \citenamefont {Ecker}, \citenamefont {Isidori},\ and\
  \citenamefont {Portoles}}]{DAmbrosio:1998gur}%
  \BibitemOpen
  \bibfield  {author} {\bibinfo {author} {\bibfnamefont {G.}~\bibnamefont
  {D'Ambrosio}}, \bibinfo {author} {\bibfnamefont {G.}~\bibnamefont {Ecker}},
  \bibinfo {author} {\bibfnamefont {G.}~\bibnamefont {Isidori}},\ and\ \bibinfo
  {author} {\bibfnamefont {J.}~\bibnamefont {Portoles}},\ }\bibfield  {title}
  {\bibinfo {title} {{The decays $K\to\pi\ell^+\ell^-$ beyond leading order in
  the chiral expansion}},\ }\href
  {https://doi.org/10.1088/1126-6708/1998/08/004} {\bibfield  {journal}
  {\bibinfo  {journal} {J. High Energy Phys.}\ }\textbf {\bibinfo {volume}
  {08}},\ \bibinfo {pages} {004 (1998)}},\ \Eprint
  {https://arxiv.org/abs/hep-ph/9808289} {arXiv:hep-ph/9808289} \BibitemShut
  {NoStop}%
\bibitem [{\citenamefont {D'Ambrosio}\ \emph {et~al.}(2019)\citenamefont
  {D'Ambrosio}, \citenamefont {Greynat},\ and\ \citenamefont
  {Knecht}}]{DAmbrosio:2018ytt}%
  \BibitemOpen
  \bibfield  {author} {\bibinfo {author} {\bibfnamefont {G.}~\bibnamefont
  {D'Ambrosio}}, \bibinfo {author} {\bibfnamefont {D.}~\bibnamefont
  {Greynat}},\ and\ \bibinfo {author} {\bibfnamefont {M.}~\bibnamefont
  {Knecht}},\ }\bibfield  {title} {\bibinfo {title} {{On the amplitudes for the
  CP-conserving $K^\pm(K_S)\to\pi^\pm(\pi^0)\ell^+\ell^-$ rare decay modes}},\
  }\href {https://doi.org/10.1007/JHEP02(2019)049} {\bibfield  {journal}
  {\bibinfo  {journal} {J. High Energy Phys.}\ }\textbf {\bibinfo {volume}
  {02}},\ \bibinfo {pages} {049 (2019)}},\ \Eprint
  {https://arxiv.org/abs/1812.00735} {arXiv:1812.00735 [hep-ph]} \BibitemShut
  {NoStop}%
\bibitem [{\citenamefont {Cirigliano}\ \emph {et~al.}(2012)\citenamefont
  {Cirigliano}, \citenamefont {Ecker}, \citenamefont {Neufeld}, \citenamefont
  {Pich},\ and\ \citenamefont {Portoles}}]{Cirigliano:2011ny}%
  \BibitemOpen
  \bibfield  {author} {\bibinfo {author} {\bibfnamefont {V.}~\bibnamefont
  {Cirigliano}}, \bibinfo {author} {\bibfnamefont {G.}~\bibnamefont {Ecker}},
  \bibinfo {author} {\bibfnamefont {H.}~\bibnamefont {Neufeld}}, \bibinfo
  {author} {\bibfnamefont {A.}~\bibnamefont {Pich}},\ and\ \bibinfo {author}
  {\bibfnamefont {J.}~\bibnamefont {Portoles}},\ }\bibfield  {title} {\bibinfo
  {title} {{Kaon Decays in the Standard Model}},\ }\href
  {https://doi.org/10.1103/RevModPhys.84.399} {\bibfield  {journal} {\bibinfo
  {journal} {Rev. Mod. Phys.}\ }\textbf {\bibinfo {volume} {84}},\ \bibinfo
  {pages} {399} (\bibinfo {year} {2012})},\ \Eprint
  {https://arxiv.org/abs/1107.6001} {arXiv:1107.6001 [hep-ph]} \BibitemShut
  {NoStop}%
\bibitem [{\citenamefont {Batley}\ \emph {et~al.}(2003)\citenamefont {Batley}
  \emph {et~al.}}]{Batley:2003mu}%
  \BibitemOpen
  \bibfield  {author} {\bibinfo {author} {\bibfnamefont {J.~R.}\ \bibnamefont
  {Batley}} \emph {et~al.} (\bibinfo {collaboration} {NA48/1}),\ }\bibfield
  {title} {\bibinfo {title} {{Observation of the rare decay $K_S \to\pi^0 e^+
  e^-$}},\ }\href {https://doi.org/10.1016/j.physletb.2003.10.001} {\bibfield
  {journal} {\bibinfo  {journal} {Phys. Lett.}\ }\textbf {\bibinfo {volume}
  {B576}},\ \bibinfo {pages} {43} (\bibinfo {year} {2003})},\ \Eprint
  {https://arxiv.org/abs/hep-ex/0309075} {arXiv:hep-ex/0309075 [hep-ex]}
  \BibitemShut {NoStop}%
\bibitem [{\citenamefont {Batley}\ \emph {et~al.}(2004)\citenamefont {Batley}
  \emph {et~al.}}]{Batley:2004wg}%
  \BibitemOpen
  \bibfield  {author} {\bibinfo {author} {\bibfnamefont {J.~R.}\ \bibnamefont
  {Batley}} \emph {et~al.} (\bibinfo {collaboration} {NA48/1}),\ }\bibfield
  {title} {\bibinfo {title} {{Observation of the rare decay $K_S\to \pi^0 \mu^+
  \mu^-$}},\ }\href {https://doi.org/10.1016/j.physletb.2004.08.058} {\bibfield
   {journal} {\bibinfo  {journal} {Phys. Lett.}\ }\textbf {\bibinfo {volume}
  {B599}},\ \bibinfo {pages} {197} (\bibinfo {year} {2004})},\ \Eprint
  {https://arxiv.org/abs/hep-ex/0409011} {arXiv:hep-ex/0409011 [hep-ex]}
  \BibitemShut {NoStop}%
\bibitem [{\citenamefont {Batley}\ \emph {et~al.}(2009)\citenamefont {Batley}
  \emph {et~al.}}]{NA482:2009pfe}%
  \BibitemOpen
  \bibfield  {author} {\bibinfo {author} {\bibfnamefont {J.~R.}\ \bibnamefont
  {Batley}} \emph {et~al.} (\bibinfo {collaboration} {NA48/2}),\ }\bibfield
  {title} {\bibinfo {title} {{Precise measurement of the $K^\pm\to\pi^\pm
  e^+e^-$ decay}},\ }\href {https://doi.org/10.1016/j.physletb.2009.05.040}
  {\bibfield  {journal} {\bibinfo  {journal} {Phys. Lett. B}\ }\textbf
  {\bibinfo {volume} {677}},\ \bibinfo {pages} {246} (\bibinfo {year}
  {2009})},\ \Eprint {https://arxiv.org/abs/0903.3130} {arXiv:0903.3130
  [hep-ex]} \BibitemShut {NoStop}%
\bibitem [{\citenamefont {Cortina~Gil}\ \emph {et~al.}(2022)\citenamefont
  {Cortina~Gil} \emph {et~al.}}]{NA62:2022qes}%
  \BibitemOpen
  \bibfield  {author} {\bibinfo {author} {\bibfnamefont {E.}~\bibnamefont
  {Cortina~Gil}} \emph {et~al.} (\bibinfo {collaboration} {NA62}),\ }\bibfield
  {title} {\bibinfo {title} {{A measurement of the $K^{+} \to \pi^{+} \mu^{+}
  \mu^{-}$ decay}},\ }\href {https://doi.org/10.1007/JHEP11(2022)011}
  {\bibfield  {journal} {\bibinfo  {journal} {J. High Energy Phys.}\ }\textbf
  {\bibinfo {volume} {11}},\ \bibinfo {pages} {011 (2022)}},\ \bibinfo {note}
  {[Addendum: J. High Energy Phys. 06, 40 (2023)]},\ \Eprint
  {https://arxiv.org/abs/2209.05076} {arXiv:2209.05076 [hep-ex]} \BibitemShut
  {NoStop}%
\bibitem [{\citenamefont {Chen}\ \emph {et~al.}(2003)\citenamefont {Chen},
  \citenamefont {Geng},\ and\ \citenamefont {Ho}}]{Chen:2003nz}%
  \BibitemOpen
  \bibfield  {author} {\bibinfo {author} {\bibfnamefont {C.-H.}\ \bibnamefont
  {Chen}}, \bibinfo {author} {\bibfnamefont {C.~Q.}\ \bibnamefont {Geng}},\
  and\ \bibinfo {author} {\bibfnamefont {I.-L.}\ \bibnamefont {Ho}},\
  }\bibfield  {title} {\bibinfo {title} {{Forward backward asymmetry in
  $K^+\to\pi^+l^+l^-$}},\ }\href {https://doi.org/10.1103/PhysRevD.67.074029}
  {\bibfield  {journal} {\bibinfo  {journal} {Phys. Rev. D}\ }\textbf {\bibinfo
  {volume} {67}},\ \bibinfo {pages} {074029} (\bibinfo {year} {2003})},\
  \Eprint {https://arxiv.org/abs/hep-ph/0302207} {arXiv:hep-ph/0302207}
  \BibitemShut {NoStop}%
\bibitem [{\citenamefont {Crivellin}\ \emph {et~al.}(2016)\citenamefont
  {Crivellin}, \citenamefont {D'Ambrosio}, \citenamefont {Hoferichter},\ and\
  \citenamefont {Tunstall}}]{Crivellin:2016vjc}%
  \BibitemOpen
  \bibfield  {author} {\bibinfo {author} {\bibfnamefont {A.}~\bibnamefont
  {Crivellin}}, \bibinfo {author} {\bibfnamefont {G.}~\bibnamefont
  {D'Ambrosio}}, \bibinfo {author} {\bibfnamefont {M.}~\bibnamefont
  {Hoferichter}},\ and\ \bibinfo {author} {\bibfnamefont {L.~C.}\ \bibnamefont
  {Tunstall}},\ }\bibfield  {title} {\bibinfo {title} {{Violation of lepton
  flavor and lepton flavor universality in rare kaon decays}},\ }\href
  {https://doi.org/10.1103/PhysRevD.93.074038} {\bibfield  {journal} {\bibinfo
  {journal} {Phys. Rev. D}\ }\textbf {\bibinfo {volume} {93}},\ \bibinfo
  {pages} {074038} (\bibinfo {year} {2016})},\ \Eprint
  {https://arxiv.org/abs/1601.009701.166} {arXiv:1601.009701.166 [hep-ph]}
  \BibitemShut {NoStop}%
\bibitem [{\citenamefont {Aaij}\ \emph {et~al.}(2020)\citenamefont {Aaij} \emph
  {et~al.}}]{LHCb:2020ycd}%
  \BibitemOpen
  \bibfield  {author} {\bibinfo {author} {\bibfnamefont {R.}~\bibnamefont
  {Aaij}} \emph {et~al.} (\bibinfo {collaboration} {LHCb}),\ }\bibfield
  {title} {\bibinfo {title} {{Constraints on the $K^0_S \rightarrow \mu^+
  \mu^-$ Branching Fraction}},\ }\href
  {https://doi.org/10.1103/PhysRevLett.125.231801} {\bibfield  {journal}
  {\bibinfo  {journal} {Phys. Rev. Lett.}\ }\textbf {\bibinfo {volume} {125}},\
  \bibinfo {pages} {231801} (\bibinfo {year} {2020})},\ \Eprint
  {https://arxiv.org/abs/2001.10354} {arXiv:2001.10354 [hep-ex]} \BibitemShut
  {NoStop}%
\bibitem [{\citenamefont {Alavi-Harati}\ \emph {et~al.}(2000)\citenamefont
  {Alavi-Harati} \emph {et~al.}}]{AlaviHarati:2000hs}%
  \BibitemOpen
  \bibfield  {author} {\bibinfo {author} {\bibfnamefont {A.}~\bibnamefont
  {Alavi-Harati}} \emph {et~al.} (\bibinfo {collaboration} {KTEV}),\ }\bibfield
   {title} {\bibinfo {title} {{Search for the Decay $K_L \to \pi^0 \mu^+
  \mu^-$}},\ }\href {https://doi.org/10.1103/PhysRevLett.84.5279} {\bibfield
  {journal} {\bibinfo  {journal} {Phys. Rev. Lett.}\ }\textbf {\bibinfo
  {volume} {84}},\ \bibinfo {pages} {5279} (\bibinfo {year} {2000})},\ \Eprint
  {https://arxiv.org/abs/hep-ex/0001006} {arXiv:hep-ex/0001006 [hep-ex]}
  \BibitemShut {NoStop}%
\bibitem [{\citenamefont {Alavi-Harati}\ \emph {et~al.}(2004)\citenamefont
  {Alavi-Harati} \emph {et~al.}}]{AlaviHarati:2003mr}%
  \BibitemOpen
  \bibfield  {author} {\bibinfo {author} {\bibfnamefont {A.}~\bibnamefont
  {Alavi-Harati}} \emph {et~al.} (\bibinfo {collaboration} {KTeV}),\ }\bibfield
   {title} {\bibinfo {title} {{Search for the Rare Decay $K_L\to\pi^0 e^+
  e^-$}},\ }\href {https://doi.org/10.1103/PhysRevLett.93.021805} {\bibfield
  {journal} {\bibinfo  {journal} {Phys. Rev. Lett.}\ }\textbf {\bibinfo
  {volume} {93}},\ \bibinfo {pages} {021805} (\bibinfo {year} {2004})},\
  \Eprint {https://arxiv.org/abs/hep-ex/0309072} {arXiv:hep-ex/0309072
  [hep-ex]} \BibitemShut {NoStop}%
\bibitem [{\citenamefont {He}\ and\ \citenamefont
  {Valencia}(2000)}]{He:1999ik}%
  \BibitemOpen
  \bibfield  {author} {\bibinfo {author} {\bibfnamefont {X.-G.}\ \bibnamefont
  {He}}\ and\ \bibinfo {author} {\bibfnamefont {G.}~\bibnamefont {Valencia}},\
  }\bibfield  {title} {\bibinfo {title} {{Constraints on $s\to d\gamma$ from
  radiative hyperon and kaon decays}},\ }\href
  {https://doi.org/10.1103/PhysRevD.61.075003} {\bibfield  {journal} {\bibinfo
  {journal} {Phys. Rev. D}\ }\textbf {\bibinfo {volume} {61}},\ \bibinfo
  {pages} {075003} (\bibinfo {year} {2000})},\ \Eprint
  {https://arxiv.org/abs/hep-ph/9908298} {arXiv:hep-ph/9908298} \BibitemShut
  {NoStop}%
\bibitem [{\citenamefont {Buras}\ \emph {et~al.}(2008)\citenamefont {Buras},
  \citenamefont {Schwab},\ and\ \citenamefont {Uhlig}}]{Buras:2004uu}%
  \BibitemOpen
  \bibfield  {author} {\bibinfo {author} {\bibfnamefont {A.~J.}\ \bibnamefont
  {Buras}}, \bibinfo {author} {\bibfnamefont {F.}~\bibnamefont {Schwab}},\ and\
  \bibinfo {author} {\bibfnamefont {S.}~\bibnamefont {Uhlig}},\ }\bibfield
  {title} {\bibinfo {title} {{Waiting for precise measurements of $K^{+} \to
  \pi^{+} \nu \bar{\nu}$ and $K_{L} \to \pi^0 \nu \bar{\nu}$}},\ }\href
  {https://doi.org/10.1103/RevModPhys.80.965} {\bibfield  {journal} {\bibinfo
  {journal} {Rev. Mod. Phys.}\ }\textbf {\bibinfo {volume} {80}},\ \bibinfo
  {pages} {965} (\bibinfo {year} {2008})},\ \Eprint
  {https://arxiv.org/abs/hep-ph/0405132} {arXiv:hep-ph/0405132} \BibitemShut
  {NoStop}%
\bibitem [{\citenamefont {D'Ambrosio}\ and\ \citenamefont
  {Isidori}(2002)}]{DAmbrosio:2001kut}%
  \BibitemOpen
  \bibfield  {author} {\bibinfo {author} {\bibfnamefont {G.}~\bibnamefont
  {D'Ambrosio}}\ and\ \bibinfo {author} {\bibfnamefont {G.}~\bibnamefont
  {Isidori}},\ }\bibfield  {title} {\bibinfo {title} {{$K^+\to\pi^+\nu\bar\nu$:
  a rising star on the stage of flavour physics}},\ }\href
  {https://doi.org/10.1016/S0370-2693(02)01328-X} {\bibfield  {journal}
  {\bibinfo  {journal} {Phys. Lett. B}\ }\textbf {\bibinfo {volume} {530}},\
  \bibinfo {pages} {108} (\bibinfo {year} {2002})},\ \Eprint
  {https://arxiv.org/abs/hep-ph/0112135} {arXiv:hep-ph/0112135} \BibitemShut
  {NoStop}%
\bibitem [{\citenamefont {Buras}\ and\ \citenamefont
  {Girrbach}(2014)}]{Buras:2013ooa}%
  \BibitemOpen
  \bibfield  {author} {\bibinfo {author} {\bibfnamefont {A.~J.}\ \bibnamefont
  {Buras}}\ and\ \bibinfo {author} {\bibfnamefont {J.}~\bibnamefont
  {Girrbach}},\ }\bibfield  {title} {\bibinfo {title} {{Towards the
  identification of new physics through quark flavour violating processes}},\
  }\href {https://doi.org/10.1088/0034-4885/77/8/086201} {\bibfield  {journal}
  {\bibinfo  {journal} {Rept. Prog. Phys.}\ }\textbf {\bibinfo {volume} {77}},\
  \bibinfo {pages} {086201} (\bibinfo {year} {2014})},\ \Eprint
  {https://arxiv.org/abs/1306.3775} {arXiv:1306.3775 [hep-ph]} \BibitemShut
  {NoStop}%
\bibitem [{\citenamefont {Buras}\ \emph {et~al.}(2015)\citenamefont {Buras},
  \citenamefont {Buttazzo},\ and\ \citenamefont {Knegjens}}]{Buras:2015yca}%
  \BibitemOpen
  \bibfield  {author} {\bibinfo {author} {\bibfnamefont {A.~J.}\ \bibnamefont
  {Buras}}, \bibinfo {author} {\bibfnamefont {D.}~\bibnamefont {Buttazzo}},\
  and\ \bibinfo {author} {\bibfnamefont {R.}~\bibnamefont {Knegjens}},\
  }\bibfield  {title} {\bibinfo {title} {{$K\to\pi\nu\overline\nu$ and
  $\varepsilon'/\varepsilon$ in simplified new physics models}},\ }\href
  {https://doi.org/10.1007/JHEP11(2015)166} {\bibfield  {journal} {\bibinfo
  {journal} {J. High Energy Phys.}\ }\textbf {\bibinfo {volume} {11}},\
  \bibinfo {pages} {166 (2015)}},\ \Eprint {https://arxiv.org/abs/1507.08672}
  {arXiv:1507.08672 [hep-ph]} \BibitemShut {NoStop}%
\bibitem [{\citenamefont {Fajfer}\ \emph {et~al.}(2018)\citenamefont {Fajfer},
  \citenamefont {Ko\v{s}nik},\ and\ \citenamefont
  {Vale~Silva}}]{Fajfer:2018bfj}%
  \BibitemOpen
  \bibfield  {author} {\bibinfo {author} {\bibfnamefont {S.}~\bibnamefont
  {Fajfer}}, \bibinfo {author} {\bibfnamefont {N.}~\bibnamefont {Ko\v{s}nik}},\
  and\ \bibinfo {author} {\bibfnamefont {L.}~\bibnamefont {Vale~Silva}},\
  }\bibfield  {title} {\bibinfo {title} {{Footprints of leptoquarks: from $
  R_{K^{(*)}} $ to $K\to\pi\nu\bar\nu$}},\ }\href
  {https://doi.org/10.1140/epjc/s10052-018-5757-5} {\bibfield  {journal}
  {\bibinfo  {journal} {Eur. Phys. J. C}\ }\textbf {\bibinfo {volume} {78}},\
  \bibinfo {pages} {275} (\bibinfo {year} {2018})},\ \Eprint
  {https://arxiv.org/abs/1802.00786} {arXiv:1802.00786 [hep-ph]} \BibitemShut
  {NoStop}%
\bibitem [{\citenamefont {Aebischer}\ \emph {et~al.}(2020)\citenamefont
  {Aebischer}, \citenamefont {Buras},\ and\ \citenamefont
  {Kumar}}]{Aebischer:2020mkv}%
  \BibitemOpen
  \bibfield  {author} {\bibinfo {author} {\bibfnamefont {J.}~\bibnamefont
  {Aebischer}}, \bibinfo {author} {\bibfnamefont {A.~J.}\ \bibnamefont
  {Buras}},\ and\ \bibinfo {author} {\bibfnamefont {J.}~\bibnamefont {Kumar}},\
  }\bibfield  {title} {\bibinfo {title} {{Another SMEFT story: $Z^\prime$
  facing new results on $\epsilon'/\epsilon$, $\Delta M_{K}$ and
  $K\to\pi\nu\overline{\nu} $}},\ }\href
  {https://doi.org/10.1007/JHEP12(2020)097} {\bibfield  {journal} {\bibinfo
  {journal} {J. High Energy Phys.}\ }\textbf {\bibinfo {volume} {12}},\
  \bibinfo {pages} {097 (2020)}},\ \Eprint {https://arxiv.org/abs/2006.01138}
  {arXiv:2006.01138 [hep-ph]} \BibitemShut {NoStop}%
\end{thebibliography}%

\end{document}